\documentclass[prd,twocolumn,showpacs,nofootinbib,floats]{revtex4-1}

\usepackage{graphicx}
\usepackage{amsmath,amsfonts}
\usepackage{dcolumn}
\usepackage{longtable}
\usepackage[caption=false]{subfig}

\mathchardef\minus = "002D
\newcommand{\swY}[4][]{{}_{{}_{#2}}\!Y^{#1}_{#3}(#4)}
\newcommand{\swSH}[5][]{{}_{{}_{#2}}S^{#1}_{#3}(#4;#5)}
\newcommand{\swS}[5][]{{}_{{}_{#2}}S^{#1}_{#3}(#4;#5)}
\newcommand{\scA}[4][]{{}_{{}_{#2}}A^{#1}_{#3}(#4)}

\newtheorem{theorem}{Theorem}

\newcolumntype{f}[1]{D{.}{.}{#1}}
\newcommand{\Chead}[1]{\multicolumn{1}{c}{#1}}
\newcommand{\Cempty}{\multicolumn{1}{c}{---}}

\begin{document}

\title{Gravitational perturbations of the {K}err geometry: 
  {H}igh-accuracy study}

\author{Gregory B. Cook}\email{cookgb@wfu.edu}
\affiliation{Department of Physics, Wake Forest University,
		 Winston-Salem, North Carolina 27109}
\author{Maxim Zalutskiy}\email{zalump8@wfu.edu}
\affiliation{Department of Physics, Wake Forest University,
		 Winston-Salem, North Carolina 27109}

\date{\today}

\begin{abstract}
We present results from a new code for computing gravitational
perturbations of the Kerr geometry.  This new code carefully maintains
high precision to allow us to obtain high-accuracy solutions for the
gravitational quasinormal modes of the Kerr space-time.  Part of this
new code is an implementation of a spectral method for solving the
angular Teukolsky equation that, to our knowledge, has not been used
before for determining quasinormal modes.  We focus our attention on
two main areas.  First, we explore the behavior of these quasinormal
modes in the extreme limit of Kerr, where the frequency of certain
modes approaches accumulation points on the real axis.  We compare our
results with recent analytic predictions of the behavior of these
modes near the accumulation points and find good agreement.  Second,
we explore the behavior of solutions of modes that approach the
special frequency $M\omega=-2i$ in the Schwarzschild limit.  Our
high-accuracy methods allow us to more closely approach the
Schwarzschild limit than was possible with previous numerical studies.
Unlike previous work, we find excellent agreement with analytic
predictions of the behavior near this special frequency.  We include a
detailed description of our methods, and make use of the theory of
confluent Heun differential equations throughout.  In particular, we
make use of confluent Heun polynomials to help shed some light on the
controversy of the existence, or not, of quasinormal and
total-transmission modes at certain special frequencies in the
Schwarzschild limit.
\end{abstract}


\maketitle

\section{Introduction}
\label{sec:introduction}

Perturbations of the Kerr geometry have received considerable
attention.  They are important because astrophysical black holes are
expected to have some angular momentum $J$, perhaps even approaching
the Kerr limit where $a\equiv J/M\to M$.  Perturbed black holes will
ring down to a quiescent state by emitting gravitational waves that
may be observed in current or future gravitational wave observatories.
Extremal black holes, those where $a\to M$, are also of interest for
quantum field theory.  Because of these, and other reasons,
perturbations of Kerr continue to receive extensive study.

Of primary interest to this paper are the Quasinormal Modes (QNMs) of
the Kerr space-time (see
Refs.~\cite{berti-QNM-2009,nollert-qnm-1999} for excellent reviews of
black-hole QNMs).  These are the physical modes of a Kerr black hole
satisfying the conditions that waves are not allowed to propagate out
from the black-hole event horizon or in from infinity.  These modes
have been explored using various numerical
techniques\cite{chandra75,piranstark86,leaver-1985,leaver-1986,nollert-1993,onozawa-1997},
and analytically using, for example, WKB
methods\cite{Mashhoon-1983,ValeriaMashhoon-1084,SchutzWill-1985,Yang-et-al-2012}
and in the eikonal limit\cite{Dolan-2010,Hod-2012}.  The standard
method for solving these equations numerically is via a
continued-fraction technique referred to, in the relativity community,
as Leaver's\cite{leaver-1985} method.  This approach is applied to the
coupled radial and angular Teukolsky equations\cite{teukolsky-1973}
that result from a separation of variables procedure applied to the
equations describing linear perturbations of fields propagating in the
Kerr geometry.  We have implemented an independent, high-precision
version of Leaver's method for solving the radial equation, but have
employed a spectral-type method for solution of the angular equation.
Using these tools, we have performed an extensive, high-accuracy
survey of the gravitational perturbation modes of Kerr.  We have paid
particular attention to the Kerr limit for the so-called zero damped
modes (ZDMs) and to the Schwarzschild limit ($a\to0$) for modes that
approach the negative imaginary axis (NIA).

It is well known that, in the Kerr limit, some modes of Kerr approach
the real axis\cite{detweiler80,cardoso04}.  As the complex frequency
$\omega$ of a mode approaches the real axis, the vanishing of the
imaginary part of the frequency implies that the damping time of the
mode diverges, hence the label ZDM for these modes.  Yang {\em et
  al}\cite{Yang-et-al-2013a,Yang-et-al-2013b} have recently performed a
WKB analysis of these modes.  Below we will compare our results with
their predictions.

In addition to the QNMs of Kerr, there are also classes of modes known
as total transmission modes (TTMs).  These correspond to cases where
the boundary conditions at the event horizon and at infinity are such
as to allow modes to travel only from the left or from the right.
Often referred to as {\em algebraically special} modes because of a
particular set of solutions\cite{chandra-1984}, the two cases are
distinguished as follows.  If perturbations are allowed to flow out at
infinity, but only modes flowing {\em out} of the black hole are
allowed at the horizon, then the mode is referred to as TTM-left
(TTM${}_{\rm L}$) since the modes are traveling from the left
(assuming a picture where the black hole is to our left and spatial
infinity to our right).  When these boundary conditions are reversed,
the mode is TTM-right (TTM${}_{\rm R}$) since the modes are traveling
from the right.  In the context of QNMs these modes are important
because there seems to be a correlation with the behavior of the QNMs
at the NIA.

In the Schwarzschild limit ($a\to0$), the TTM frequencies become
purely imaginary.  These modes also seem to have some correlation to
certain overtones of the QNMs.  It appears that certain QNMs approach
these special frequencies in the Schwarzschild limit; however, it is
not possible to use existing numerical techniques to evaluate the mode
equations at $a=0$ in these cases.  Thus, there has been some
controversy over the existence of QNMs (and the TTM${}_{\rm L}$s) at
these special
frequencies\cite{andersson-1994,onozawa-1997,van_den_brink-2000}.  We
have performed high-accuracy surveys in the neighborhood of the NIA to
further clarify the behavior of QNMs in this regime.

It is known (see Ref.~\cite{Fiziev-2010} and references within), but
historically somewhat underappreciated in the physics literature,
that both the radial and angular Teukolsky equations are examples of
the {\em confluent Heun equation}\cite{Heun-eqn}.  In this paper, we
will treat the mode equations as confluent Heun equations, making use
of techniques for solving these equations when it is useful.  In
Sec.~\ref{sec:Solveing_Teukolsky_eqns}, we present the Teukolsky
equations and a brief introduction to the confluent Heun equations.
We will present enough of the theory of the Heun equations so that
readers can follow our discussions.  Additional details and references
can be found in Ref.~\cite{Heun-eqn}.  We will next transform each
mode equation into an appropriate standard form of the confluent Heun
equation.  These transformations are not new, and our work closely
follows that of Fiziev\cite{Fiziev-2010,Fiziev-2009b}.  For the radial
equation, we give a detailed outline of the theory of its solution via
Leaver's method.  Again, this work is not new (see
Refs.\cite{leaver-1986,nollert-1993}) but we present this in the
context of the confluent Heun equation for completeness.  For the
angular mode equation, we implement and test a highly efficient
spectral method\cite{hughes-2000} for finding both the eigenvalues and
eigenfunctions.

In Sec.~\ref{sec:Kerr_QNMs}, we give details of our numerical methods
for solving the mode equations and present a small set of our results
for comparison with prior solutions.  While no novel behavior is seen
in any of our solutions, they have been found to unprecedented
accuracy.  Our solutions have been found using at least 24 decimal
digits of precision and our numerical methods have used an absolute
tolerance of $10^{-12}$ conservatively giving us confidence in our
results to at least $\pm10^{-10}$.  We compare our highly accurate data
sets with the expected
behavior\cite{Yang-et-al-2013a,Yang-et-al-2013b,Hod08} for ZDMs, and
we reexamine the behavior of QNMs near the algebraically special
solution at $M\omega=-2i$.  In the latter case, prior numerical
work\cite{berticardoso-2003} has not found agreement with predicted
behavior\cite{van_den_brink-2000}.  However, our high-accuracy results
do find good agreement between theory and numerical solutions.  Next,
we rederive the algebraically special perturbations of Kerr (the TTMs)
as examples of finding polynomial solutions of the confluent Heun
equation.  The fact that some Heun polynomial solutions of the radial
equation correspond to the TTMs is not new\cite{Fiziev-2009c}, but we
feel it is important to include these examples for completeness.
Finally, we make use of these Heun polynomial solutions for the TTMs
in the limit that $a=0$ to reexamine the existence of QNMs and
TTM${}_{\rm R}$s at special frequencies on the NIA.

\section{Solving The Teukolsky equations}
\label{sec:Solveing_Teukolsky_eqns}

\subsection{The Teukolsky Equations}
\label{sec:Teukolsky_eqns}
In terms of Boyer-Lindquist coordinates, the line element of the Kerr
metric is given by
\begin{subequations} \label{eq:Kerr_Metric}
\begin{align}
  {\rm d}s^2 &= -\left(1-\frac{2Mr}{\Sigma}\right)\rm{d}t^2
\nonumber\\ &\mbox{}\hspace{0.2in}
       -\frac{4Mra\sin^2\theta}{\Sigma}\rm{d}t\rm{d}\phi 
       + \frac{\Sigma}{\Delta}\rm{d}r^2 + \Sigma\rm{d}\theta^2
\\ &\mbox{}\hspace{0.2in}
       + \left(r^2+a^2+\frac{2Mra^2\sin^2\theta}{\Sigma}\right)
       \sin^2\theta\rm{d}\phi^2, \nonumber \\
\Sigma &\equiv r^2 + a^2\cos^2\theta, \\
\Delta &\equiv r^2-2Mr+a^2.
\end{align}
\end{subequations}
Making use of the Newman-Penrose formalism,
Teukolsky\cite{teukolsky-1973} showed that perturbations of several
kinds of fields are governed by a single master equation
\begin{align}\label{eq:Teukolsky_Master}
\left[\frac{(r^2+a^2)^2}{\Delta} - a^2\sin^2\theta\right]
   \frac{\partial^2{}_s\psi}{\partial{t}^2}
 &\nonumber \\
+ \frac{4Mar}{\Delta}\frac{\partial^2{}_s\psi}{\partial{t}\partial\phi}
+ \left[\frac{a^2}{\Delta} - \frac1{\sin^2\theta}\right]
   \frac{\partial^2{}_s\psi}{\partial\phi^2}
 &\nonumber \\
- \Delta^{-s}\frac{\partial}{\partial{r}}\left(
    \Delta^{s+1}\frac{\partial{}_s\psi}{\partial{r}}\right)
-  \frac1{\sin\theta}\frac{\partial}{\partial\theta}\left(
   \sin\theta\frac{\partial{}_s\psi}{\partial\theta}\right)
 &\\
- 2s\left[\frac{a(r-M)}{\Delta} + \frac{i\cos\theta}{\sin^2\theta}\right]
     \frac{\partial{}_s\psi}{\partial\phi}
 &\nonumber \\
- 2s\left[\frac{M(r^2-a^2)}{\Delta} - r - ia\cos\theta\right]
     \frac{\partial{}_s\psi}{\partial{t}}
 &\nonumber \\
+ (s^2\cot^2\theta - s){}_s\psi
&= 4\pi\Sigma T,\nonumber 
\end{align}
where $T$ represents source terms and ${}_s\psi$ is a scalar of spin
weight $s$.  Scalar fields are represented by $s=0$,
spin-$\mathfrak{\frac12}$ fields by $s=\pm\mathfrak{\frac12}$,
electromagnetic fields by $s=\pm1$, and gravitational perturbations by
$s=\pm2$.  Assuming the vacuum case ($T=0$),
Eq.~(\ref{eq:Teukolsky_Master}) separates if we let
\begin{equation}\label{eq:Teukolsky_separation_form}
  {}_s\psi(t,r,\theta,\phi) = e^{-i\omega{t}} e^{im\phi}S(\theta)R(r).
\end{equation}
With $x\equiv\cos\theta$, the function
$S(\theta)=\swS{s}{\ell{m}}{x}{a\omega}$ is the spin-weighted
spheroidal function satisfying
\begin{align}\label{eqn:swSF_DiffEqn}
\partial_x \Big[ (1-x^2)\partial_x [\swS{s}{\ell{m}}{x}{c}]\Big] 
& \nonumber \\ 
    + \bigg[(cx)^2 - 2 csx + s& + \scA{s}{\ell m}{c} 
 \\ 
      & - \frac{(m+sx)^2}{1-x^2}\bigg]\swS{s}{\ell{m}}{x}{c} = 0,
\nonumber
\end{align}
where $c\ (=a\omega)$ is the oblateness parameter, $m$ the azimuthal
separation constant, and $\scA{s}{\ell m}{c}$ is the angular
separation constant.  The radial function $R(r)$ must satisfy
\begin{subequations}
\begin{align}\label{eqn:radialR:Diff_Eqn}
\Delta^{-s}\frac{d}{dr}&\left[\Delta^{s+1}\frac{dR(r)}{dr}\right]
 \\
&+ \left[\frac{K^2 -2is(r-M)K}{\Delta} + 4is\omega{r} - \lambdabar\right]R(r)=0,
\nonumber
\end{align}
where
\begin{align}
  K &\equiv (r^2+a^2)\omega - am, \\
  \lambdabar &\equiv \scA{s}{\ell{m}}{a\omega} + a^2\omega^2 - 2am\omega.
\end{align}
\end{subequations}
Equations (\ref{eqn:swSF_DiffEqn}) and (\ref{eqn:radialR:Diff_Eqn})
are coupled.  Given a value for $\scA{s}{\ell{m}}{a\omega}$,
Eq.~(\ref{eqn:radialR:Diff_Eqn}) is solved for the complex frequency
$\omega$.  Given a complex frequency $\omega$,
Eq.~(\ref{eqn:swSF_DiffEqn}) is an eigenvalue problem for
$\scA{s}{\ell{m}}{a\omega}$.  In both cases, appropriate boundary
conditions must be supplied.

\subsection{The confluent Heun equation}
\label{sec:Confluent_Heun}
A Heun equation is a second-order linear differential equation with
four regular singular points.  The {\em confluent Heun equation}
occurs when one regular singular point is lost by confluence with
another and the point at infinity becomes irregular.  Both
Eqs.~(\ref{eqn:swSF_DiffEqn}) and (\ref{eqn:radialR:Diff_Eqn}) are
examples of the confluent Heun equation.  Reference~\cite{Heun-eqn} contains
a very useful summary of the various forms of Heun's differential
equations and their solutions, and this section represents a summary
of the important points relevant to this paper.

A useful form of the confluent Heun equation is the so-called {\em
  nonsymmetrical canonical form}:
\begin{equation}\label{eq:Heun_NSCF}
\frac{d^2H(z)}{dz^2} 
+ \left(4p + \frac\gamma{z}+\frac\delta{z-1}\right)\frac{dH(z)}{dz}
+ \frac{4\alpha pz -\sigma}{z(z-1)}H(z)=0.
\end{equation}
Much of the theory of solutions to the confluent Heun equation is
based on this form of the equation.  It is defined by five parameters:
$p$, $\alpha$, $\gamma$, $\delta$, and $\sigma$.  It has a regular
singular point at $z=0$ with characteristic exponents
$\{0,1-\gamma\}$ so solutions in general behave like
\begin{equation}
  \lim_{z\to0}H(z)\sim1 \quad\mbox{or}\quad z^{1-\gamma}.
\end{equation}
There is a regular singular point at $z=1$ with characteristic
exponents $\{0,1-\delta\}$ so solutions in general behave like
\begin{equation}
  \lim_{z\to1}H(z)\sim1 \quad\mbox{or}\quad (z-1)^{1-\delta}.
\end{equation}
And there is an irregular singular point at $z=\infty$ where solutions
in general behave like
\begin{equation}
  \lim_{z\to\infty}H(z)\sim z^{-\alpha} \quad\mbox{or}\quad 
  e^{-4pz}z^{\alpha-\gamma-\delta}.
\end{equation}

Other forms of the confluent Heun equation are common. If we
let $z\to1-2z$ and define
\begin{equation}
  H^{(B)}(z)\equiv(z-1)^{\frac{\mu+\nu}2}(z+1)^{\frac{\mu-\nu}2}e^{-pz}H((1-z)/2),
\end{equation}
we obtain the {\em B\^{o}cher symmetrical form}:
\begin{align}\label{eq:Heun_Bocher}
\frac{d}{dz}\left((z^2-1)\frac{dH^{(B)}(z)}{dz}\right)& \nonumber \\ \mbox{}
+\Biggl(-p^2(z^2-1)+ & 2p\beta z-\lambda \\ \mbox{}
    - &\frac{\mu^2+\nu^2+2\mu\nu z}{z^2-1}\Biggr)H^{(B)}(z)=0, \nonumber
\end{align}
where the new parameters $\beta$, $\lambda$, $\mu$, and $\nu$ are
defined by
\begin{align}
 \gamma=\mu+\nu+1,&\qquad \delta=\mu-\nu+1,\qquad \alpha=-\beta+\mu+1, 
\nonumber \\
\sigma = \lambda - & 2p(\beta-\mu-\nu-1)-\mu(\mu+1),
\end{align}
and the regular singular points are now at $z=\pm1$.
Equation~(\ref{eq:Heun_Bocher}) is also referred to as the {\em
  generalized spheroidal equation}.  Finally, the {\em normal
  symmetrical form}, or the {\em normal form of the generalized
  spheroidal equation}, is obtained by the transformation
\begin{equation}
  H^{(B)}(z)\equiv(1-z^2)^{-\frac12}H^{(N)}(z)
\end{equation}
and takes the form
\begin{align}\label{eq:Heun_Normal}
(z^2-1)\frac{d^2H^{(N)}(z)}{dz^2} & \nonumber \\ \mbox{}
+\Biggl(-p^2(z^2-1)& + 2p\beta z-\lambda \\ \mbox{}
    & -\frac{\mu^2+\nu^2-1+2\mu\nu z}{z^2-1}\Biggr)H^{(N)}(z)=0. \nonumber
\end{align}

\subsubsection{Local solutions}
\label{sec:Heun_local}
Power series solutions local to each of the singular points can be
defined in terms of two functions:
$Hc^{(a)}(p,\alpha,\gamma,\delta,\sigma;z)$ and
$Hc^{(r)}(p,\alpha,\gamma,\delta,\sigma;z)$.  These are defined by
\begin{subequations}\label{eq:local_a_sol}
\begin{align}
\label{eq:local_a_sol_limit}
 Hc^{(a)}(p,\alpha,\gamma,\delta,\sigma;0)&=1, \\
\label{eq:local_a_sol_series}
 Hc^{(a)}(p,\alpha,\gamma,\delta,\sigma;z)&= \sum_{k=0}^\infty{c^{(a)}_kz^k},
\end{align}
\end{subequations}
and
\begin{subequations}\label{eq:local_r_sol}
\begin{align}
\label{eq:local_r_sol_limit}
 \lim_{z\to\infty}z^\alpha Hc^{(r)}(p,\alpha,\gamma,\delta,\sigma;z)&=1, \\
\label{eq:local_r_sol_series}
 Hc^{(r)}(p,\alpha,\gamma,\delta,\sigma;z) = \sum_{k=0}^\infty{c^{(r)}_kz^{-\alpha-k}}.
\end{align}
\end{subequations}  
The coefficients $c^{(a)}_k$ and $c^{(r)}_k$ are defined by three-term
recurrence relations
\begin{subequations}
\label{eq:local_a_defs}
\begin{align}
\label{eq:local_a_3term}
  0 =& f^{(a)}_kc^{(a)}_{k+1}+ g^{(a)}_kc^{(a)}_k+ h^{(a)}_kc^{(a)}_{k-1}, \\
  &\mbox{with}\qquad c^{(a)}_{-1}=0, \qquad c^{(a)}_0=1, \nonumber \\
\label{eq:local_a_g}
  g^{(a)}_k =& k(k-4p+\gamma+\delta-1)-\sigma, \\
\label{eq:local_a_f}
  f^{(a)}_k =& -(k+1)(k+\gamma), \\
\label{eq:local_a_h}
  h^{(a)}_k =& 4p(k+\alpha-1),
\end{align}
\end{subequations} 
and
\begin{subequations}
\label{eq:local_r_defs}
\begin{align}
\label{eq:local_r_3term}
  0 =& f^{(r)}_kc^{(r)}_{k+1}+ g^{(r)}_kc^{(r)}_k+ h^{(r)}_kc^{(r)}_{k-1},\\
  &\mbox{with}\qquad c^{(r)}_{-1}=0, \qquad c^{(r)}_0=1, \nonumber \\
\label{eq:local_r_g}
  g^{(r)}_k =& (k+\alpha)(k+4p+\alpha-\gamma-\delta+1)-\sigma, \\
\label{eq:local_r_f}
  f^{(r)}_k =& -4p(k+1), \\
\label{eq:local_r_h}
  h^{(r)}_k =& -(k+\alpha-1)(k+\alpha-\gamma).
\end{align}
\end{subequations} 

The solutions local to the regular singular points can be written
in terms of $Hc^{(a)}$.  The two solutions local to $z=0$ are given
by
\begin{subequations}
\label{eq:all_local_sol}
\begin{align}
\label{eq:local_sol_z0a}
  Hc^{(a)}(p,\alpha,\gamma,\delta,\sigma;z&), \\
\label{eq:local_sol_z0b}
  z^{1-\gamma}Hc^{(a)}(p,\alpha+1-\gamma,&2-\gamma,\delta, \\
 &\sigma+(1-\gamma)(4p-\delta);z). \nonumber
\intertext{Local to $z=1$ we have}
\label{eq:local_sol_z1a}
  Hc^{(a)}(-p,\alpha,\delta,&\gamma,\sigma-4p\alpha;1-z) \\
\label{eq:local_sol_z1b}
  (z-1)^{1-\delta}Hc^{(a)}(-p,\alpha+&1-\delta,2-\delta,\gamma, \\
   \sigma-(&1-\delta)\gamma-4p(\alpha+1-\delta);1-z). \nonumber
\intertext{The two solutions local to the irregular singular point at $z=\infty$
are}
\label{eq:local_sol_zinfa}
  Hc^{(r)}(p,\alpha,\gamma,\delta,\sigma;&z) \\
\label{eq:local_sol_zinfb}
  e^{-4pz}Hc^{(r)}(-p,-\alpha+\gamma&+\delta,\gamma,\delta,\sigma-4p\gamma;z).
\end{align}
\end{subequations}

\subsubsection{Confluent Heun functions}
\label{sec:Heun_functions}
The term {\em confluent Heun function} is used for solutions that are
simultaneously Frobenius solutions for two adjacent singular points.
Such functions naturally arise in the solution of eigenvalue problems.
In this case, boundary conditions are imposed at the two singular
points, and the value of a particular parameter in the equation is
found that allows both boundary conditions to be satisfied.  The
solutions we seek for the angular and radial Teukolsky equations,
Eqs.~(\ref{eqn:swSF_DiffEqn}) and~(\ref{eqn:radialR:Diff_Eqn}) fall
into this category.  There are many ways of finding these solutions
and we will address this further below.

\subsubsection{Polynomial solutions}
\label{sec:Heun_polynomials}
An important subset of the confluent Heun functions are the {\em
  confluent Heun polynomials}.  These solutions are simultaneously
Frobenius solutions of all three singular points.  This requires that
the series solutions terminate yielding polynomial solutions.  A
necessary, but not sufficient condition for this to occur is for the
second parameter, $\alpha$, of either Eq.~(\ref{eq:local_a_sol})
or~(\ref{eq:local_r_sol}) to be a nonpositive integer $-q$.  In this
case, we find from either Eq.~(\ref{eq:local_a_h})
or~(\ref{eq:local_r_h}) that $h_{q+1}^{(a,r)}=0$.  If
$c_{q+1}^{(a,r)}=0$ also holds, then the infinite series will
terminate, yielding a polynomial solution.  This second condition,
referred to as a $\Delta_{q+1}=0$ condition, is satisfied by the
vanishing of the determinant
\begin{equation}\label{eq:Delta_q_cond}
  \left|\begin{array}{ccccc}
  g_0 & f_0 & 0 & \cdots & 0 \\
  h_1 & g_1 & f_1 & \ddots & 0 \\
  0 & h_2 & g_2 & \ddots & 0 \\
  0 & 0 & \ddots & \ddots & \ddots \\
  0 & 0 & \cdots & h_q & g_q
  \end{array}\right|.
\end{equation}
Noting the form of the $g_k^{(a,r)}$ coefficients in
Eqs.~(\ref{eq:local_a_g}) and~(\ref{eq:local_r_g}), we see that the
$\Delta_{q+1}=0$ condition can be viewed as an eigenvalue problem for
the allowed values of $\sigma$.  The $q+1$ eigenvalues are labeled
$\sigma_{nq}$, and can be ordered such that
$\sigma_{0q}<\sigma_{1q}<\cdots<\sigma_{qq}$.  So, given a local
solution defined by $Hc^{(a,r)}(p,-q,\gamma,\delta,\sigma;z)$, a
polynomial solution $Hc_{nq}(p,\gamma,\delta;z)$ exists if
$\sigma_{nq}=\sigma$.

\subsection{The radial Teukolsky equation}
\label{sec:Radial_TE}
The radial Teukolsky equation, Eq.~(\ref{eqn:radialR:Diff_Eqn}), has
regular singular points at the inner and outer horizons.  These are 
located at the roots, $r_\pm$, of $\Delta=0$:
\begin{equation}
  r_\pm = M\pm\sqrt{M^2-a^2}.
\end{equation}
The outer or event horizon is labeled by $r_+$ and the inner or Cauchy
horizon by $r_\minus$.  $r=\infty$ is an irregular singular point.  We
can rewrite Eq.~(\ref{eqn:radialR:Diff_Eqn}) in nonsymmetrical canonical
form by making the following transformation\cite{Fiziev-2009b}:
\begin{equation}
R(r) = (r-r_+)^\xi(r-r_\minus)^\eta e^{\zeta r}H(r),
\end{equation}
where the parameters $\zeta$, $\xi$, and $\eta$ must be fixed to be
\begin{subequations}\label{eq:Teukolsky_Heun_parameters}
\begin{align}
  \zeta&=\pm i\omega\equiv\zeta_\pm, \\
  \xi&=\frac{-s\pm(s + 2i\sigma_+)}2\equiv\xi_\pm, \\
  \eta&=\frac{-s\pm(s - 2i\sigma_\minus)}2\equiv\eta_\pm,
\end{align}
\end{subequations}
where
\begin{equation}
 \sigma_\pm\equiv\frac{2\omega{M}r_\pm - ma}{r_+-r_\minus}.
\end{equation}
Before proceeding further, we define the following dimensionless
variables:
\begin{subequations}
\begin{align}
  \bar{r} &\equiv \frac{r}{M}, \\
  \bar{a} &\equiv \frac{a}{M}, \\
  \bar\omega &\equiv M\omega, \\
  \bar\zeta &\equiv M\zeta,  
\end{align}
\end{subequations}
It is also useful to transform the radial coordinate.  There are many 
useful ways to do this\cite{Fiziev-2009b}, but here we will focus on
\begin{equation}
  z \equiv \frac{r-r_\minus}{r_+-r_\minus} 
  = \frac{\bar{r}-\bar{r}_\minus}{\bar{r}_+-\bar{r}_\minus}.
\end{equation}
All eight possible combinations of the parameters in
Eq.~(\ref{eq:Teukolsky_Heun_parameters}) reduce
Eq.~(\ref{eqn:radialR:Diff_Eqn}) to the form of
Eq.~(\ref{eq:Heun_NSCF}) when we define 
\begin{subequations}
\begin{align}
  p &= (\bar{r}_+-\bar{r}_\minus)\frac{\bar\zeta}2 \\
  \alpha &= 1+s+\xi+\eta - 2\bar\zeta + s\frac{i\bar\omega}{\bar\zeta} \\
  \gamma &= 1+s+2\eta \\
  \delta &= 1+s+2\xi \\
  \sigma &= \scA{s}{\ell{m}}{\bar{a}\bar\omega} + \bar{a}^2\bar\omega^2 
   - 8\bar\omega^2 + p(2\alpha+\gamma-\delta) \\
& \mbox{}\hspace{0.75in}
   +\left(1+s-\frac{\gamma+\delta}2\right)\left(s+\frac{\gamma+\delta}2\right).
\nonumber
\end{align}
\end{subequations}
Each of these eight forms produce the same pair of local solutions at each
singular point.  With \\$R(z)\sim z^\eta(z-1)^\xi
e^{(\bar{r}_+-\bar{r}_\minus)\bar\zeta z} H(z)$, we find
\begin{subequations} \label{eq:R_local_zall}
\begin{align}
\label{eq:R_local_z0}
  \lim_{z\to0}R(z)&\sim z^{-s+i\sigma_\minus} \quad\mbox{or}\quad z^{-i\sigma_\minus}, \\
\label{eq:R_local_z1}
  \lim_{z\to1}R(z)&\sim(z-1)^{-s-i\sigma_+} \quad\mbox{or}\quad (z-1)^{i\sigma_+},\\
\label{eq:R_local_zinf}
  \lim_{z\to\infty}R(z)&\sim
  z^{-1-2s+2i\bar\omega}e^{i(\bar{r}_+-\bar{r}_\minus)\bar\omega z} \\
& \hspace{0.5in}  \mbox{or}\qquad
  z^{-1-2i\bar\omega}e^{-i(\bar{r}_+-\bar{r}_\minus)\bar\omega z}.\nonumber
\end{align}
\end{subequations}

Given the sign choice made on $\omega$ in
Eq.~(\ref{eq:Teukolsky_separation_form}), we see that the first form
of Eq.~(\ref{eq:R_local_z1}) corresponds to waves propagating into the
black-hole event horizon, and the first form of
Eq.~(\ref{eq:R_local_zinf}) corresponds to waves propagating out at
infinity, the appropriate boundary conditions for QNMs (see
the Appendix for details).  Each of the eight possible
choices for the parameters $\{\zeta,\xi,\eta\}$ corresponds to
assigning each of the pairs of physical solutions in
Eq.~(\ref{eq:R_local_zall}) to a corresponding pair from
Eq.~(\ref{eq:all_local_sol}).  In particular, $\xi$ fixes the
correspondence at the event horizon.  The choice of $\xi=\xi_\minus$
means that Eq.~(\ref{eq:local_sol_z1a}) corresponds to the case of
waves propagating into the black hole, and
Eq.~(\ref{eq:local_sol_z1b}) corresponds to waves propagating out of
the black hole.  The choice of $\zeta$ fixes the correspondence at
infinity.  The choice $\zeta=\zeta_+$ means that
Eq.~(\ref{eq:local_sol_zinfa}) corresponds to the case of waves
propagating out at infinity, and Eq.~(\ref{eq:local_sol_zinfb})
corresponds to waves propagating in from infinity.  The choice of
$\eta$ fixes the correspondence at the Cauchy horizon.  These mappings
become particularly important when exploring the polynomial solutions
of the radial equation below.

\subsubsection{Confluent Heun functions}
\label{sec:Radial_Hc_functions}
In this paper, our main goal in solving the radial Teukolsky equation
is to determine the QNMs of Kerr.  These solutions will not, in
general, correspond to polynomial solutions, but will be confluent
Heun functions.  For QNMs, we want the boundary condition at the
horizon to allow waves to propagate into the black hole, and waves to
propagate out at infinity.  If we choose $\xi=\xi_\minus$, then the
characteristic exponent at $z=1$ is zero, so
Eq.~(\ref{eq:local_sol_z1a}) represents the desired boundary
condition.  With the choice $\zeta=\zeta_+$,
Eq.~(\ref{eq:local_sol_zinfa}) represents the desired boundary
condition at $z=\infty$.  From Eq.~(\ref{eq:local_r_sol_limit}) we see
that $\lim_{z\to\infty}{H(z)}=z^{-\alpha}$.  If we choose
\begin{equation}
  H(z) = z^{-\alpha}\bar{R}(z),
\end{equation}
then we guarantee that $\bar{R}(z)$ tends to a finite number at both
boundaries.

The confluent Heun function we seek will be simultaneously a Frobenius
solution for both $z=1$ and $z=\infty$.  The radius of convergence for
the local solutions is in general the distance to the next singular
point.  Thus, the series solution around $z=1$ has a radius of
convergence no larger than $1$, far short of infinity.  We closely
follow Leaver's approach\cite{leaver-1985,leaver-1986} to circumvent
this problem.

Let $z\to\frac{z-1}{z}$ so that the regular singular point at
$\bar{r}_\minus$ is now pushed to $z=-\infty$.  The regular singular
point at $\bar{r}_+$ is now located at $z=0$, and the irregular
singular point at infinity is now located at $z=1$.  The differential
equation now becomes
\begin{align}\label{eq:Leaver_radDE_form}
  z(1-z)^2\frac{d^2\hat{R}(z)}{dz^2} 
     + \left[D_0 + D_1z + D_2z^2\right]\frac{d\hat{R}(z)}{dz} & \\
     + \left[D_3 + D_4z\right]\hat{R}(z)& = 0, \nonumber
\end{align}
with $\bar{R}(z)=\hat{R}((z-1)/z)$ and
\begin{subequations}\label{eq:Leaver_D_coefs}
\begin{align}
  D_0 &= \delta = 1+s+2\xi, \\
  D_1 &= 4p-2\alpha+\gamma-\delta-2, \\
  D_2 &= 2\alpha-\gamma+2, \\
  D_3 &= \alpha(4p-\delta)-\sigma, \\
  D_4 &= \alpha(\alpha-\gamma+1).
\end{align}
\end{subequations}
Expanding $\hat{R}(z)$ in Eq.~(\ref{eq:Leaver_radDE_form}) in terms of
the Taylor series $\hat{R}(z)=\sum_{n=0}^\infty{a_nz^n}$, we obtain
the three-term recurrence relation where the coefficients, $a_n$, satisfy
\begin{subequations}
\begin{align}
\label{eq:Leaver_rad_2-term}
  0 &= a_0\beta_0 + a_1\alpha_0, \\ 
\label{eq:Leaver_rad_3-term}
  0 &= a_{n+1}\alpha_n + a_n\beta_n + a_{n-1}\gamma_n, \\
\intertext{with}
\label{eq:Leaver_rad_alpha}
  \alpha_n &\equiv n^2 + (D_0+1)n + D_0, \\
\label{eq:Leaver_rad_beta}
  \beta_n &\equiv -2n^2 + (D_1+2)n + D_3, \\
\label{eq:Leaver_rad_gamma}
  \gamma_n &\equiv n^2 + (D_2-3)n +D_4 - D_2 +2.
\end{align}
\end{subequations}
With $a_0=1$, we are guaranteed to obey the boundary condition at the
event horizon ($z=0$), but must find the solution that yields a finite
answer at infinity ($z=1$).  A three-term recurrence relation such as
Eq.~(\ref{eq:Leaver_rad_3-term}) will have two linearly independent
solutions.  To understand their behaviors, consider the radius of
convergence $\rho_a\equiv\lim_{n\to\infty}\frac{a_{n+1}}{a_n}$.  From
Eq.~(\ref{eq:Leaver_rad_3-term}) we find
\begin{equation}\label{eqn:a_ratio_characteristic}
  \lim_{n\to\infty}\left\{\frac{a_{n+1}}{a_n}+\frac{\beta_n}{\alpha_n}+
  \frac{a_{n-1}}{a_n}\frac{\gamma_n}{\alpha_n}\right\}=\rho_a-2+\frac1{\rho_a}=0,
\end{equation}
which has the double root $\rho_a=1$.  Since we do not have distinct
roots, we must consider higher-order behavior in
$\lim_{n\to\infty}\frac{a_{n+1}}{a_n}$.  We assume
\begin{equation}\label{eqn:a_ratio_expansion}
  \lim_{n\to\infty}\frac{a_{n+1}}{a_n} = 1 + \frac{u_1}{\sqrt{n}}
      + \frac{u_2}{n} + \frac{u_3}{n^{\frac32}} +\cdots.
\end{equation}
If we consider
\begin{align}
  \left(\lim_{n\to\infty}\frac{ \frac{d}{dn}a_{n+1}}{a_n}\right) &\approx
   \lim_{n\to\infty}\left(\frac{a_{n+1}-a_n}{a_n}\right) \\
   &= \frac{u_1}{\sqrt{n}} + \frac{u_2}{n} + \frac{u_3}{n^{\frac32}} +\cdots,
   \nonumber
\end{align}
then
\begin{equation}\label{eq:residual_def}
  \lim_{n\to\infty}\ln(a_n) \approx \int\left(\frac{u_1}{\sqrt{n}} 
    + \frac{u_2}{n} + \frac{u_3}{n^{\frac32}} +\cdots\right)dn.
\end{equation}
The result is
\begin{equation}\label{eq:asymptotic_a}
  \lim_{n\to\infty}a_n \propto n^{u_2} e^{2u_1\sqrt{n}}.
\end{equation}

Inserting Eq.~(\ref{eqn:a_ratio_expansion}) into
Eq.~(\ref{eqn:a_ratio_characteristic}), we find
\begin{subequations}
\begin{align}
  u_1 &=\pm\sqrt{-4p}, \\
  u_2 &= -\frac14(8p-4\alpha+2\gamma+2\delta+3), \\
  u_3 &= \frac1{32u_1}\bigl[32p(2p-4\alpha+\gamma+3\delta+4) \\
    & \mbox{}\hspace{0.5in}
                +4(\gamma+\delta)(\gamma+\delta-2) +16\sigma+3\bigr],
  \nonumber \\
   & \ \vdots \quad. \nonumber
\end{align}
\end{subequations}
The asymptotic behavior of $a_n$ is dominated by the exponential term
in Eq.~(\ref{eq:asymptotic_a}), and the two solutions for $u_1$
correspond to the two possible asymptotic behaviors for $a_n$.  The
solution where $\rm{Re}(u_1)<0$ we define as $a_n\equiv f_n$, the
other solution is labeled $g_n$.  Since these two solutions have the
property $\lim_{n\to\infty}\frac{f_n}{g_n}=0$, when it exists, $f_n$
will be the {\em minimal} solution of Eq.~(\ref{eq:Leaver_rad_3-term})
while $g_n$ will be a dominant solution.  With
$\bar\zeta=\bar\zeta_+=i\bar\omega$, so long as
$\rm{Im}(\bar\omega)<0$, $\rm{Re}(\sqrt{-4p})<0$ and $u_1=+\sqrt{-4p}$
is associated with the minimal solution and $u_1=-\sqrt{-4p}$ with a
dominant solution.

Defining the ratio $r_n\equiv\frac{a_{n+1}}{a_n}$,
Eq.~(\ref{eq:Leaver_rad_3-term}) can be written as
\begin{equation}\label{eq:Leaver_ratio_CF}
  \frac{a_{n+1}}{a_n} \equiv r_n 
  = \frac{-\gamma_{n+1}}{\beta_{n+1}+ \alpha_{n+1}r_{n+1}}
  \qquad n=0,1,2,\ldots,
\end{equation}
which is a recursive form equivalent to the continued fraction
\begin{equation}\label{eq:Leaver_rad_CF_dom}
r_n 
  = \frac{-\gamma_{n+1}}{\beta_{n+1}-} 
    \frac{\alpha_{n+1}\gamma_{n+2}}{\beta_{n+2}-}
    \frac{\alpha_{n+2}\gamma_{n+3}}{\beta_{n+3}-}\ldots.
\end{equation}
The usefulness of this continued fraction comes from a theorem due to
Pincherle (see Ref.~\cite{Gautschi-1967}).
\begin{theorem}[Pincherle]\label{th:Pincherle}
  The continued fraction $r_0$ converges if and only
  if the recurrence relation Eq.~(\ref{eq:Leaver_rad_3-term}) possesses
  a minimal solution $a_n=f_n$, with $f_0\ne0$.  In case of
  convergence, moreover, one has $\frac{f_{n+1}}{f_n}=r_n$ with
  $n=0,1,2,\ldots$ provided $f_n\ne0$.
\end{theorem}
Because the recurrence terminates, we have the special condition given
by Eq.~(\ref{eq:Leaver_rad_2-term}).  This tells us that
\begin{equation}\label{eq:termination_cond}
  r_0=\frac{f_1}{f_0}=-\frac{\beta_0}{\alpha_0}.
\end{equation}
Therefore, the continued fraction $r_0$ does converge, and by
Theorem~\ref{th:Pincherle}, the coefficients $a_n=f_n$ are the minimal
solution of Eq.~(\ref{eq:Leaver_rad_3-term}).  Moreover, the continued
fraction cannot converge to any arbitrary value, but must converge to 
the value of Eq.~(\ref{eq:termination_cond}).  This will only be possible 
when $\bar\omega$ takes on certain values, the QNMs.

Combining Eqs.~(\ref{eq:Leaver_ratio_CF})
and~(\ref{eq:termination_cond}), we obtain a general continued
fraction equation
\begin{equation}\label{eq:Leaver_Cf_inf}
  0=\beta_0 -\frac{\alpha_0\gamma_1}{\beta_1-}
             \frac{\alpha_1\gamma_2}{\beta_2-}
             \frac{\alpha_2\gamma_3}{\beta_3-}\ldots,
\end{equation}
that is considered a function of $\bar\omega$.  For fixed $s$, $m$,
and $\bar{a}$, and for a given value of
$\scA{s}{\ell{m}}{\bar{a}\bar\omega}$, we can solve for the
eigenvalues $\bar\omega$ of Eq.~(\ref{eq:Leaver_radDE_form}) by
finding the roots of Eq.~(\ref{eq:Leaver_Cf_inf}).  This approach for
solving the radial Teukolsky equation is referred to in the literature
as {\em Leaver's method}.

In practice, the continued fraction must be truncated by some means.
We define a truncated version of this continued fraction as
\begin{equation}\label{eq:Leaver_Cf_N}
  \rm{Cf}(N) \equiv \beta_0 -\frac{\alpha_0\gamma_1}{\beta_1-}
             \frac{\alpha_1\gamma_2}{\beta_2-}
             \frac{\alpha_2\gamma_3}{\beta_3-}\ldots
             \frac{\alpha_{N-1}\gamma_N}{\beta_N+\alpha_N r_N}.
\end{equation}
It is also useful to define the $n$th inversion of this
truncated continued fraction:
\begin{align}\label{eq:Leaver_Cf_N_inv}
  \rm{Cf}(n;N) \equiv &\beta_n-\frac{\alpha_{n-1}\gamma_n}{\beta_{n-1}-}
                         \frac{\alpha_{n-2}\gamma_{n-1}}{\beta_{n-2}-}
                         \ldots\frac{\alpha_0\gamma_1}{\beta_0} \\
                         &\mbox{}
                         -\frac{\alpha_n\gamma_{n+1}}{\beta_{n+1}-}
                          \frac{\alpha_{n+1}\gamma_{n+2}}{\beta_{n+2}-}
                          \ldots\frac{\alpha_{N-1}\gamma_N}{\beta_N+\alpha_N r_N},
                          \nonumber
\end{align}
with $\rm{Cf}(0;N)\equiv\rm{Cf}(N)$ and $0\le n<N$.

In order to find roots of $\rm{Cf}(n;N)$, the truncated continued
fraction must be explicitly evaluated for some appropriate value of
$N$.  Using Lentz's method\cite{numrec_c}, $N$ is determined
dynamically by demanding that the change in $\rm{Cf}(n;N)$ caused by
adding additional terms is below some tolerance.  This is a
``top-down'' approach, evaluating the continued fraction as a ratio of
two infinite series.  With this method, a value for $r_N$ is never
needed.  An alternate method is to choose a value for $N$, assume
$r_N=0$, and evaluate $\rm{Cf}(n;N)$ from the ``bottom up''.  This
method is less prone to rounding errors, but one must test various
values of $N$ to ensure that the desired tolerance in $\rm{Cf}(n;N)$
has been reached.  An even better approach, first used in the context
of QNMs by Nollert\cite{nollert-1993}, is to use
Eq.~(\ref{eqn:a_ratio_expansion}) with $n\mapsto N$ to approximate
$r_N$.  We use this approach to solve the radial Teukolsky equation,
and use terms out through $u_5$ in our approximation of $r_N$.

\subsection{The angular Teukolsky equation}
\label{sec:Angular_TE}
The angular Teukolsky equation, Eq.~(\ref{eqn:swSF_DiffEqn}), has
regular singular points at $x=\pm1$ and an irregular singular point 
at infinity.  It is an example of the confluent Heun equation in
B\^ocher symmetrical form with
\begin{subequations}
\begin{align}
  p &=c, \\
  \beta &=s, \\
  \lambda &=\scA{s}{\ell m}{c}+s(s+1),
\intertext{and any of the 8 possible combinations of}
  \mu = &\pm m\quad\hbox{and}\quad\nu=\pm s \\
\intertext{or}
  \mu = &\pm s\quad\hbox{and}\quad\nu=\pm m.
\end{align}
\end{subequations}

Using exactly the same methods discussed in Sec.~\ref{sec:Radial_TE},
Leaver's method can be used to solve a continued fraction equation
derived from Eq.~(\ref{eqn:swSF_DiffEqn}) for its eigenvalues
$\scA{s}{\ell{m}}{c}$.  Following Leaver\cite{leaver-1985}, it is
common to expand $\swS{s}{\ell{m}}{x}{c}$ as
\begin{equation}
  \swS{s}{\ell{m}}{x}{c} = \sum_{n=0}^\infty{b_n (1+x)^n}.
\end{equation}
Other approaches are possible\cite{sasakinakamura-1982}, and several
others have used an expansion in terms of Jacobi
polynomials\cite{fackerell-1976,leaver-1986} instead of powers of
$1+x$.  The latter approach also leads to a continued fraction
equation, the roots of which yield the eigenvalues associated with a
minimal solution of Eq.~(\ref{eqn:swSF_DiffEqn}).  However, we have
chosen to solve the angular Teukolsky equation using a direct spectral
eigenvalue approach\cite{hughes-2000}.  This method does not seem to
have been used previously in the computation of QNMs and we developed
this method independently, only learning of its prior use by
Hughes\cite{hughes-2000} after submission of this manuscript.  His
derivation is slightly different than ours, and contains a few errors.
We present a full derivation below.

\subsubsection{The spectral eigenvalue method}
\label{sec:sectral_EV_method}
The spin-weighted spheroidal harmonics
$\swSH{s}{\ell{m}}{\theta,\phi}{c}$ are generalizations of the
spin-weighted spherical harmonics
$\swY{s}{\ell{m}}{\theta,\phi}=\swSH{s}{\ell{m}}{\theta,\phi}{0}$,
where $s$ is the spin weight of the harmonic and $c$ is the
oblateness parameter.  The angular dependence separates as
\begin{equation}
  \swSH{s}{\ell{m}}{\theta,\phi}{c} \equiv 
  \frac1{2\pi}\,\swS{s}{\ell{m}}{\cos\theta}{c}e^{im\phi}.
\end{equation}
The spin-weighted spheroidal {\em function} $\swS{s}{\ell{m}}{x}{c}$
satisfies the angular Teukolsky equation (\ref{eqn:swSF_DiffEqn}).
These functions are direct generalization of both the {\em angular
  spheroidal functions} $\swS{}{\ell{m}}{x}{c} =
\swS{0}{\ell{m}}{x}{c}$ and the {\em spin-weighted spherical
  functions} $\swS{s}{\ell{m}}{x}{0}$.

The basic symmetries inherent in the spin-weighted spheroidal
functions follow from Eq.~(\ref{eqn:swSF_DiffEqn}) via the three
transformations: $\{s\to-s,x\to-x\}$, $\{m\to-m,x\to-x,c\to-c\}$, and
complex conjugation.  Additional sign and normalization conditions can
be chosen for consistency with common sign conventions for the
angular-spheroidal and spin-weighted spherical functions.  From 
these, it follows that the spin-weighted spheroidal functions 
and the separation constants satisfy
the following conditions:
\begin{subequations}\label{eq:swSF_all_ident}
\begin{align}
\label{eq:swSF_sx_ident}
\swS{-s}{\ell{m}}{x}{c} &= (-1)^{\ell+m}\swS{s}{\ell{m}}{-x}{c}, \\
\label{eq:swSF_mxc_ident}
\swS{s}{\ell(-m)}{x}{c} &= (-1)^{\ell+s}\swS{s}{\ell{m}}{-x}{-c}, \\
\label{eq:swSF_cc_ident}
\swS[*]{s}{\ell{m}}{x}{c} &= \swS{s}{\ell{m}}{x}{c^*}, \\
\intertext{and}
\label{eq:swSF_sA_ident}
\scA{-s}{\ell{m}}{c} &= \scA{s}{\ell{m}}{c} + 2s, \\
\label{eq:swSF_mcA_ident}
\scA{s}{\ell(-m)}{c} &= \scA{s}{\ell{m}}{-c}, \\
\label{eq:swSF_cA_ident}
\scA[*]{s}{\ell{m}}{c} &= \scA{s}{\ell{m}}{c^*}.
\end{align}
\end{subequations}

We can expand the spin-weighted spheroidal functions in terms of the
spin-weighted spherical functions,
\begin{equation}
  \swS{s}{\ell{m}}{x}{c} = \sum_{\ell^\prime=\ell_{\mbox{\tiny min}}}^\infty
      C_{\ell^\prime\ell{m}}(c)\swS{s}{\ell^\prime{m}}{x}{0},
\end{equation}
where $\ell_{\mbox{min}}\equiv\max(|m|,|s|)$.  Using this expansion in
Eq.~(\ref{eqn:swSF_DiffEqn}), and using the fact that
\begin{equation}\label{eqn:scA_spherical}
\scA{s}{\ell{m}}{0} = l(l+1) - s(s+1),
\end{equation}
we obtain
\begin{align}
   \sum_{\ell^\prime=\ell_{\mbox{\tiny min}}}^\infty C_{\ell^\prime\ell{m}}(c)
   \bigl[\ell^\prime(&\ell^\prime+1) - s(s+1) - (cx)^2  \\
&\mbox{}
     + 2csx -\scA{s}{\ell{m}}{c} \bigr]
   \swS{s}{\ell^\prime{m}}{x}{0} = 0. \nonumber
\end{align}
To proceed, we need to remove the $x$ dependence.  This can be accomplished
using the recurrence relation\cite{blanco-1997}
\begin{subequations}
\begin{align}
  \label{eqn:swS:xrecur}
  x \swS{s}{\ell{m}}{x}{0} = 
       & \mathcal{F}_{s\ell{m}} \swS{s}{(\ell+1){m}}{x}{0} \\
& \mbox{}
       + \mathcal{G}_{s\ell{m}} \swS{s}{(\ell-1){m}}{x}{0}
       + \mathcal{H}_{s\ell{m}} \swS{s}{\ell{m}}{x}{0}, \nonumber
\end{align}
where
\begin{align}
  \label{eqn:x1coefF}
  \mathcal{F}_{s\ell{m}} &= 
       \sqrt{\frac{(\ell+1)^2-m^2}{(2\ell+3)(2\ell+1)}}
       \sqrt{\frac{(\ell+1)^2-s^2}{(\ell+1)^2}}, \\
  \label{eqn:x1coefG}
  \mathcal{G}_{s\ell{m}} &= \left\{\begin{array}{rcl}
       \mbox{if\ }\ell\ne0 &:& 
       \sqrt{\frac{\ell^2-m^2}{4\ell^2-1}}
       \sqrt{\frac{\ell^2-s^2}{\ell^2}}, \\
       \mbox{if\ }\ell=0 &:& 0, 
       \end{array}\right. \\
  \label{eqn:x1coefH}
  \mathcal{H}_{s\ell{m}} &= \left\{\begin{array}{rcl}
       \mbox{if\ }\ell\ne0 \mbox{\ and\ } s\ne0 &:& 
       -\frac{ms}{\ell(\ell+1)}, \\
       \mbox{if\ }\ell=0 \mbox{\ or\ } s=0 &:& 0.
       \end{array}\right.
\end{align}
\end{subequations}
For the terms that arise from the $x^2$ term, we define the following
quantities:
\begin{subequations}
\begin{align}
  \label{eqn:x2coefA}
  \mathcal{A}_{s\ell{m}} &= \mathcal{F}_{s\ell{m}}\mathcal{F}_{s(\ell+1)m}, \\
  \label{eqn:x2coefD}
  \mathcal{D}_{s\ell{m}} &= \mathcal{F}_{s\ell{m}}
         (\mathcal{H}_{s(\ell+1)m} + \mathcal{H}_{s\ell{m}}) , \\
  \label{eqn:x2coefB}
  \mathcal{B}_{s\ell{m}} &= \mathcal{F}_{s\ell{m}}\mathcal{G}_{s(\ell+1)m}
     + \mathcal{G}_{s\ell{m}}\mathcal{F}_{s(\ell-1)m} 
     + \mathcal{H}^2_{s\ell{m}}, \\
  \label{eqn:x2coefE}
  \mathcal{E}_{s\ell{m}} &= \mathcal{G}_{s\ell{m}}
         (\mathcal{H}_{s(\ell-1)m} + \mathcal{H}_{s\ell{m}}) , \\
  \label{eqn:x2coefC}
  \mathcal{C}_{s\ell{m}} &= \mathcal{G}_{s\ell{m}}\mathcal{G}_{s(\ell-1)m}.
\end{align}
\end{subequations}
Using $\mathcal{F}_{s(\ell_{\mbox{\tiny min}}-1){m}}=0$, together with
$\swS{s}{(\ell_{\mbox{\tiny min}}-1){m}}{x}{0} =
\swS{s}{(\ell_{\mbox{\tiny min}}-2){m}}{x}{0} = 0$, yields the five-term
recurrence relation
\begin{align}\label{eq:5-term_recurrence}
- c^2 \mathcal{A}_{s(\ell^\prime-2)m} C_{(\ell^\prime-2)\ell{m}}(c)&
\nonumber \\ \mbox{}
-  \left[c^2\mathcal{D}_{s(\ell^\prime-1)m} 
            - 2cs\mathcal{F}_{s(\ell^\prime-1)m}\right] C_{(\ell^\prime-1)\ell{m}}(c)&
\nonumber \\ \mbox{}
\mbox{} + \Bigl[\ell^\prime(\ell^\prime+1) -s(s+1) 
       - c^2 \mathcal{B}_{s\ell^\prime{m}} \hspace{0.7in}& 
  \\ \mbox{}
       + 2cs\mathcal{H}_{s\ell^\prime{m}} - \scA{s}{\ell{m}}{c} 
       \Bigr] C_{\ell^\prime\ell{m}}(c)&
\nonumber  \\ \mbox{}
\mbox{} -  \left[c^2\mathcal{E}_{s(\ell^\prime+1)m} 
            - 2cs\mathcal{G}_{s(\ell^\prime+1)m}\right] C_{(\ell^\prime+1)\ell{m}}(c)&
\nonumber  \\ \mbox{}
- c^2 \mathcal{C}_{s(\ell^\prime+2)m} 
       C_{(\ell^\prime+2)\ell{m}}(c)
& = 0. \nonumber
\end{align}

If we truncate the series at $\ell^\prime=\ell_{\mbox{\tiny max}}$
then we have a finite-dimensional spectral approximation to the
eigenvalue problem for the spin-weighted spheroidal harmonics.  With
$N=\ell_{\mbox{\tiny max}} - \ell_{\mbox{\tiny min}} + 1$, then for
given values of $s$ and $m$, we have an $N\times{N}$ symmetric
pentediagonal matrix $\mathbb{M}$ whose elements are given by
\begin{equation}
  M_{\ell\ell^\prime} = \left\{\begin{array}{lcl}
  \mbox{if\ }\ell^\prime=\ell-2 &:& -c^2 \mathcal{A}_{s\ell^\prime{m}}, \\
  \mbox{if\ }\ell^\prime=\ell-1 &:& -c^2 \mathcal{D}_{s\ell^\prime{m}} 
                                   + 2cs \mathcal{F}_{s\ell^\prime{m}} ,\\
  \mbox{if\ }\ell^\prime=\ell &:& \scA{s}{\ell^\prime{m}}{0}
                                   - c^2 \mathcal{B}_{s\ell^\prime{m}} 
                                   + 2cs \mathcal{H}_{s\ell^\prime{m}}, \\
  \mbox{if\ }\ell^\prime=\ell+1 &:& -c^2 \mathcal{E}_{s\ell^\prime{m}} 
                                   + 2cs \mathcal{G}_{s\ell^\prime{m}}, \\
  \mbox{if\ }\ell^\prime=\ell+2 &:& -c^2 \mathcal{C}_{s\ell^\prime{m}}, \\
  \mbox{otherwise} &:& 0.
  \end{array}\right.
\end{equation}
The truncated version of Eq.~(\ref{eq:5-term_recurrence}) is then
simply the eigenvalue equation
\begin{equation}\label{eqn::eigenvalue_eqn:fin}
  \mathbb{M}\cdot\vec{C}_{\ell{m}}(c) = \scA{s}{\ell{m}}{c}\vec{C}_{\ell{m}}(c).
\end{equation}
For given $s$, $m$, and $c$, the matrix $\mathbb{M}$ is constructed
and its $N$ eigenvalues are the $\scA{s}{\ell{m}}{c}$, where $\ell\in
\{\ell_{\mbox{\tiny min}},...,\ell_{\mbox{\tiny max}}\}$, and the
elements of the corresponding eigenvector $\vec{C}_{\ell{m}}(c)$ are the
expansion coefficients $C_{\ell^\prime\ell{m}}(c)$.

Because of the exponential convergence of spectral expansions, for
$\ell \ge \ell_{\mbox{\tiny min}}$ but $\ell \ll \ell_{\mbox{\tiny
    max}}$, we expect $\scA{s}{\ell{m}}{c}$ and $\vec{C}_{\ell{m}}(c)$
to have extremely high accuracy.  The largest value of $\ell$ for
which we can expect an accurate eigenvalue and eigenvector can be
determined by looking at the resulting values of the expansion
coefficients $C_{\ell^\prime\ell{m}}(c)$.  For given $\ell$ and $m$,
only coefficients where $\ell^\prime$ is reasonably close to $\ell$
will have significant weight.  As long as the coefficients
$C_{\ell^\prime\ell{m}}(c)$ are negligible for $\ell^\prime \ge
\ell_{\mbox{\tiny max}}$, then the solution associated with
$\scA{s}{\ell{m}}{c}$ will be highly accurate.

In practice, given values for $s$, $m$, and $c$ we choose a value for
$N$ and solve the eigenvalue problem of
Eq.~(\ref{eqn::eigenvalue_eqn:fin}) for its $N$ eigenvalues and
associated eigenvectors.  While many of these eigensolutions will be
accurate, we are typically only interested in the solution associated
with a particular value of $\ell$.  What is meant by the
$\ell$th eigenvalue is not uniquely defined.  For $c=0$, the
eigenvalues $\scA{s}{\ell{m}}{0}$ increase monotonically with $\ell$.
While this is a logical choice, it is simply a labeling scheme.  We
choose to identify the $\ell$th eigenvalue via continuity
along some sequence of solutions connected to the well-defined value
of $\scA{s}{\ell{m}}{0}$.  In practice, this means that we also need
the eigenvalue $\scA{s}{\ell{m}}{c-\epsilon}$ from a neighboring
solution along a sequence.  Here, $\epsilon$ is complex but
$|\epsilon|$ is small.  Then, we choose the eigenvalue
$\scA{s}{\ell{m}}{c}$ as the element, from the set of eigenvalues,
that minimizes $|\scA{s}{\ell{m}}{c}-\scA{s}{\ell{m}}{c-\epsilon}|$.
The necessity of picking one solution from a set of possible solutions
is not unique to this new spectral method for solving
Eq.~(\ref{eqn:swSF_DiffEqn}).  Rather, it is common to all methods.

In order to determine if the solution is accurate, we examine the
eigenvector, $\vec{C}_{\ell{m}}(c)$, associated with the chosen
eigenvalue.  The elements, $C_{\ell^\prime\ell{m}}(c)$, of
$\vec{C}_{\ell{m}}(c)$ are indexed by the spin-weighted {\em
  spherical} function index $\ell^\prime$.  When $\ell^\prime$ is
close to the value of the {\em spheroidal} label $\ell$, the
coefficients will have significant weight.  However, once the number
of values of $\ell^\prime$ used becomes sufficiently large (the
convergent regime) then $|C_{\ell^\prime\ell{m}}(c)|$ will decrease
exponentially with increasing $\ell^\prime$.  In practice, once a
solution is obtained, we determine $|C_{\ell^\prime\ell{m}}(c)|$ with
$\ell^\prime=\ell_{\mbox{\tiny max}}$ and demand that it be smaller
than some tolerance. If it is not, we increase $N$ and repeat the
process.  Because of exponential convergence, the magnitude of this
last element of $\vec{C}_{\ell{m}}(c)$ places a robust upper bound on
the error of the solution.

If the spin-weighted spheroidal eigenfunctions are needed, we must fix
their phase and normalization.  These are fixed by the corresponding
spin-weighted spherical function at $c=0$.  To fix the phase of the
solutions, we demand that $C_{\ell^\prime\ell{m}}(c)$ with
$\ell^\prime=\ell$ is real.  To maintain proper normalization, we
enforce $\sum_{\ell^\prime=\ell_{\mbox{\tiny min}}}^{\ell_{\mbox{\tiny
      max}}}{|C_{\ell^\prime\ell{m}}(c)|^2}=1$.

This spectral eigenvalue approach for solving
Eq.~(\ref{eqn:swSF_DiffEqn}) has several clear advantages over
approaches that use one of the variants of Leaver's method.  Compared
to an expansion in terms of powers of $(1+x)$ used in the traditional
Leaver approach, the size of the matrix $N$ is vastly smaller than the
number of terms required in the continued fraction in order to
determine the eigenvalue with high accuracy.  The use of an expansion
in terms of Jacobi polynomials provides a significant improvement.
This is easily seen because the spin-weighted spherical functions are
proportional to the Wigner $d$ function $d_{mn}^\ell(\theta)$ which
itself is the product of powers of both $\cos\frac\theta2$ and
$\sin\frac\theta2$, and a Jacobi polynomial.  However, the expansion
in terms of spin-weighted spherical functions is still more efficient.
More importantly, very sophisticated tools exist for solving matrix
eigenvalue problems and these can be used with the spectral eigenvalue
approach to simultaneously yield both the desired eigenvalue and the
coefficients of the series expansion of the spin-weighted spheroidal
function.  When the expansion coefficients are required for a solution
based on one of the variants of Leaver's method, it is important to
remember that it is a {\em minimal} solution that is being
constructed.  Care must be taken in constructing the coefficients so
that a dominant solution is not found instead.

The size of the matrix needed in order to meet a given tolerance tends to
increase with both $c$ and $\ell$.  For the QNM examples we have
considered, in the extreme Kerr limit of $a\to M$ and values of $\ell$
up to $12$, achieving a tolerance of $10^{-14}$ required a matrix no 
larger than $N=25$.
\begin{figure}
\vspace{0.1in}
\includegraphics[width=\linewidth,clip]{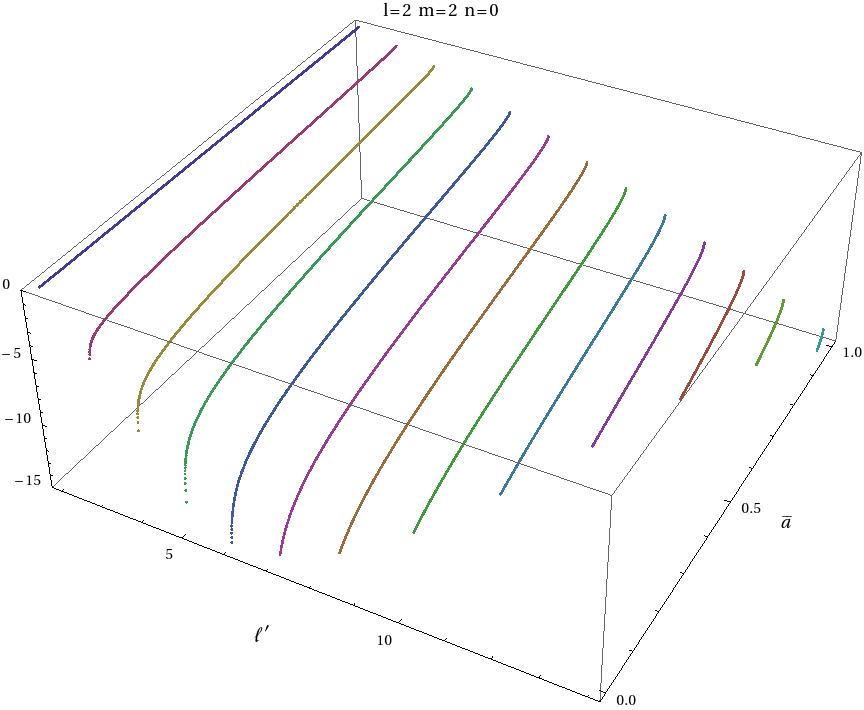}
\includegraphics[width=\linewidth,clip]{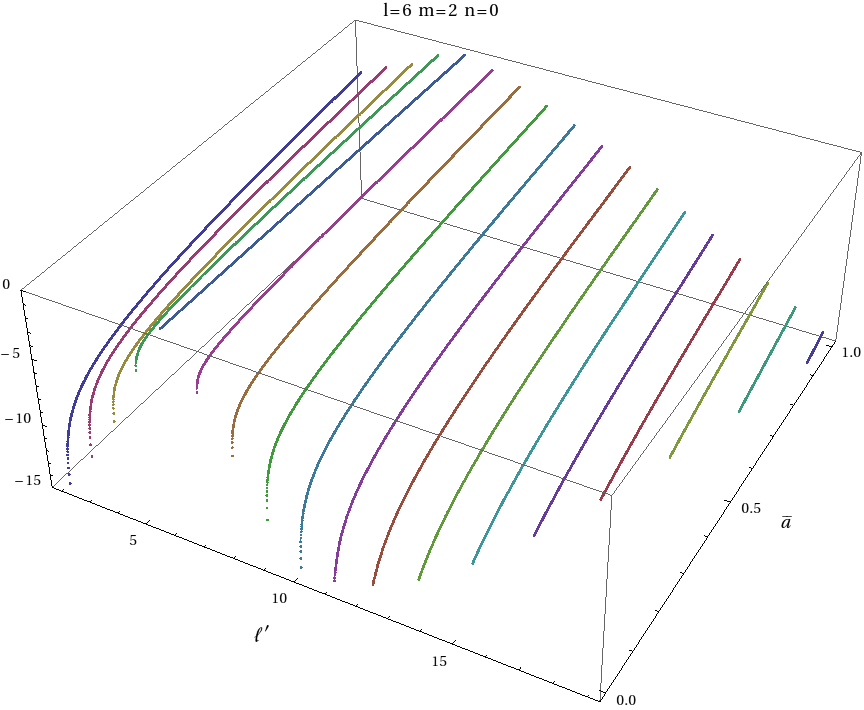}
\caption{\label{fig:Spectral_Coefs} Magnitude of the spectral
  coefficients $C_{\ell^\prime\ell{m}}(c)$ along Kerr QNM sequences.
  Each line plots $\log|C_{\ell^\prime\ell{m}}(c)|$, for a particular
  $\ell^\prime$ and with $c=\bar{a}\bar\omega$ along a particular QNM
  sequence with $0\le\bar{a}<1$.  The upper plot shows the
  coefficient of the fundamental mode for $\ell=2$, $m=2$ (see
  Fig.~\ref{fig:AOmega_l2m2}).  The lower plot shows the coefficients
  of the fundamental mode for $\ell=6$, $m=2$ (see
  Fig.~\ref{fig:AOmega_l6m2}).}
\end{figure}
Figure~\ref{fig:Spectral_Coefs} displays the behavior of the spectral
expansion coefficients, $C_{\ell^\prime\ell{m}}(c)$, for two
representative examples.  The upper plot of
Fig.~\ref{fig:Spectral_Coefs} corresponds to the fundamental mode of
$\ell=2$, $m=2$.  In this case, $\ell^\prime=2,\ldots,6$ for small
$\bar{a}$, and $\ell^\prime=2,\ldots,14$ in the extreme limit.  Along
the curve for each $\ell^\prime$, $\bar{a}$ varies from 0 to nearly 1,
and $c=\bar{a}\bar\omega$ where $\bar\omega$ is the complex QNM
frequency for this mode, the value of which also depends on $\bar{a}$.
The vertical axis gives the base-10 log of
$|C_{\ell^\prime\ell{m}}(c)|$.  We can clearly see that $C_{220}(c)$
contains the dominant amount of spectral power in this case, and that
the convergence of the expansion is exponential as $\ell^\prime$
increases.  The lower plot of Fig.~\ref{fig:Spectral_Coefs}
corresponds to the fundamental mode of $\ell=6$, $m=2$.  In this case,
$\ell^\prime=2,\ldots,10$ for small $\bar{a}$,
$\ell^\prime=2,\ldots,19$ in the extreme limit, and it is $C_{660}(c)$
that contains the dominant amount of spectral power in the expansion.
Again, we can see the exponential falloff in the importance of the
other spectral elements as $|\ell^\prime-\ell|$ increases.

\section{QNM{s} of Kerr}
\label{sec:Kerr_QNMs}

As discussed in Sec.~\ref{sec:Teukolsky_eqns}, to compute the QNMs of
Kerr requires that we find a solution of the coupled angular and
radial Teukolsky equations, Eqs.~(\ref{eqn:swSF_DiffEqn})
and~(\ref{eqn:radialR:Diff_Eqn}).  For given $\ell$ and $m$, there are
an infinite number of solutions to the coupled equations.  These correspond to
the fundamental mode and its overtones, labeled by the overtone
index $n$.  At $\bar{a}=0$, the fundamental mode has the slowest decay
rate [smallest $|\rm{Im}(\bar\omega)|$] and is labeled $n=0$.  The
overtones are labeled by increasing values for $n$ in order of
successively faster decay rates at $\bar{a}=0$.  At sufficiently
large values of $\bar{a}$, this ordering can change, but we
maintain the overtone index at $\bar{a}=0$ along a QNM sequence
as $\bar{a}$ increases.

The QNM spectrum is thus labeled by the triplet $\{\ell,m,n\}$ and the
complex frequency is a function of $\bar{a}$.  Because of a
fundamental symmetry to the equations, the Kerr QNM spectrum has two
distinct sets of modes for any given value of $\bar{a}$: the
``positive-frequency'' modes $\bar\omega_{\ell{m}n}$ with
$|\rm{Re}(\bar\omega)|\ge0$, and the ``negative-frequency''
modes $\bar\omega^\prime_{\ell{m}n}$.  The two sets are directly
related to each other by
\begin{equation}\label{eq:neg_freq_modes}
  \bar\omega^\prime_{\ell{m}n}(\bar{a})=-\bar\omega^*_{\ell(-m)n}(\bar{a}).
\end{equation}
The corresponding separation constants are related by
\begin{equation}
  \scA[\prime]{s}{\ell{m}}{\bar{a}\bar\omega^\prime_{\ell{m}n}} =
  \scA[*]{s}{\ell(-m)}{\bar{a}\bar\omega_{\ell(-m)n}}.
\end{equation}
Because of this symmetry, it is only necessary to compute either the
positive- or negative-frequency modes.  We will always compute the
positive-frequency modes.

In general, computation of these modes must be done numerically via an
iterative process.  The routines to do this were coded in {\tt
  Mathematica} in order to make use of its ability to perform extended
precision arithmetic.  Calculations were performed using 24 {\em
  decimal digit} precision (as opposed to the 16 digits of machine
precision).  In a few cases where even higher precision might be
required, we have used up to 32 digit precision.

In this paper, we will be concerned primarily with gravitational modes
so that $s=-2$; however the methods we discuss apply to any fixed
value of $s$.  For a given triplet $\{\ell,m,n\}$, a ``mode
sequence'' consists of computing the complex frequency
$\bar\omega_{\ell{m}n}(\bar{a})$, the separation constant
$\scA{s}{\ell{m}}{\bar{a}\bar\omega_{\ell{m}n}}$, and a set (indexed
by $\ell^\prime$) of spectral coefficients
$C_{\ell^\prime\ell{m}}(\bar{a}\bar\omega_{\ell{m}n})$ for
$0\le\bar{a}<1$.  A mode sequence is computed by providing initial
guesses for $\bar\omega_{\ell{m}n}(\bar{a})$ and
$\scA{s}{\ell{m}}{\bar{a}\bar\omega_{\ell{m}n}}$.  We {\em almost}
always start with $\bar{a}=0$, using values for the QNMs of
Schwarzschild.  Given a solution at this value of $\bar{a}$ we
increase $\bar{a}$ by at most $10^{-3}$ and use the previous solution
as an initial guess for the next.  When possible, we use quadratic
extrapolation based on three previous solutions to compute a more accurate
initial guess.  We use an adaptive step-size approach to ensure we
resolve fine details in the sequence, allowing steps as small as
$10^{-6}$ in general.  However, in order to explore the behavior near
$\bar{a}=1$, we use step sizes as small as $10^{-9}$.

Each element in a mode sequence is computed, starting with an initial
guess for $\bar\omega_{\ell{m}n}(\bar{a})$ and
$\scA{s}{\ell{m}}{\bar{a}\bar\omega_{\ell{m}n}}$, by finding
simultaneous solutions of the angular and radial Teukolsky equations,
Eqs.~(\ref{eqn:swSF_DiffEqn}) and~(\ref{eqn:radialR:Diff_Eqn}).  The
angular equation is solved using the spectral eigenvalue approach
outlined in Sec.~\ref{sec:sectral_EV_method} using the {\tt
  Mathematica} routine {\tt Eigensystem} to obtain the eigenvalues and
eigenvectors of the matrix.  The size of the matrix is at least $N=4$,
but is chosen dynamically.  The radial equation is solved using the
continued fraction method outlined in
Sec.~\ref{sec:Radial_Hc_functions}.  The desired root of the continued
fraction $\rm{Cf}(n;N)=0$ [see Eq.~(\ref{eq:Leaver_Cf_N_inv})] is
located using a two-dimensional Newton iteration with numerical
derivatives.  The ``inversion'' number of the continued fraction is
set to the overtone index $n$, and the truncation number $N$ is at
least 300, but is chosen dynamically.

An iterative approach is needed for solving the coupled pair of
equations.  The most common approach would be to solve the radial
equation for $\bar\omega$ holding
$\scA{s}{\ell{m}}{\bar{a}\bar\omega}$ fixed at the value from the
previous solution of the angular equation.  Then the angular equation
is solved for $\scA{s}{\ell{m}}{\bar{a}\bar\omega}$ holding
$\bar\omega$ fixed from its previous solution.  For most cases, an
iterative approach of this sort works well, but convergence to a
solution is not guaranteed.  In fact, there are situations where
convergence is very slow, or even fails unless some sort of
``underrelaxation'' is used.  We have even encountered extreme
situations where no amount of underrelaxation could achieve
convergence.  The source of the convergence problem, and its solution,
can be seen by considering the details of Newton's method used to
solve the radial equation.

Newton's method requires that we evaluate the derivative of the
function whose root we are seeking\cite{numrec_c}.  In our case, this
is done numerically via finite differencing.  The continued-fraction
equation derived from the radial equation is a function of both the
complex frequency $\bar\omega$ and the angular separation constant
$\scA{s}{\ell{m}}{\bar{a}\bar\omega}$ which is itself a function of
$\bar\omega$.  The derivative of the continued fraction equations
has the form
\begin{equation}
\label{eq:Newton-derivative}
  \frac{\partial f(\bar\omega,A)}{\partial\bar\omega}
  + \frac{\partial f(\bar\omega,A)}{\partial A}
    \frac{\partial A}{\partial\bar\omega}.
\end{equation}
During a standard iterative approach, where
$\scA{s}{\ell{m}}{\bar{a}\bar\omega}$ is held fixed, the second term
in Eq.~(\ref{eq:Newton-derivative}) is ignored.  If this term is
small, then it should not significantly affect the convergence of the
Newton iteration for the root of the continued fraction, and it may
not spoil the iterative convergence of the coupled equations.  However,
if this term is not small, then it can degrade the convergence of the
Newton iterations, and can spoil the convergence of the iterative
solution of the coupled equations.

The iterative convergence problem is solved by effectively including
this second term when computing the numerical derivatives of the
radial continued-fraction equation.  Our numerical derivatives are
computed by evaluating the continued-fraction equation at neighboring
values of $\bar\omega$.  Since $\bar\omega$ is complex, we must
compute derivatives with step $\delta\bar\omega$ both purely real and
purely imaginary.  Prior to evaluating the radial continued-fraction
equation for any value of $\bar\omega$, the angular equation is solved
with that value for $\bar\omega$.  The derivatives are then evaluated
as
\begin{equation}
  \frac{f(\bar\omega+\delta\bar\omega,A(a(\bar\omega+\delta\bar\omega)))
    -f(\bar\omega,A(a\bar\omega))}{\delta\bar\omega}.
\end{equation}
In this way, the coupled equations are solved simultaneously during
the Newton iteration process for finding the root of the radial
continued-fraction equation.  This process involves solving the
angular eigenvalue problem three times during each Newton iteration,
but solving the angular problem is fast in practice compared to each
evaluation of the radial continued fraction.

Once a root for the radial equation has been found, both the matrix
size and the continued fraction truncation length are checked to make
sure they provide sufficient accuracy.  If not, these are increased
and the entire process is repeated.

For the results discussed in this paper, sufficiently high precision
and sufficiently small tolerances have been used that we can expect
our results to have an absolute accuracy of at least $10^{-10}$.
This is a conservative estimate based on the possibility of
pathological cases such as the continued fraction equation having
small derivatives local to the root.  In general, we expect an
absolute accuracy of $\sim10^{-12}$.

\subsection{The gravitational QNMs}
\label{sec:Grav_QNMs}

\begin{figure*}[ht]
\includegraphics[width=\linewidth,clip]{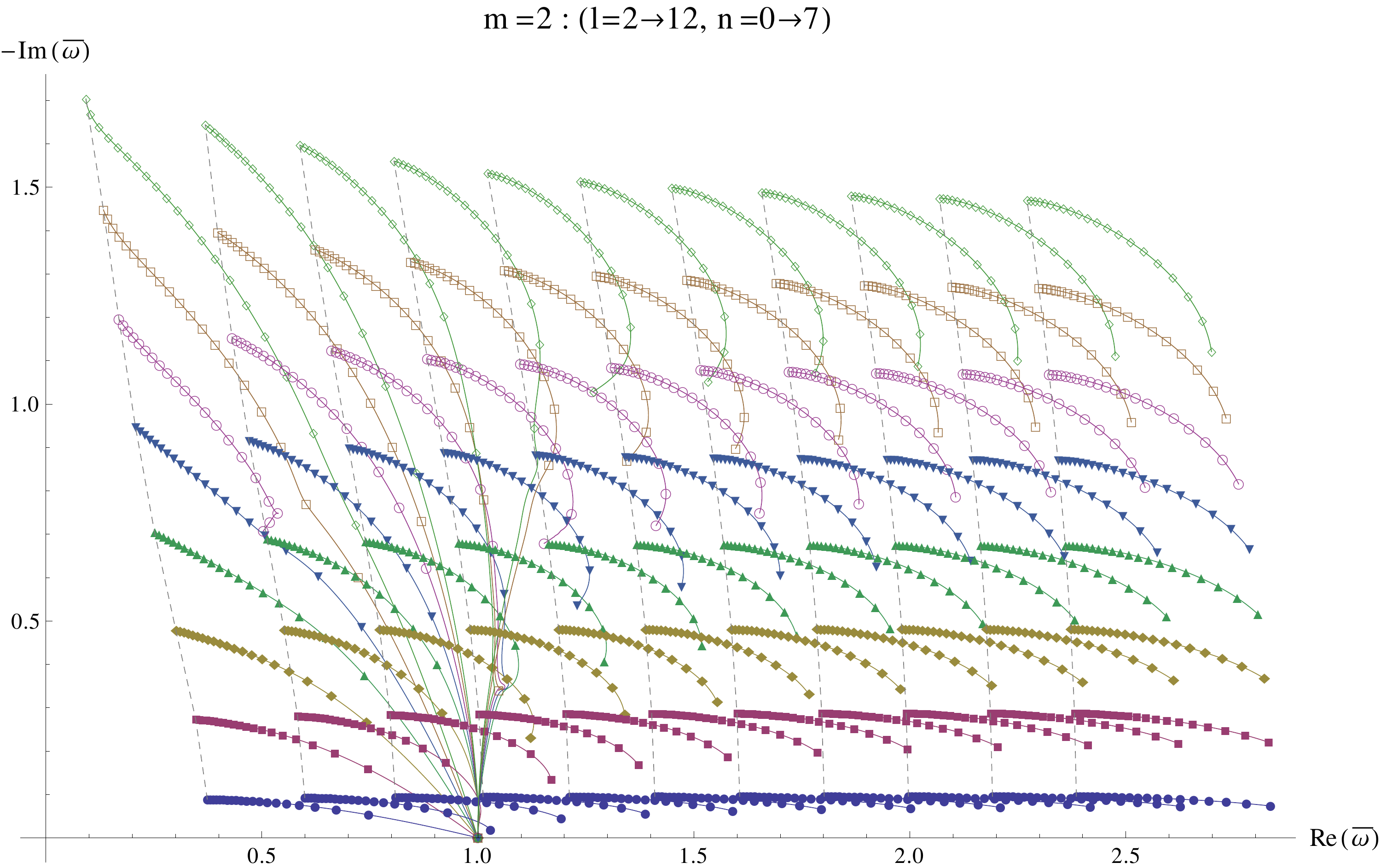}
\caption{\label{fig:AllOmega_m2} Kerr QNM mode sequences for $m=2$.
  The complex frequency $\bar\omega$ is plotted for the cases
  $\ell=2\to12$ and $n=0\to7$.  Note that the imaginary axis is
  inverted.  Each sequence covers the range $0\le\bar{a}<1$, with
  markers on each sequence denoting a change in $\bar{a}$ of $0.05$.
  The $\bar{a}=0$ element of each set of mode sequences with the same
  $\ell$ are connected by a dashed line.  $\ell$ increases
  monotonically as we move to the right in the plot.  The overtone
  index $n$ for each set of modes with the same $\ell$ increases
  monotonically as we move up along each dashed line.}
\end{figure*}

\begin{figure*}[htbp!]
\begin{tabular}{cc}
\includegraphics[width=0.5\linewidth,clip]{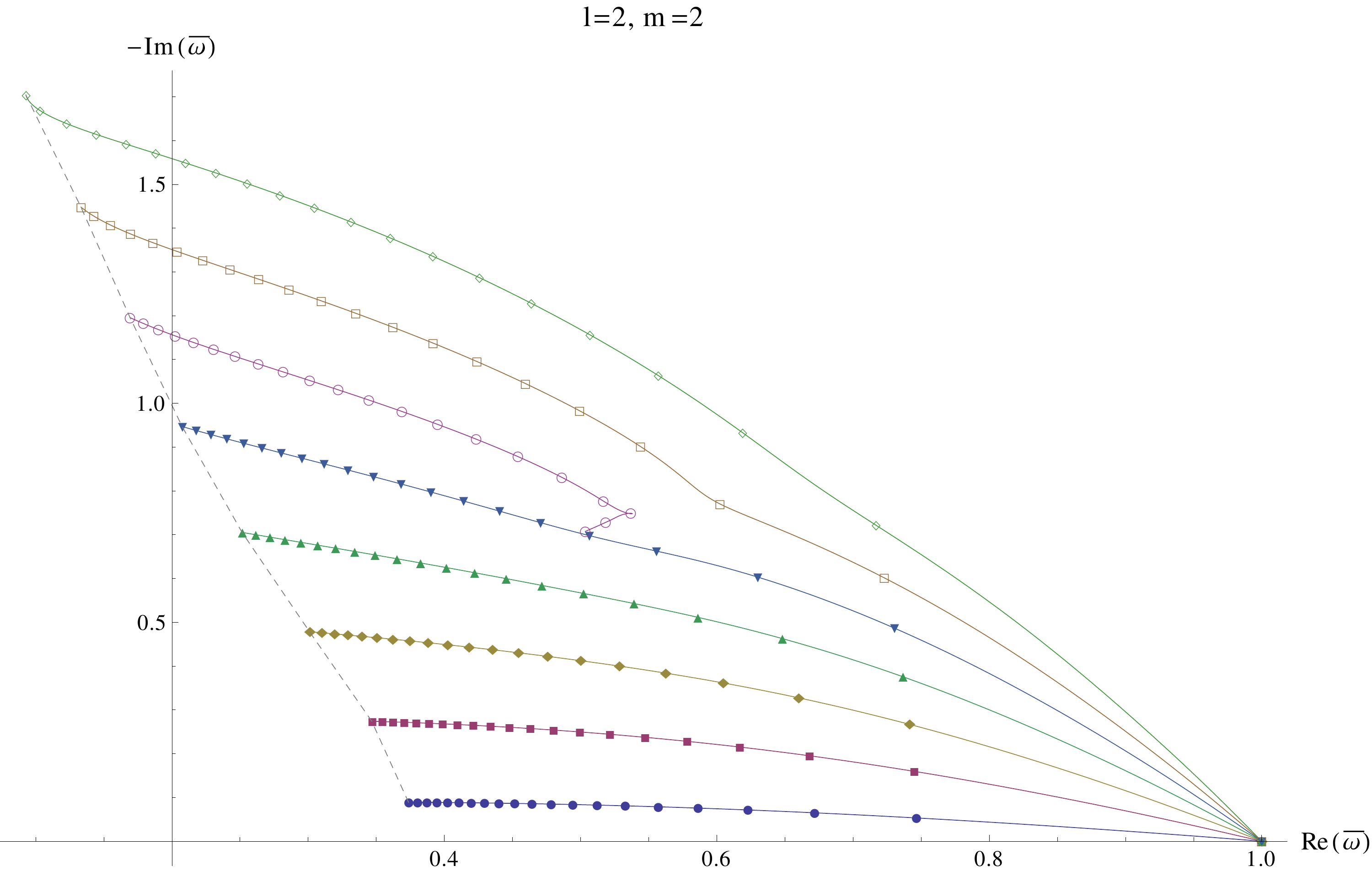} &
\includegraphics[width=0.5\linewidth,clip]{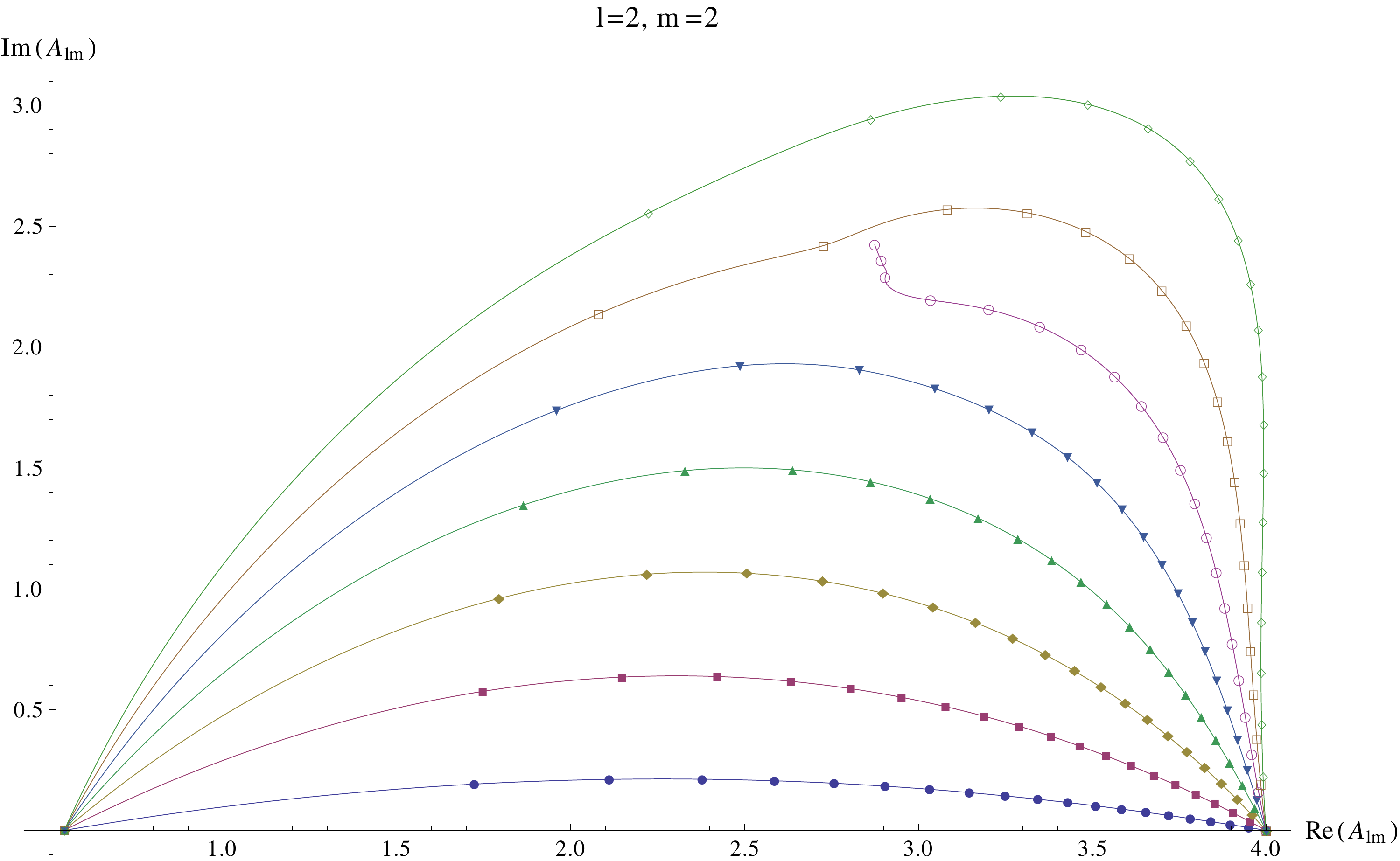}
\end{tabular}
\caption{\label{fig:AOmega_l2m2} Kerr QNM mode sequences for $\ell=2$
  and $m=2$.  The left plot displays the complex frequency
  $\bar\omega$ and the right plot displays the separation constant
  $\scA{-2}{22}{\bar{a}\bar\omega_{22n}}$ for corresponding mode
  sequences.  The left plot is as described in the caption for
  Fig.~\ref{fig:AllOmega_m2}.  For the right plot, each sequence
  begins with $\bar{a}=0$ on the real axis at $\ell(\ell+1)-2$.  The
  overtones all radiate from this point with $n$ increasing in a
  clockwise direction.  $n=0\to7$ are displayed for both plots.}
\end{figure*}
\begin{figure}[htbp!]
\includegraphics[width=\linewidth,clip]{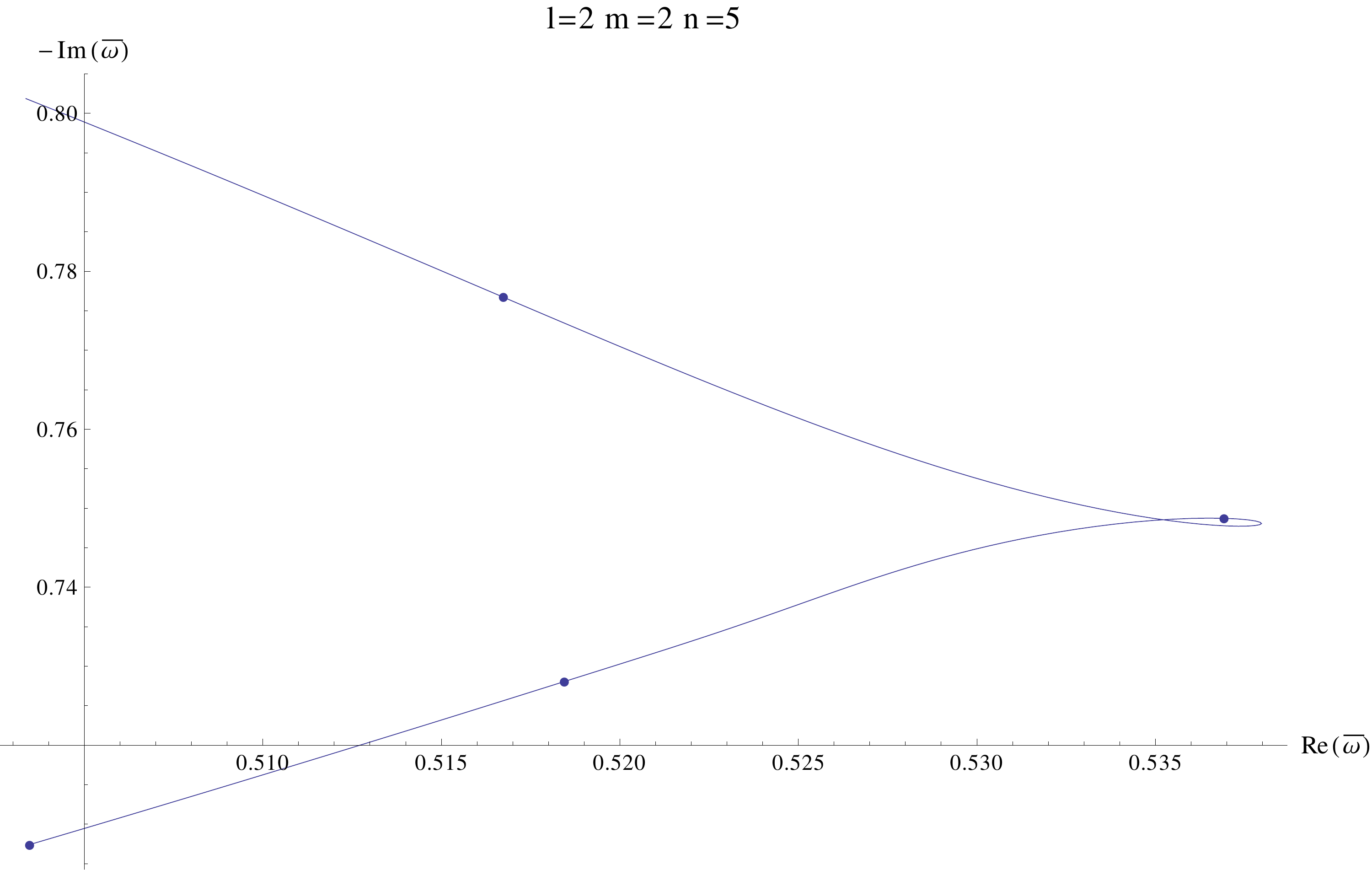}
\caption{\label{fig:Zoom_O_l2m2n5} Close-up of $\bar\omega$ for the
  $\{2,2,5\}$ sequence showing the unusual termination of this
  sequence.  Instead of approaching the accumulation point at $m/2$,
  this sequence makes a sharp loop and then terminates with finite
  damping. See Fig.~\ref{fig:AOmega_l2m2} for context.}
\end{figure}

We have performed an extensive, high-accuracy survey of the
gravitational ($s=-2$) QNMs of Kerr for $2\le\ell\le12$,
$-\ell\le{m}\le\ell$, and $0\le{n}\le7$.  In addition we have computed
the mode sequences for $n=8$ with $\ell=2$ and $-2\le{m}\le2$.  Many
of the results covered in this set have been computed and discussed in
the past\cite{leaver-1985,onozawa-1997,berticardoso-2003}.  We will
not attempt to display all of these results.  We will display some
representative examples, particularly interesting mode sequences, and
enough detail to allow our results to be confirmed.  We will then turn
to a detailed exploration of two interesting problems: the extremal
limit, and zero-frequency modes.

We begin with Fig.~\ref{fig:AllOmega_m2} which displays the complex
frequencies $\bar\omega$ for all mode sequences with $m=2$.  The plot
is rather dense but provides a useful overview of the $m=2$ modes.  We
note that in all plots of $\bar\omega$, we invert the imaginary axis.
Beginning on the left side of the figure, we find eight sequences
connected by a dashed line.  We will denote individual mode sequences
by their triplet $\{\ell,m,n\}$.  The bottommost sequence on the left
is $\{2,2,0\}$.  The topmost sequence on the left is $\{2,2,7\}$.
Between them, in order going up the figure, are the $n=1\textendash6$
modes.  The dashed line connecting these eight mode sequences passes
through the $\bar{a}=0$ mode of each sequence and $\bar{a}$ increase
along each sequence as it extends from this point.  The markers along
each sequence denote an increase in $\bar{a}$ of $0.05$.  The last
marker, on each sequence appears at $\epsilon\equiv1-\bar{a}=10^{-6}$
for most sequences, but for sequences that approach the real axis, it
is at $\epsilon=10^{-8}$.  Just to the right of the $\ell=2$ modes are
the $\ell=3$ modes.  Again for each $n$, the $\bar{a}=0$ modes are
connected by a dashed line.  Proceeding sequentially to the right
we find the remaining $\ell=4\textendash12$ mode sequences.

The left plot in Fig.~\ref{fig:AOmega_l2m2} shows the $\{2,2,n\}$
modes in isolation.  The sequences in this figure are exactly as
described above.  The right plot shows the corresponding sequences for
the separation constant $\scA{-2}{22}{\bar{a}\bar\omega_{22n}}$.  All
$\scA{-2}{\ell{m}}{}$ sequences begin on the real axis at
$\ell(\ell+1)-2$.  In the neighborhood of this point, the overtone
index $n$ of each sequence increases as we move to a new sequence in a
clockwise direction.  Markers along these sequences are at the same
values of $\bar{a}$ as their $\bar\omega$ counterparts.

Notice that all of the mode sequences, except $n=5$, approach an
accumulation point at $\bar\omega=m/2$.  This behavior is well known,
studied by Detweiler\cite{detweiler80} and
others\cite{Glampedakis01,cardoso04,Hod08,Yang-et-al-2013a,Yang-et-al-2013b}.
As the mode sequences approach the real axis, the decrease in the
imaginary part of $\bar\omega$ corresponds to a decrease in the mode
damping rate.  We will consider these accumulation points in more
detail below.  There is no clear understanding yet of why some modes,
like $\{2,2,5\}$ break the trend of neighboring modes.  As seen in
Fig.~\ref{fig:Zoom_O_l2m2n5}, the anomalous behavior is not a cusp,
but rather a loop.  The behavior of this mode was first described by
Leaver\cite{leaver-1985}.

\begin{figure*}[htbp!]
\begin{tabular}{cc}
\includegraphics[width=0.5\linewidth,clip]{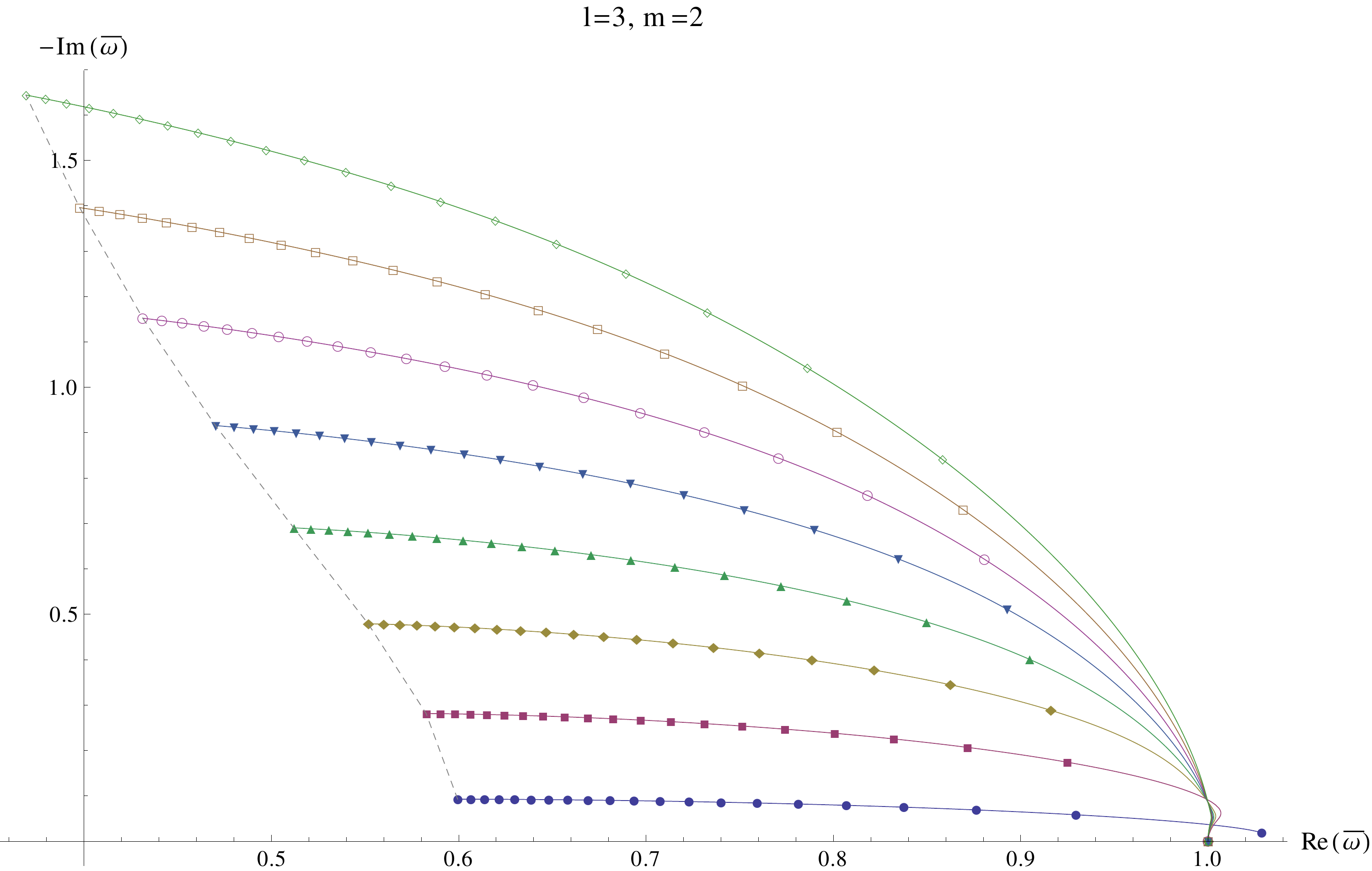} &
\includegraphics[width=0.5\linewidth,clip]{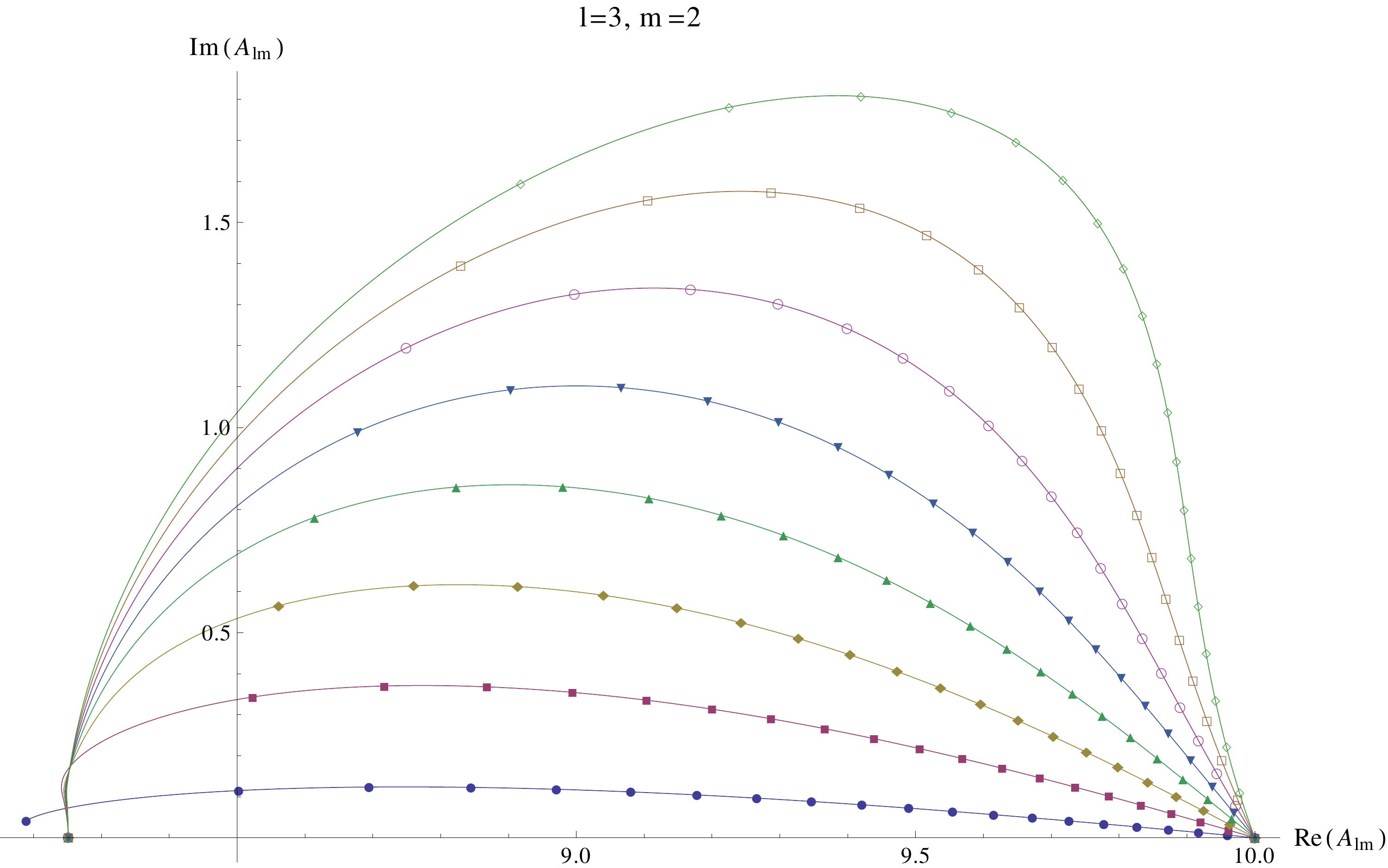}
\end{tabular}
\caption{\label{fig:AOmega_l3m2} Kerr QNM mode sequences for $\ell=3$
  and $m=2$. Plots are as described in Fig.~\ref{fig:AOmega_l2m2}.}
\end{figure*}
The plots in Fig.~\ref{fig:AOmega_l3m2} show the $\{3,2,n\}$ modes
in isolation.  In contrast to the $\{2,2,n\}$ modes, we note that
the $\{3,2,0\}$ sequence does not approach the accumulation point
at $\bar\omega=m/2$.  However, all of the displayed sequences with
$n\ge1$ do approach this accumulation point.
\begin{figure*}[htbp!]
\begin{tabular}{cc}
\includegraphics[width=0.5\linewidth,clip]{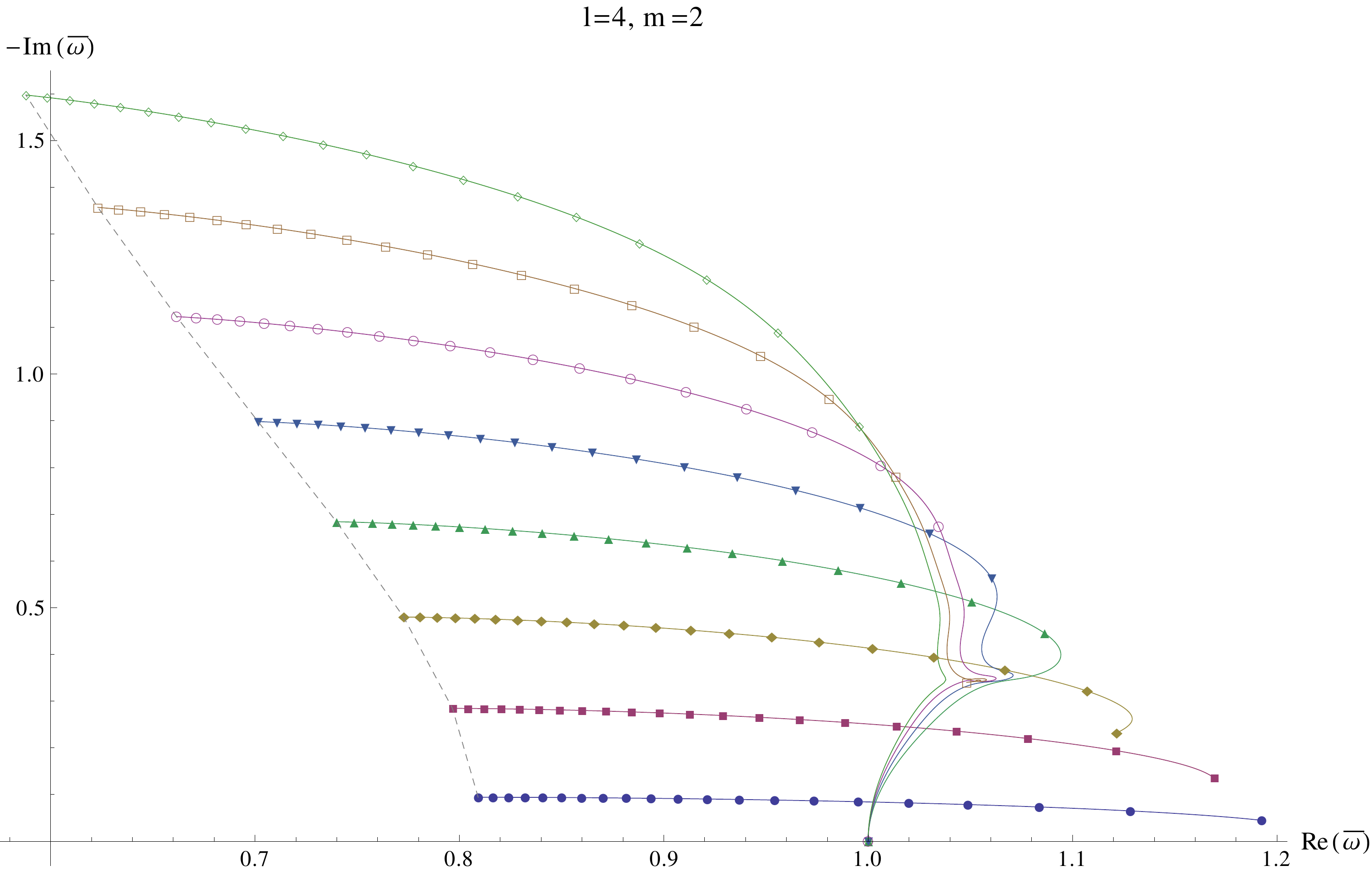} &
\includegraphics[width=0.5\linewidth,clip]{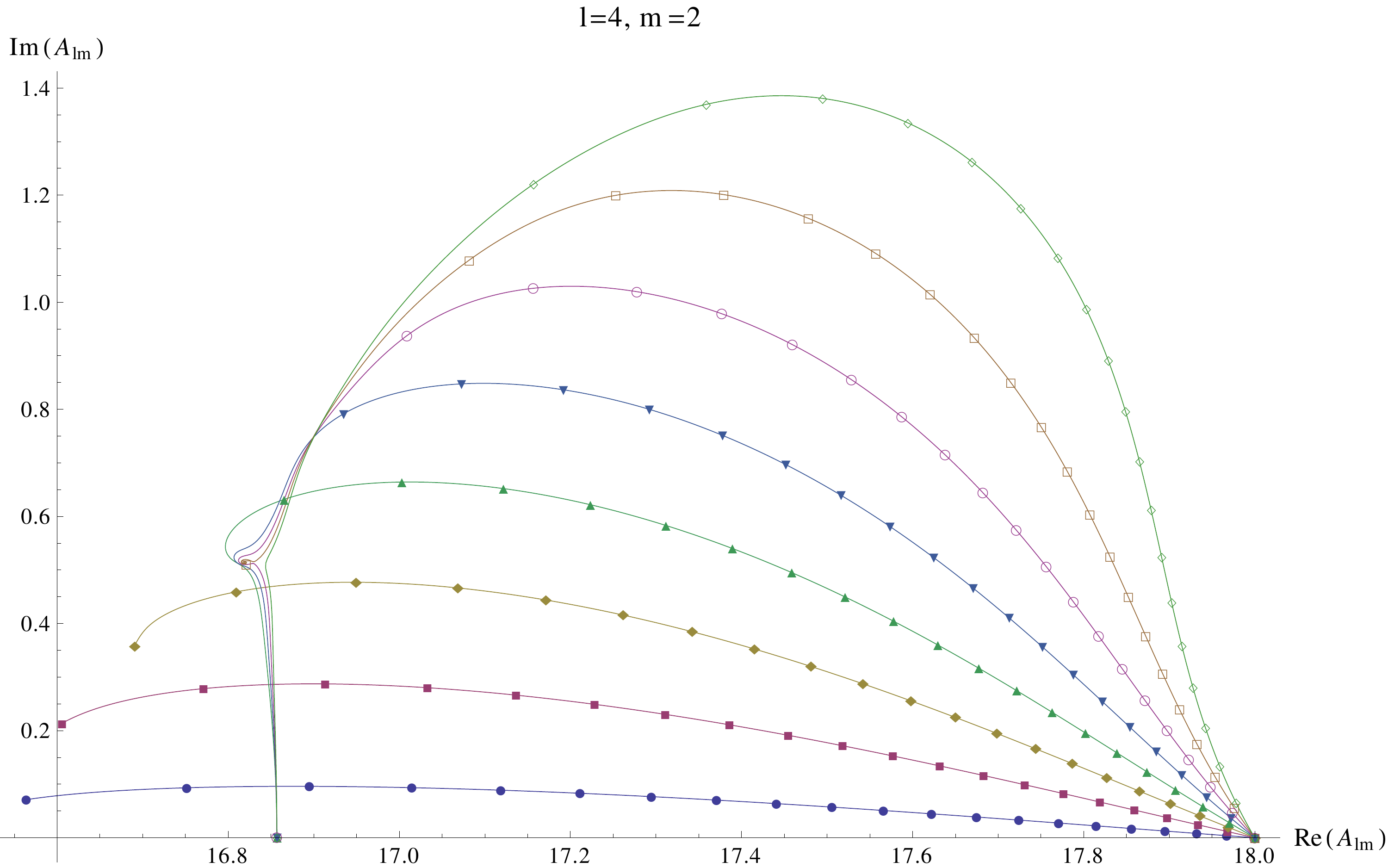}
\end{tabular}
\caption{\label{fig:AOmega_l4m2} Kerr QNM mode sequences for $\ell=4$
  and $m=2$. Plots are as described in Fig.~\ref{fig:AOmega_l2m2}.}
\end{figure*}
We see similar behavior in Fig.~\ref{fig:AOmega_l4m2} which shows the
$\{4,2,n\}$ mode sequences in isolation.  In this case, the first three
sequences ($n=0$, 1, and 2) do not approach the accumulation point.
Sequences with higher overtones do approach the accumulation point
with the exception of $n=6$.  As seen in Fig.~\ref{fig:Zoom_O_l4m2n6},
this anomalous sequence terminates with finite damping after executing
a number of spirals.
\begin{figure}[htbp!]
\includegraphics[width=\linewidth,clip]{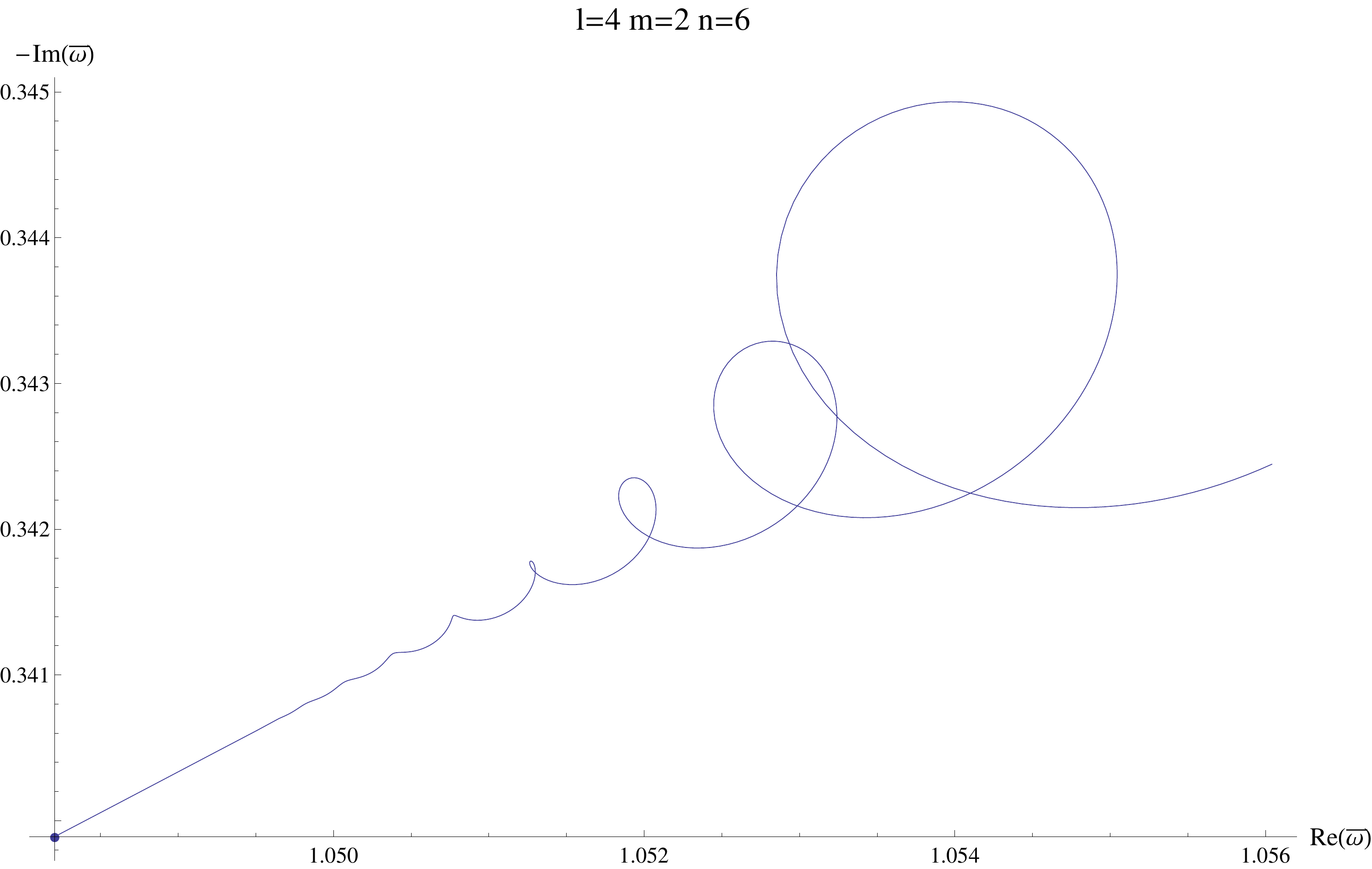}
\caption{\label{fig:Zoom_O_l4m2n6} Close-up of $\bar\omega$ for the
  $\{4,2,6\}$ sequence showing the unusual termination of this
  sequence. See Fig.~\ref{fig:AOmega_l4m2} for context.}
\end{figure}
Finally, Figs.~\ref{fig:AOmega_l5m2} and~\ref{fig:AOmega_l6m2} show,
respectively, the $\{5,2,n\}$ and $\{6,2,n\}$ mode sequences in
isolation.  For $\ell=5$, the sequences with $n=0\textendash5$ do not
approach the accumulation point while the last two do.  For $\ell=6$,
none of the displayed sequences approach the accumulation point.
However, we do expect that mode sequences with high enough overtone
numbers will terminate at the accumulation point.
\begin{figure*}[htbp!]
\begin{tabular}{cc}
\includegraphics[width=0.5\linewidth,clip]{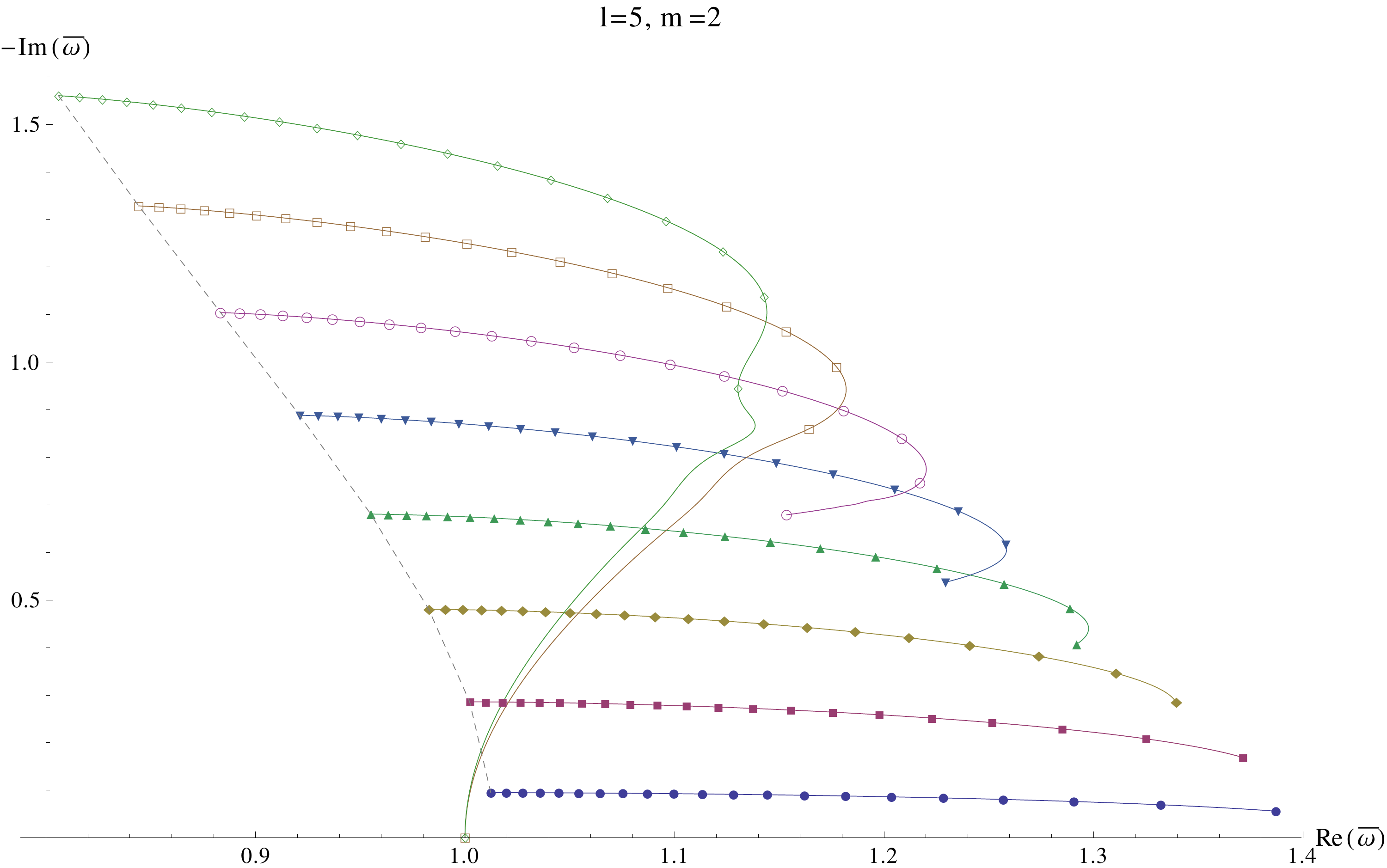} &
\includegraphics[width=0.5\linewidth,clip]{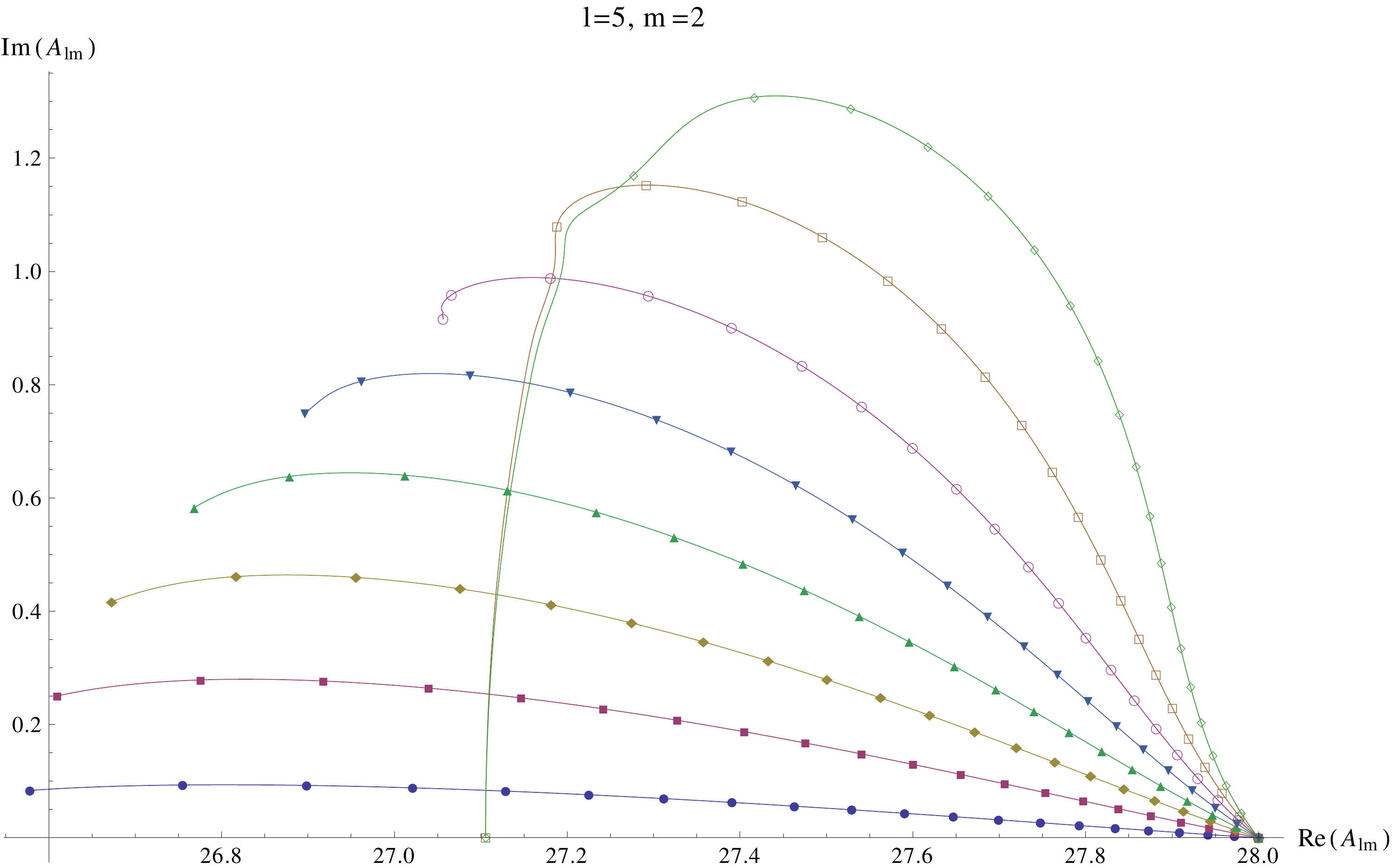}
\end{tabular}
\caption{\label{fig:AOmega_l5m2} Kerr QNM mode sequences for $\ell=5$
  and $m=2$. Plots are as described in Fig.~\ref{fig:AOmega_l2m2}.}
\end{figure*}
\begin{figure*}[htbp!]
\begin{tabular}{cc}
\includegraphics[width=0.5\linewidth,clip]{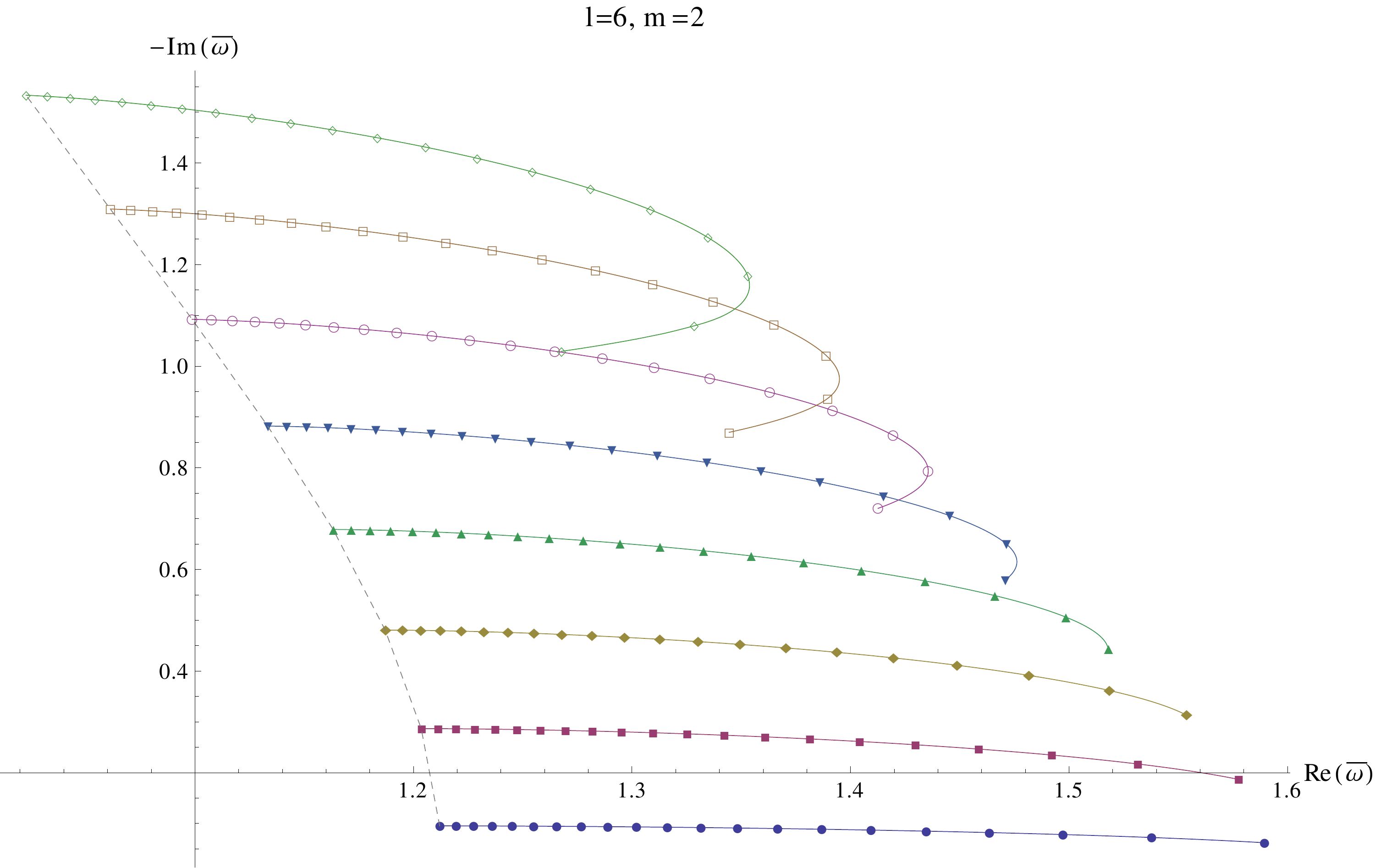} &
\includegraphics[width=0.5\linewidth,clip]{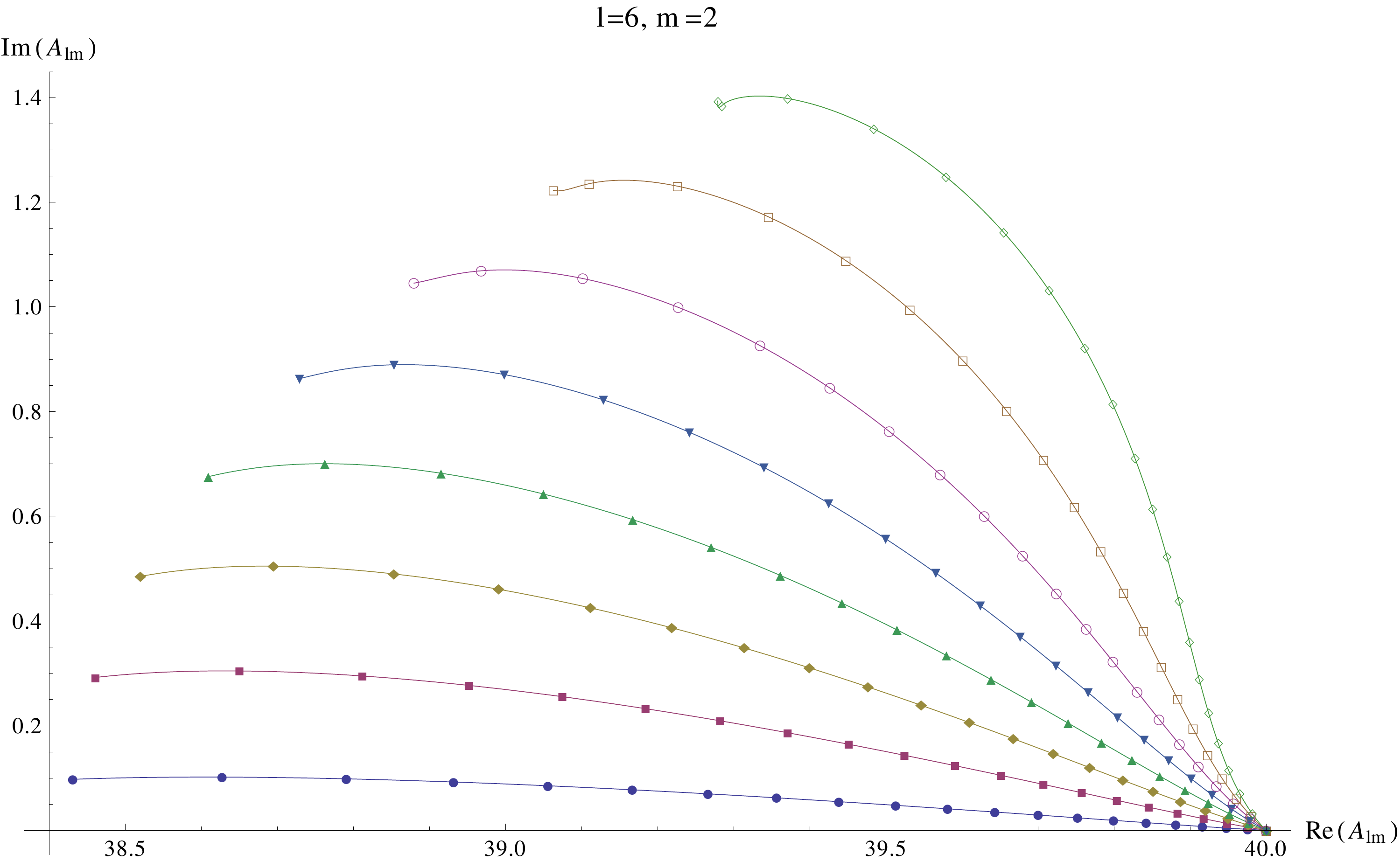}
\end{tabular}
\caption{\label{fig:AOmega_l6m2} Kerr QNM mode sequences for $\ell=6$
  and $m=2$. Plots are as described in Fig.~\ref{fig:AOmega_l2m2}.}
\end{figure*}

\begin{figure*}[htbp!]
\includegraphics[width=\linewidth,clip]{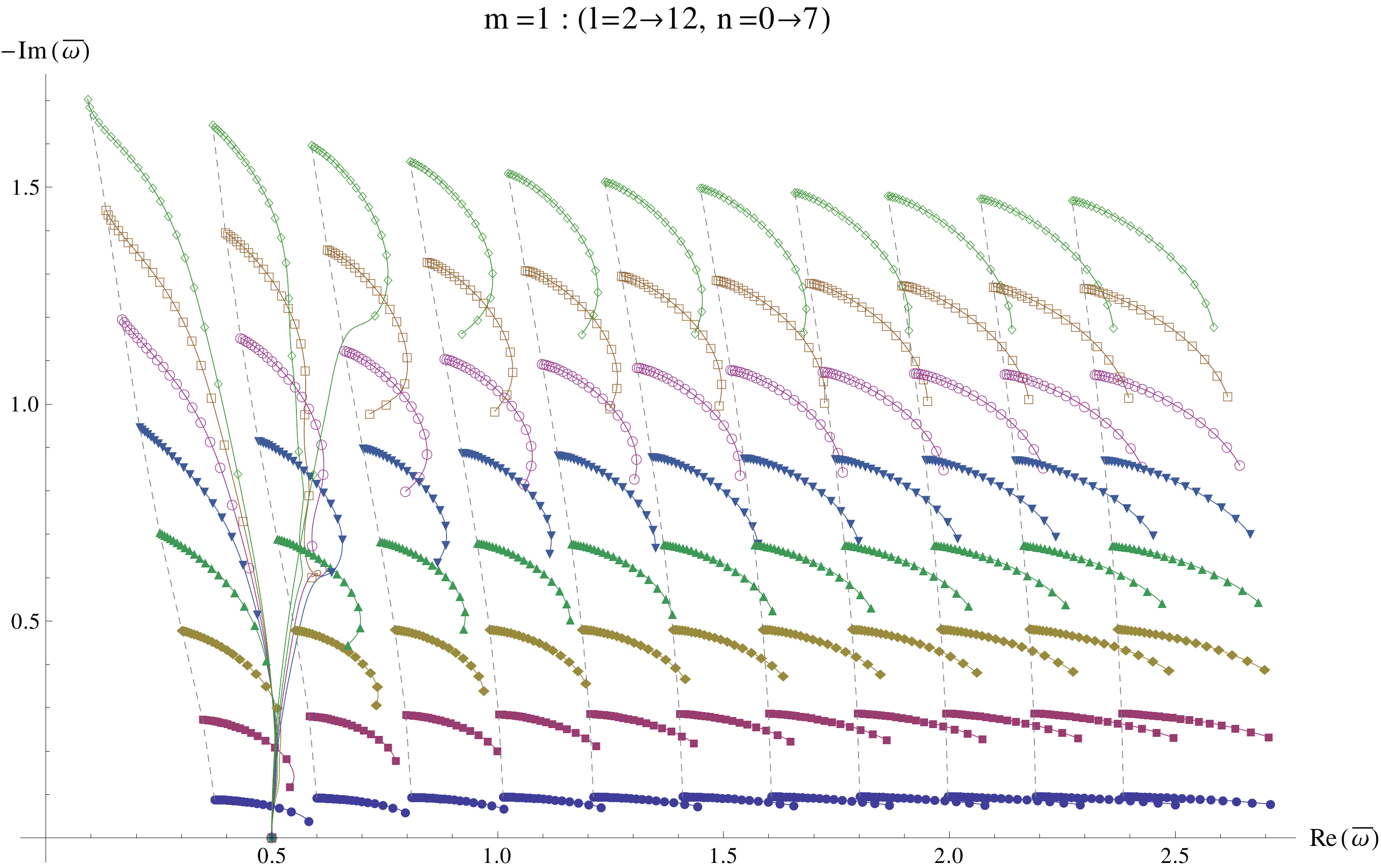}
\caption{\label{fig:AllOmega_m1} Kerr QNM mode sequences for $m=1$.
  See Fig.~\ref{fig:AllOmega_m2} for a full description.}
\end{figure*}

\begin{figure*}[htbp!]
\begin{tabular}{cc}
\includegraphics[width=0.5\linewidth,clip]{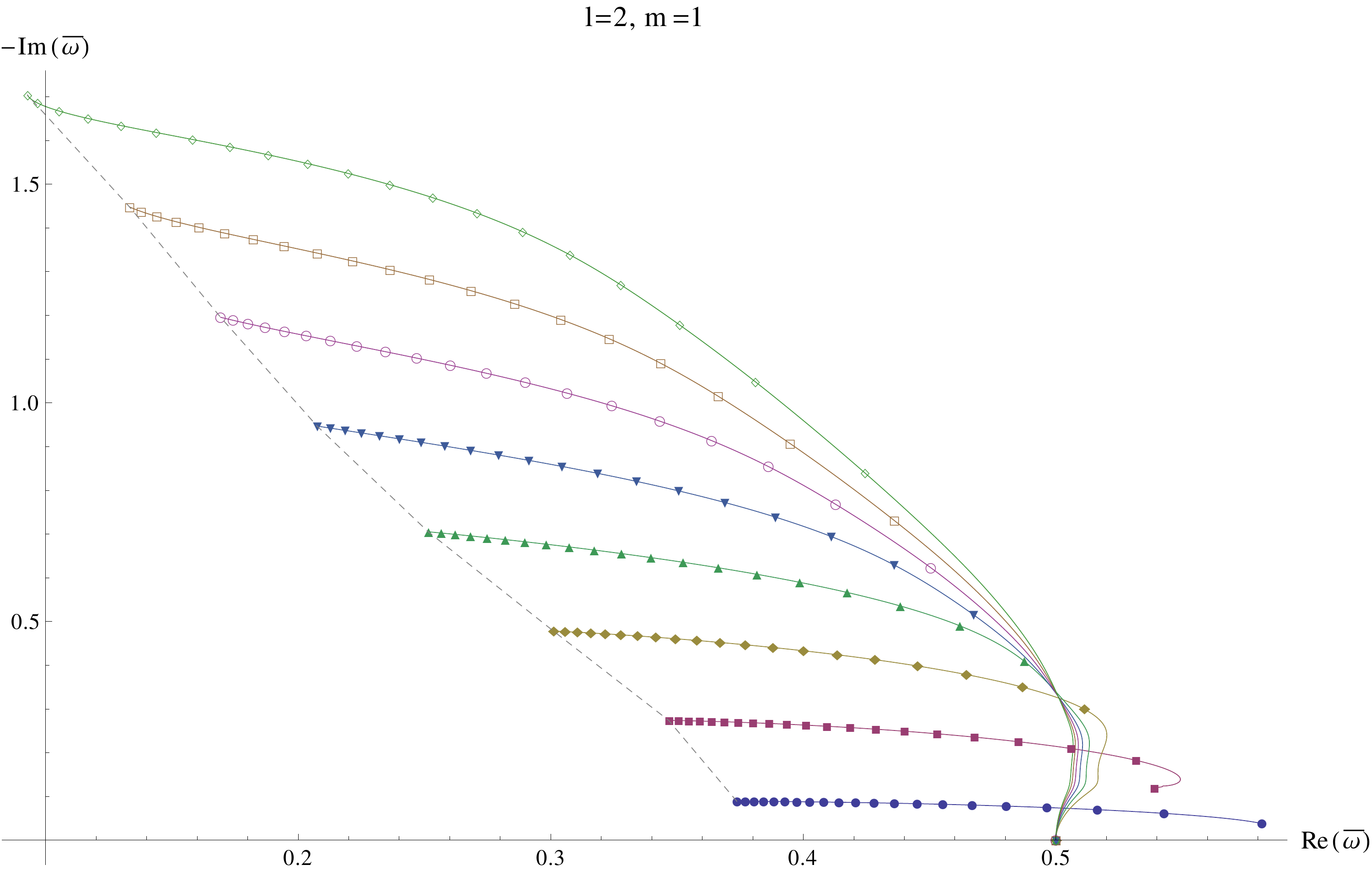} &
\includegraphics[width=0.5\linewidth,clip]{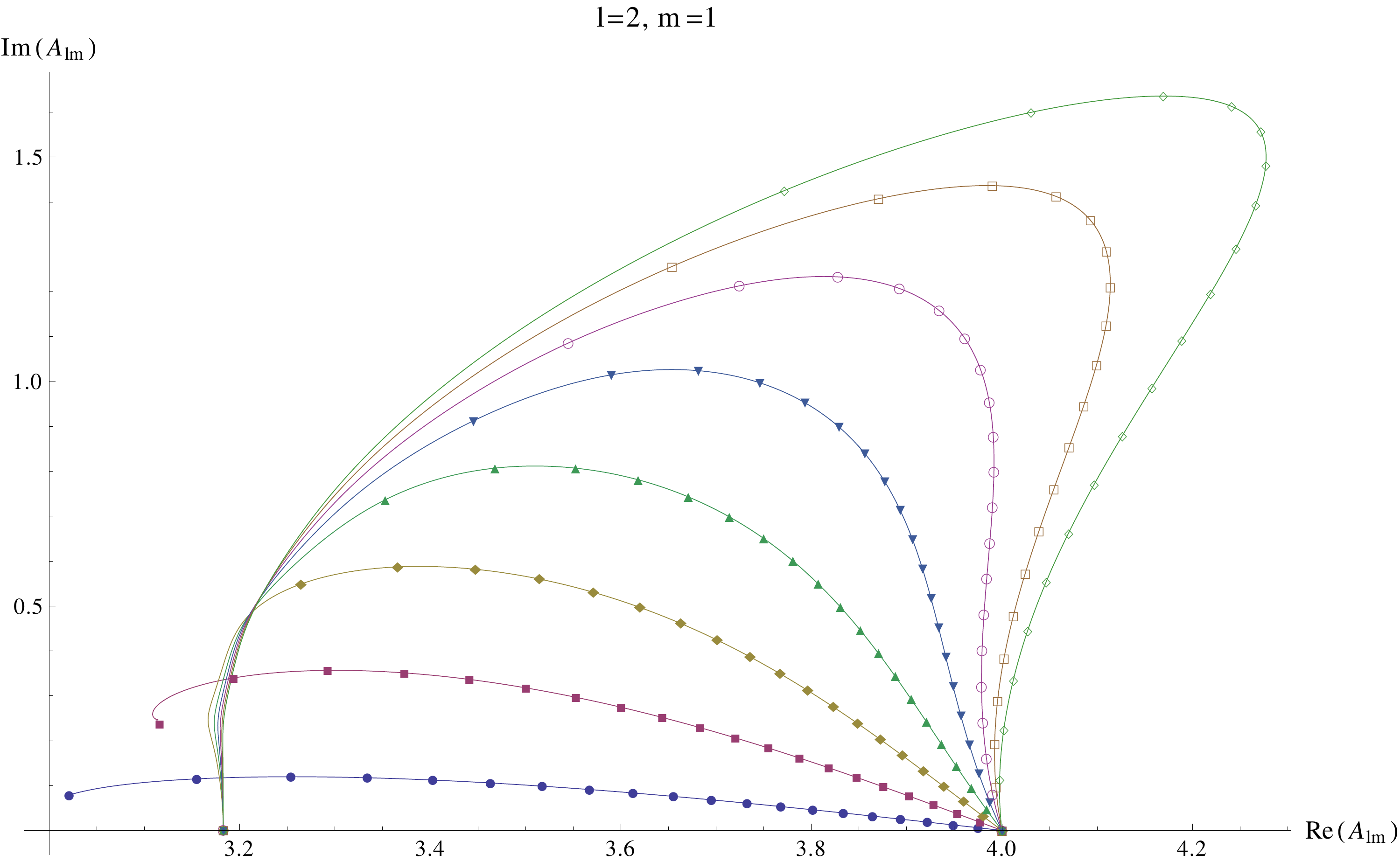}
\end{tabular}
\caption{\label{fig:AOmega_l2m1} Kerr QNM mode sequences for $\ell=2$
  and $m=1$. Plots are as described in Fig.~\ref{fig:AOmega_l2m2}.}
\end{figure*}

\begin{figure*}[htbp!]
\begin{tabular}{cc}
\includegraphics[width=0.5\linewidth,clip]{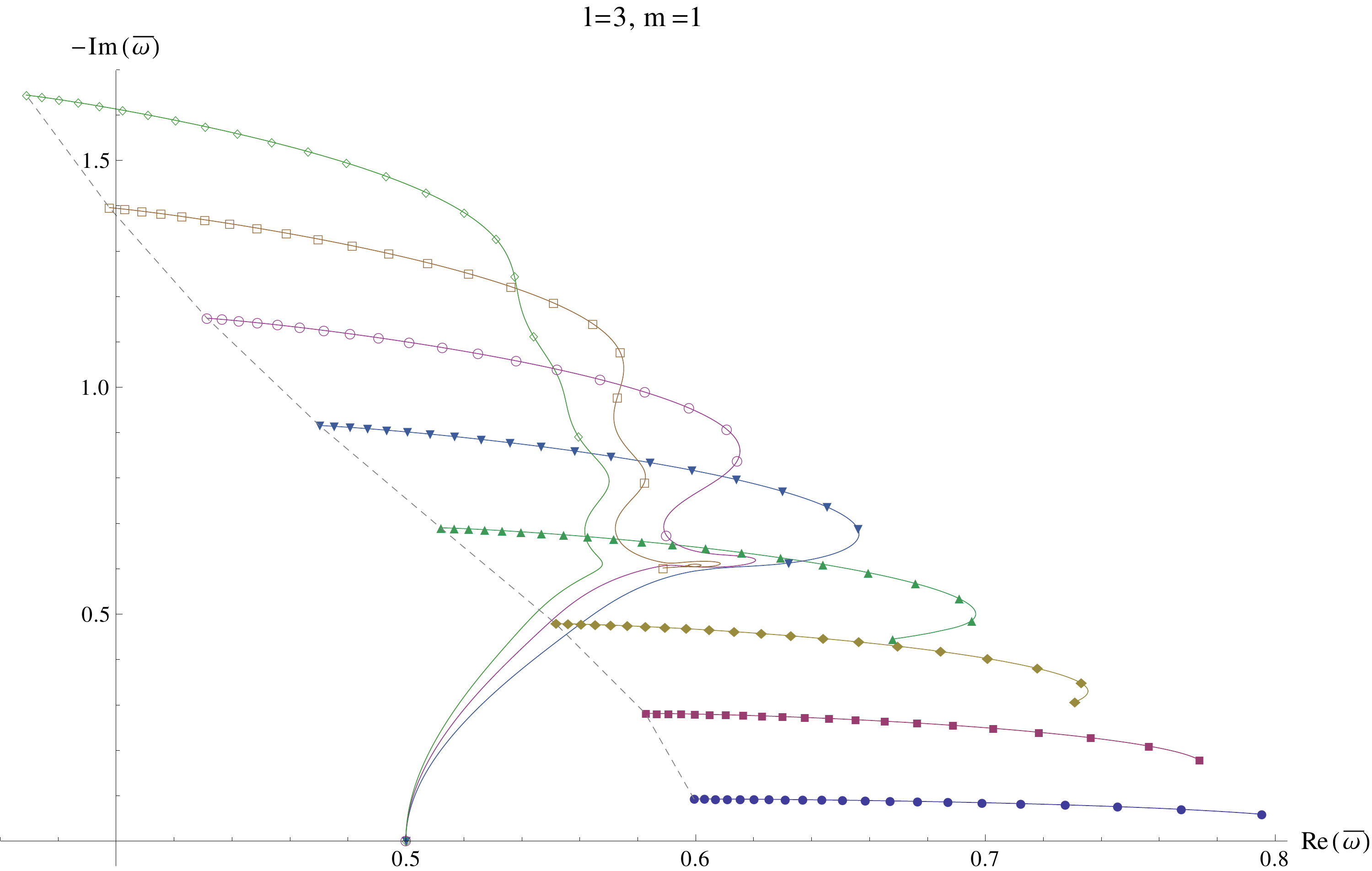} &
\includegraphics[width=0.5\linewidth,clip]{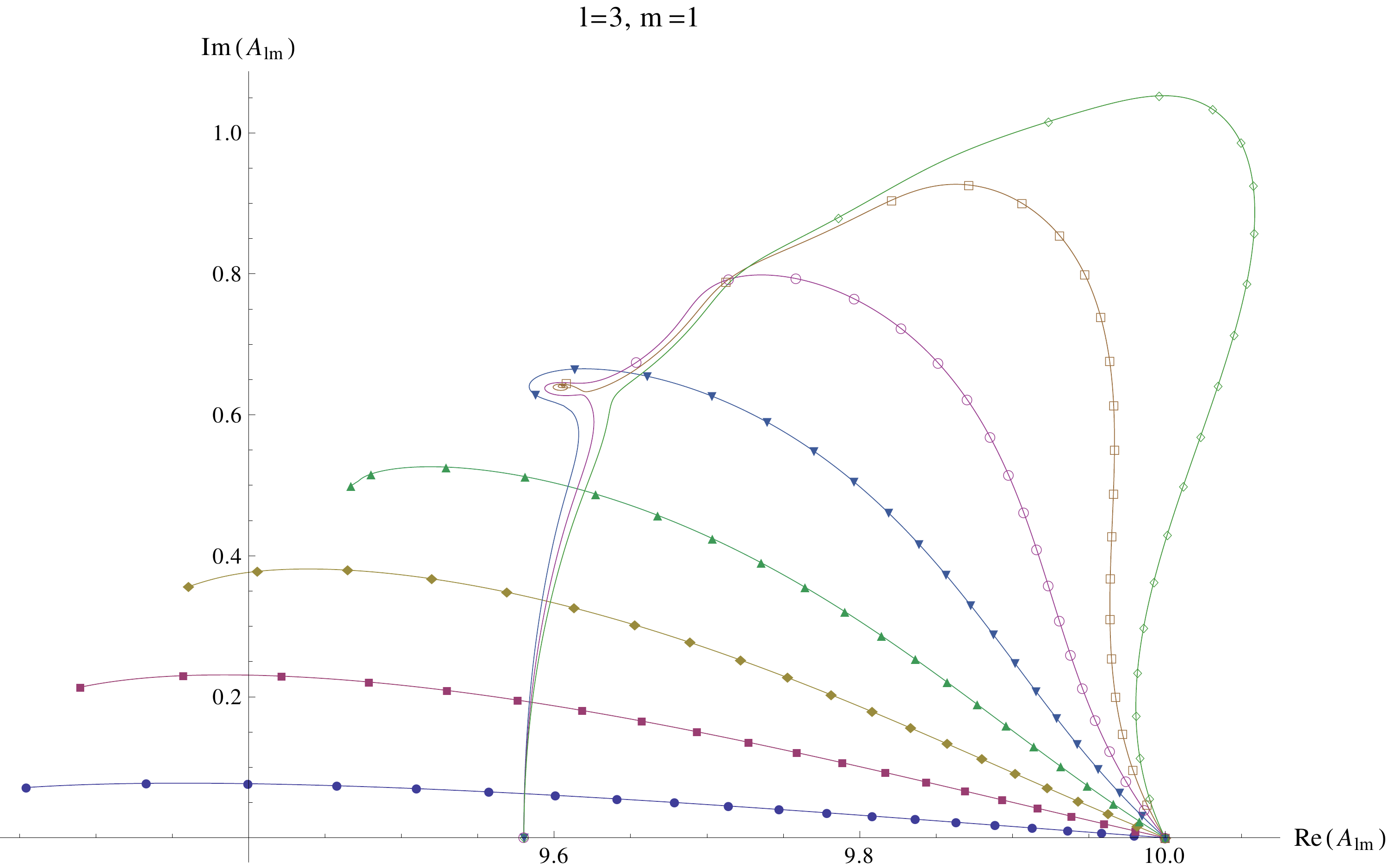}
\end{tabular}
\caption{\label{fig:AOmega_l3m1} Kerr QNM mode sequences for $\ell=3$
  and $m=1$. Plots are as described in Fig.~\ref{fig:AOmega_l2m2}.}
\end{figure*}

\begin{figure}[htbp!]
\includegraphics[width=\linewidth,clip]{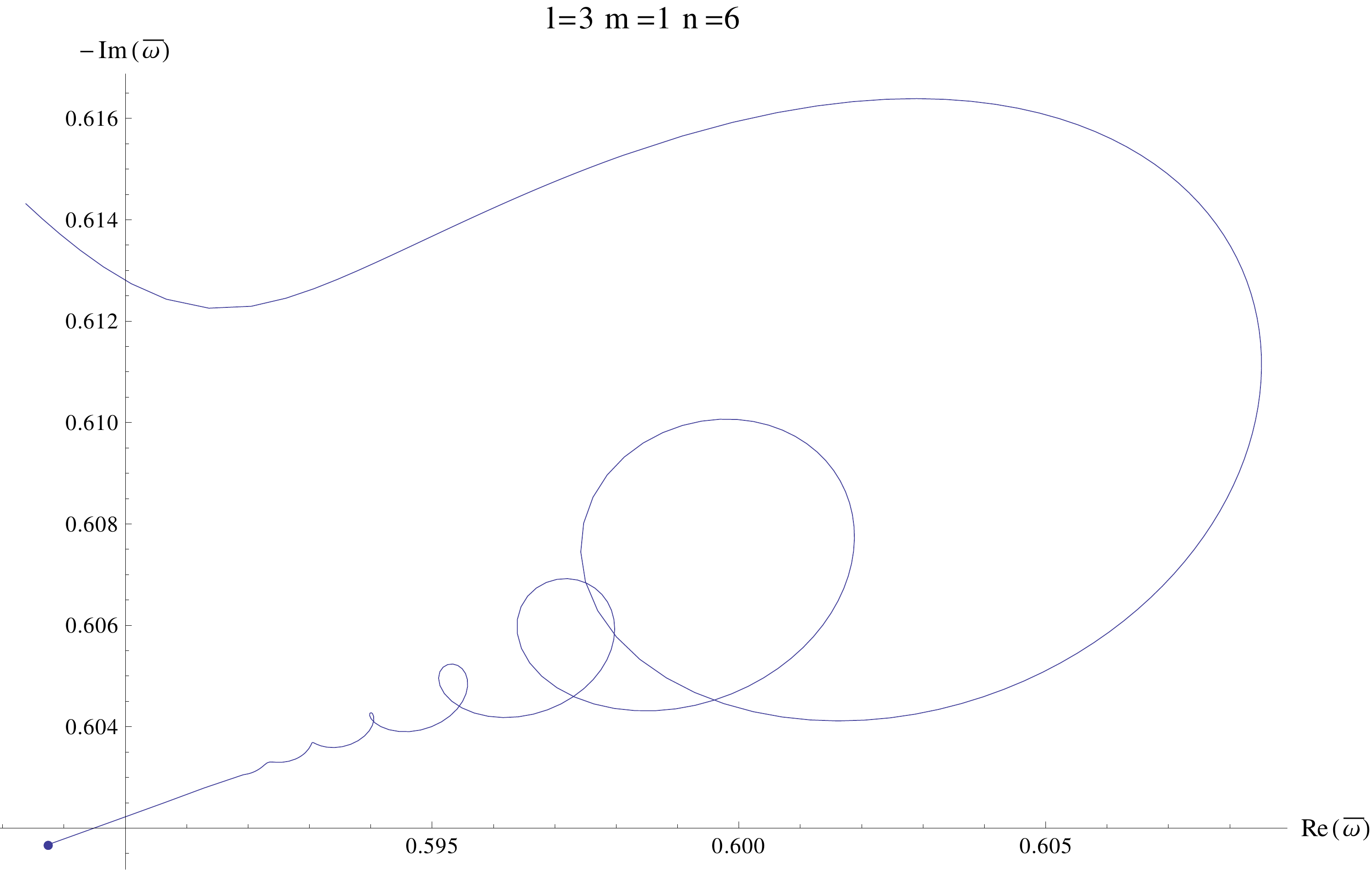}
\caption{\label{fig:Zoom_O_l3m1n6} Close-up of $\bar\omega$ for the
  $\{3,1,6\}$ sequence showing the unusual termination of this
  sequence. See Fig.~\ref{fig:AOmega_l3m1} for context.}
\end{figure}

Figure~\ref{fig:AllOmega_m1} displays all mode sequences for $m=1$.
Again, we see an accumulation point at $m/2$.
Mode sequences for $\{2,1,n\}$ and $\{3,1,n\}$ are displayed in
isolation, respectively, in Figs.~\ref{fig:AOmega_l2m1} and
\ref{fig:AOmega_l3m1}.
For $\ell=2$, we see that the $n=0$ and $1$ overtones do not approach
the accumulation point, while all subsequent displayed sequences do
terminate there.  For $\ell=3$, the $n=0\textendash3$ do not approach
the accumulation point, the next two do approach it, but $n=6$
terminates at finite damping following several spirals.  The end of
the $\{3,1,6\}$ sequence is shown in Fig.~\ref{fig:Zoom_O_l3m1n6}.

\begin{figure*}[htbp!]
\includegraphics[width=\linewidth,clip]{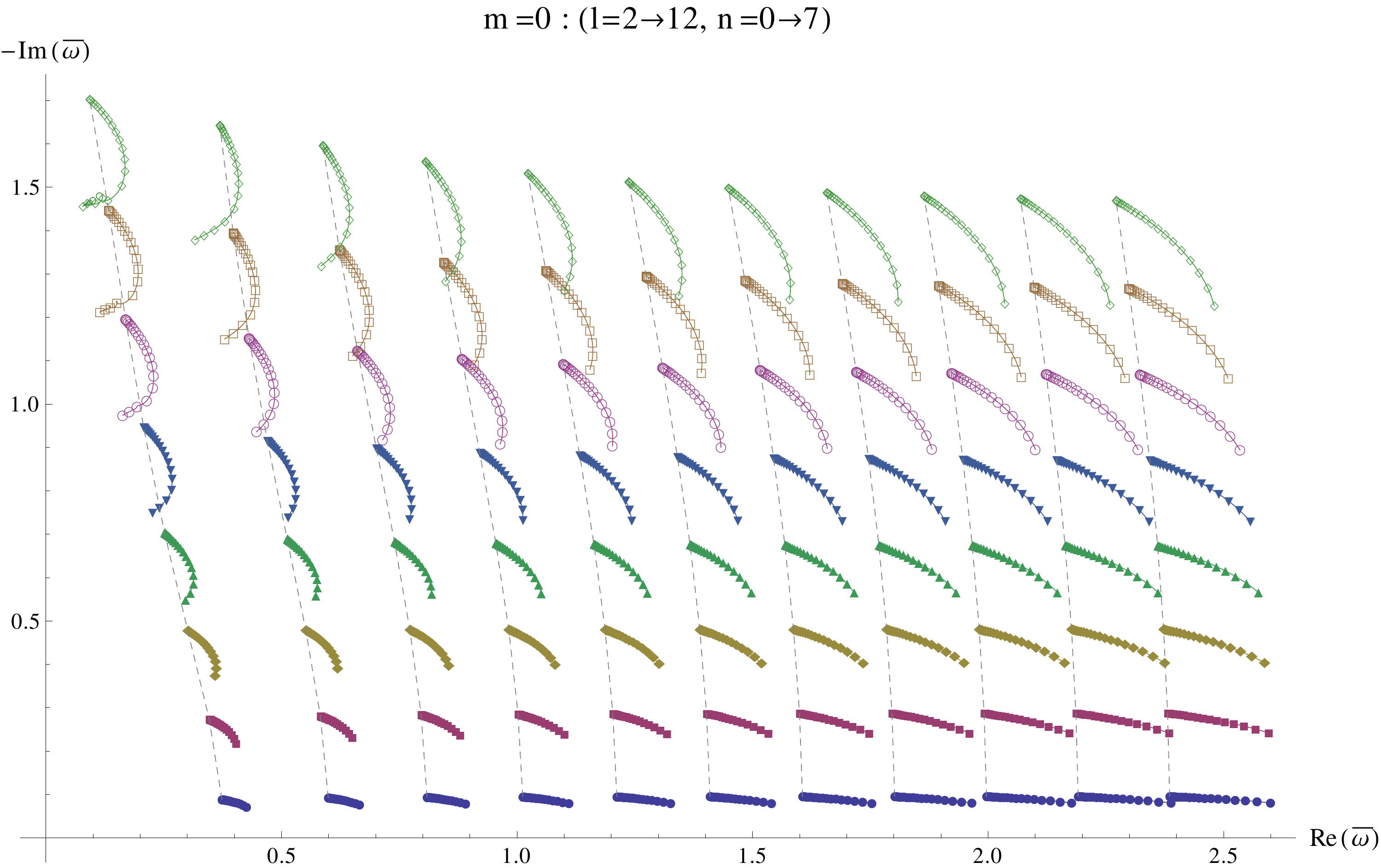}
\caption{\label{fig:AllOmega_m0} Kerr QNM mode sequences for $m=0$.
  See Fig.~\ref{fig:AllOmega_m2} for a full description.}
\end{figure*}
\begin{figure*}[htbp!]
\begin{tabular}{cc}
\includegraphics[width=0.5\linewidth,clip]{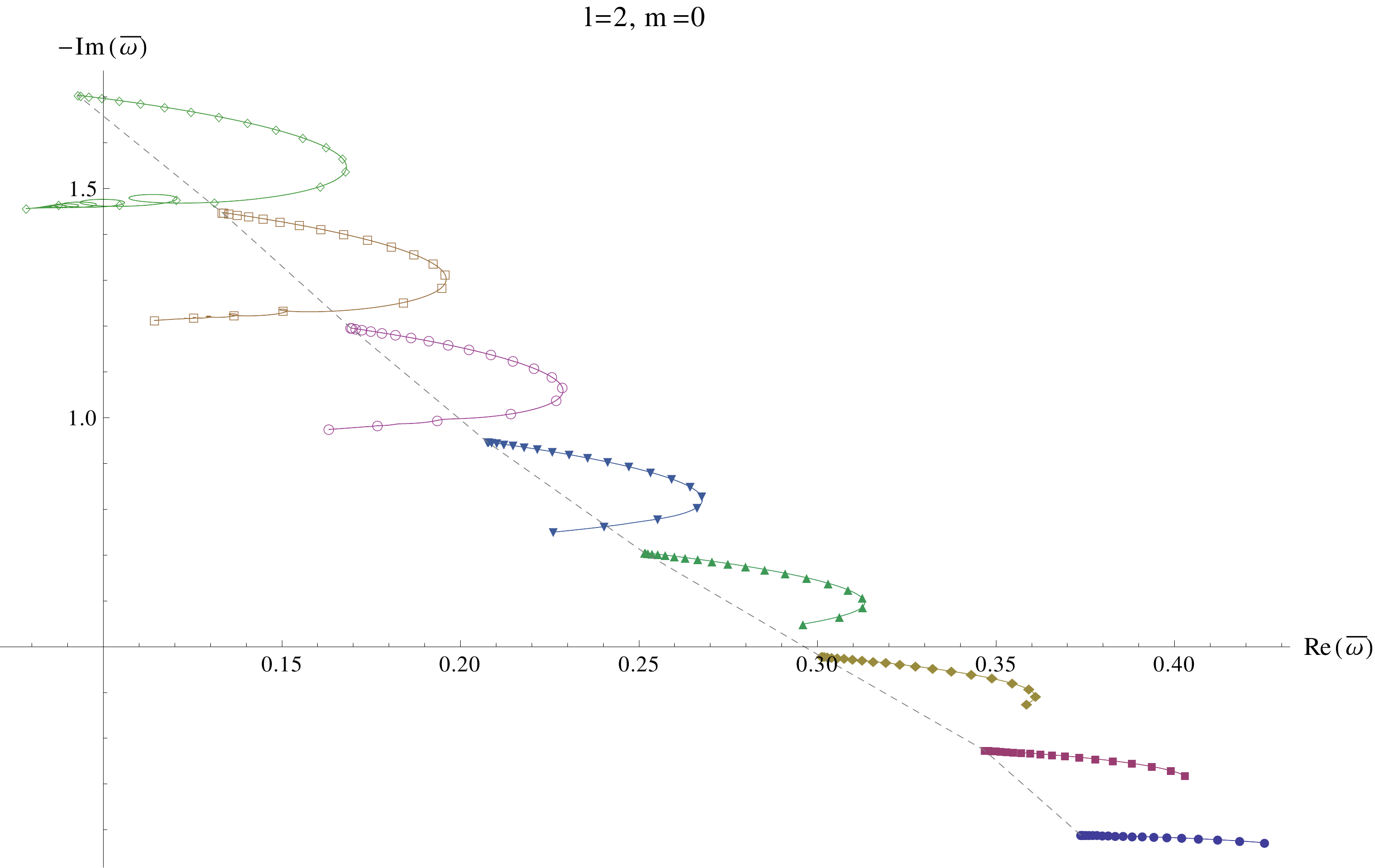} &
\includegraphics[width=0.5\linewidth,clip]{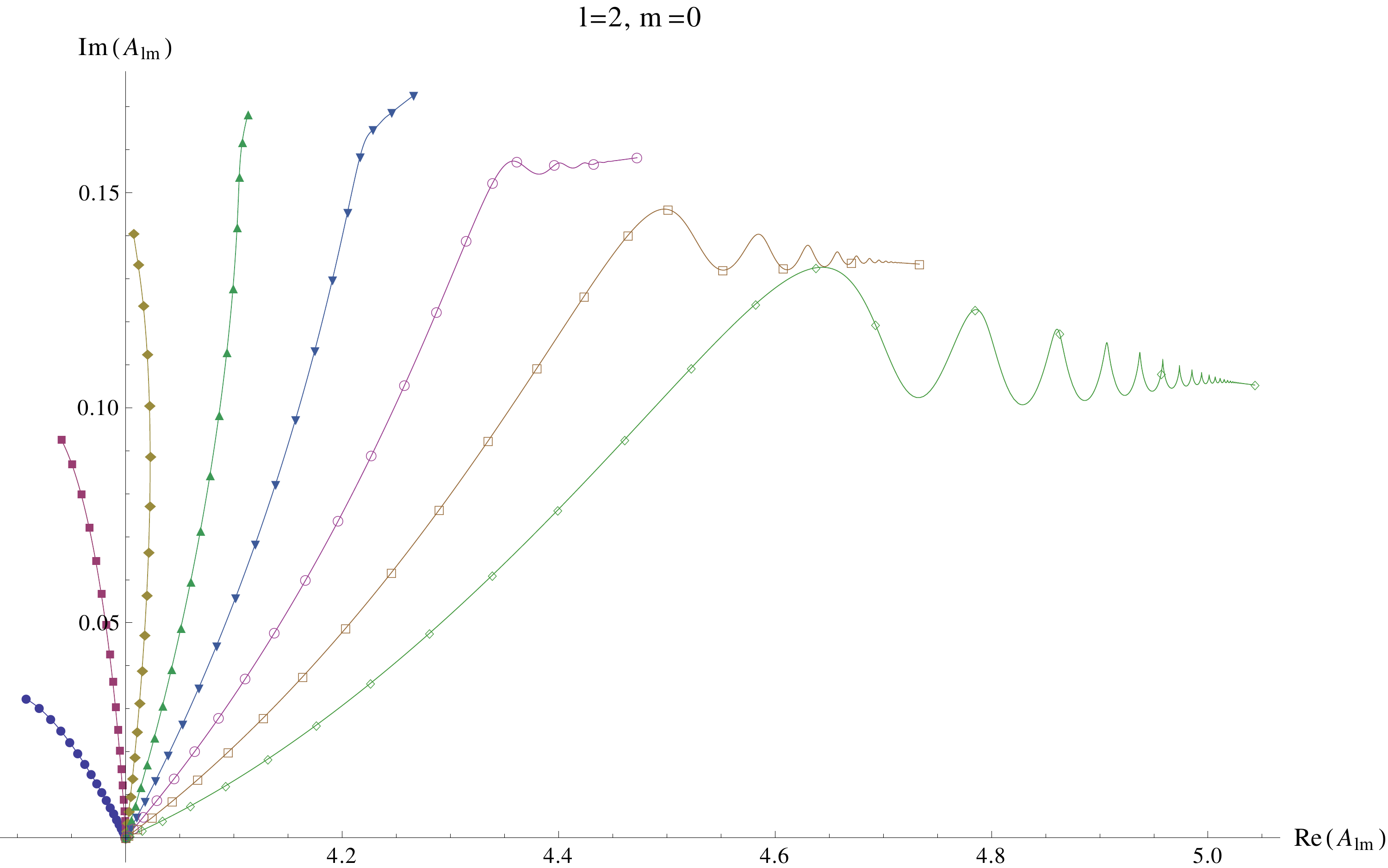}
\end{tabular}
\caption{\label{fig:AOmega_l2m0} Kerr QNM mode sequences for $\ell=2$
  and $m=0$. Plots are as described in Fig.~\ref{fig:AOmega_l2m2}.}
\end{figure*}
\begin{figure*}[htbp!]
\begin{tabular}{ccc}
\includegraphics[width=0.33\linewidth,clip]{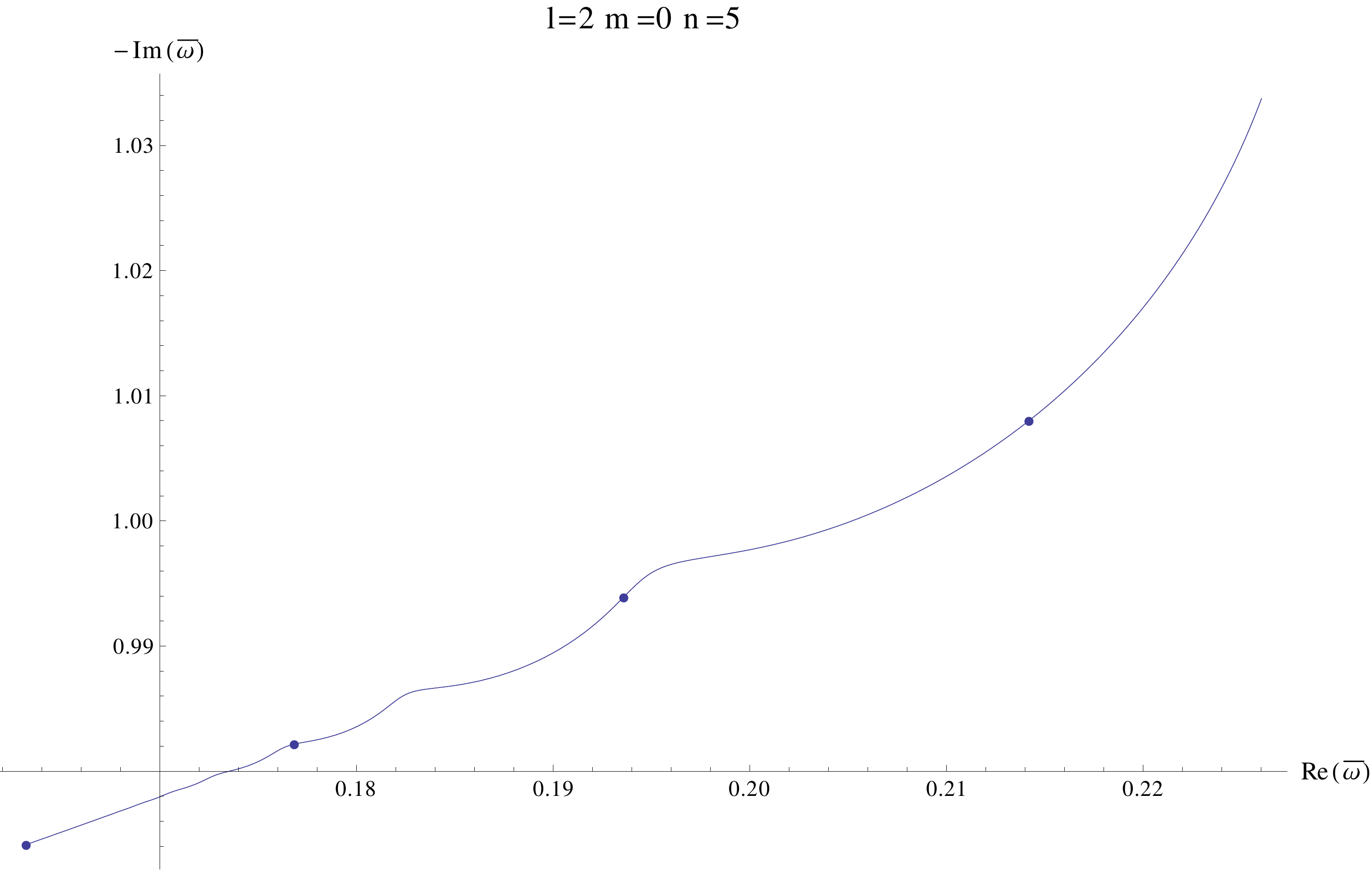} &
\includegraphics[width=0.33\linewidth,clip]{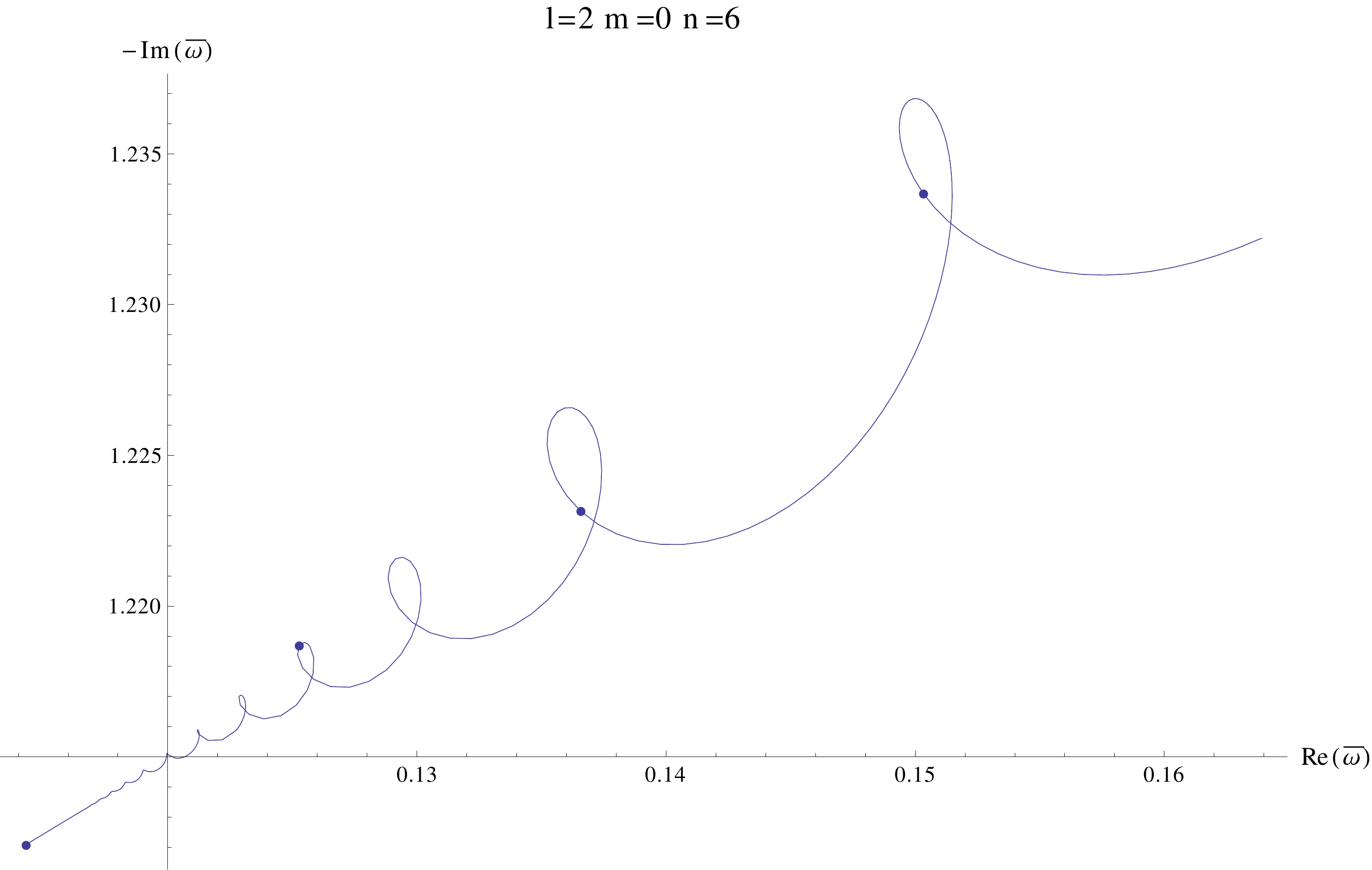} &
\includegraphics[width=0.33\linewidth,clip]{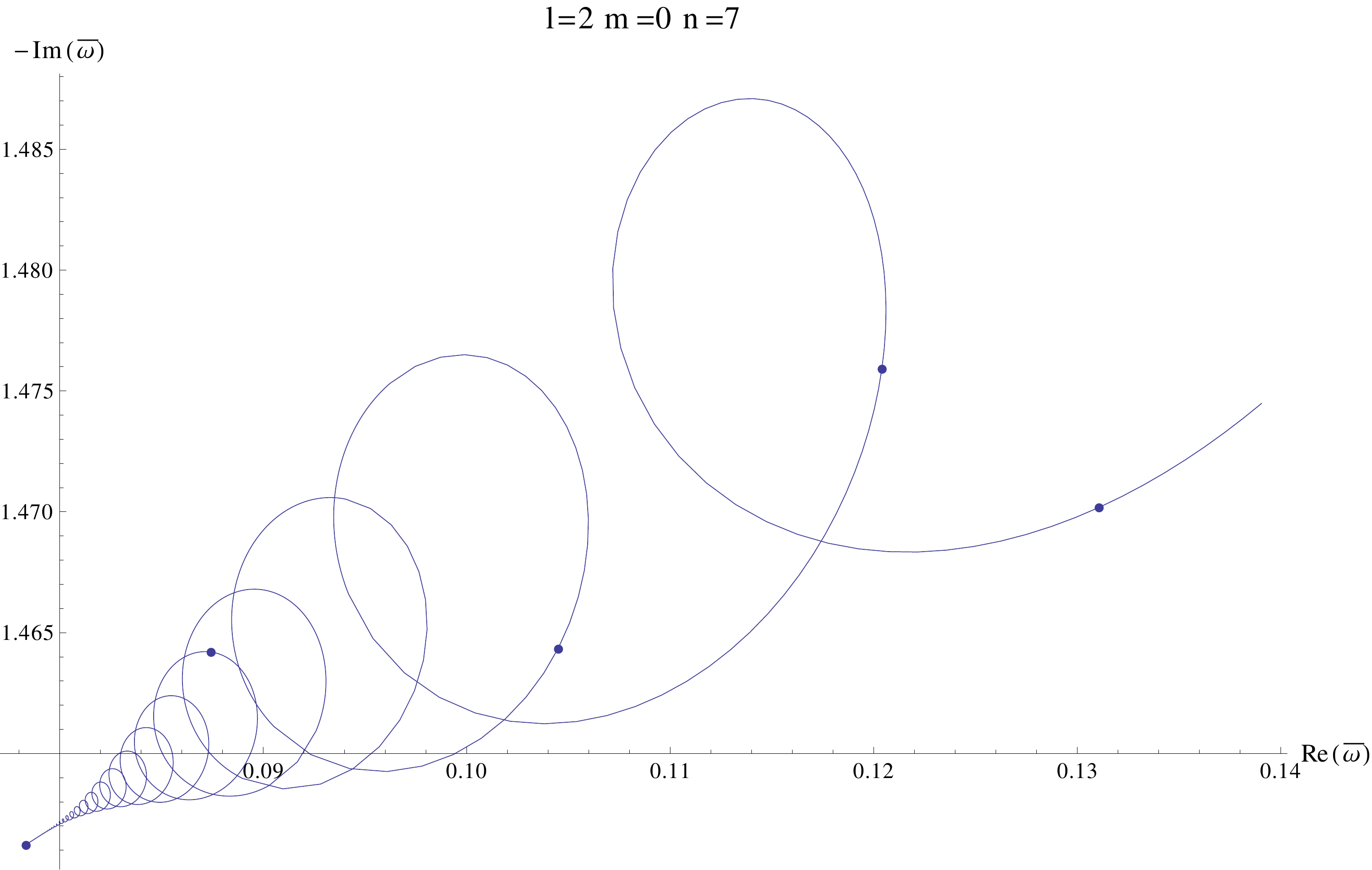}
\end{tabular}
\caption{\label{fig:Zoom_O_l2m0n567} Close-up of $\bar\omega$ for the
  $\{2,0,5\textendash7\}$ sequence showing the unusual termination of
  these sequences. See Fig.~\ref{fig:AOmega_l2m0} for context.}
\end{figure*}

Figure~\ref{fig:AllOmega_m0} displays all mode sequences for $m=0$,
and Fig.~\ref{fig:AOmega_l2m0} displays the $\{2,0,n\}$ mode sequences
in isolation.  None of the sequences shown approach an accumulation
point at $0$, and indeed accumulation points are only expected for
$m>0$ for positive-frequency modes.  For $\ell=2$, the
$n=5\textendash7$ mode sequences terminate in the unusual oscillation
or spiraling seen previously in Figs.~\ref{fig:Zoom_O_l2m2n5},
\ref{fig:Zoom_O_l4m2n6}, and~\ref{fig:Zoom_O_l3m1n6}.  The termination
of the $\{2,0,5\textendash7\}$ sequences are shown in
Fig.~\ref{fig:Zoom_O_l2m0n567}.  We also note that these examples
display the most dramatic oscillatory effects in the separation
constant $\scA{-2}{20}{\bar{a}\bar\omega_{20n}}$ as seen in
Fig.~\ref{fig:AOmega_l2m0}.

\begin{figure*}[htbp!]
\includegraphics[width=\linewidth,clip]{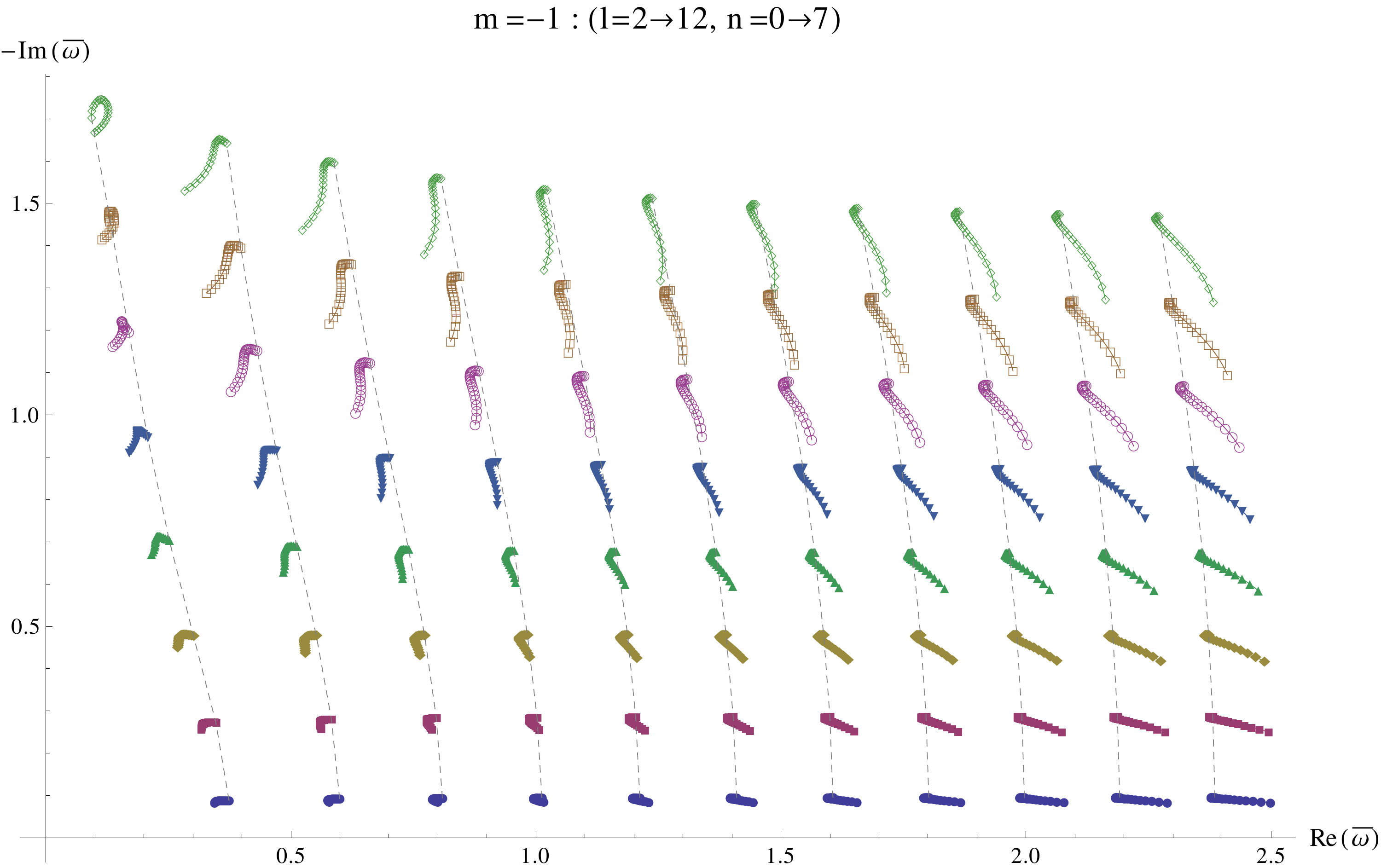}
\caption{\label{fig:AllOmega_m-1} Kerr QNM mode sequences for $m=-1$.
  See Fig.~\ref{fig:AllOmega_m2} for a full description.}
\end{figure*}
\begin{figure*}[htbp!]
\includegraphics[width=\linewidth,clip]{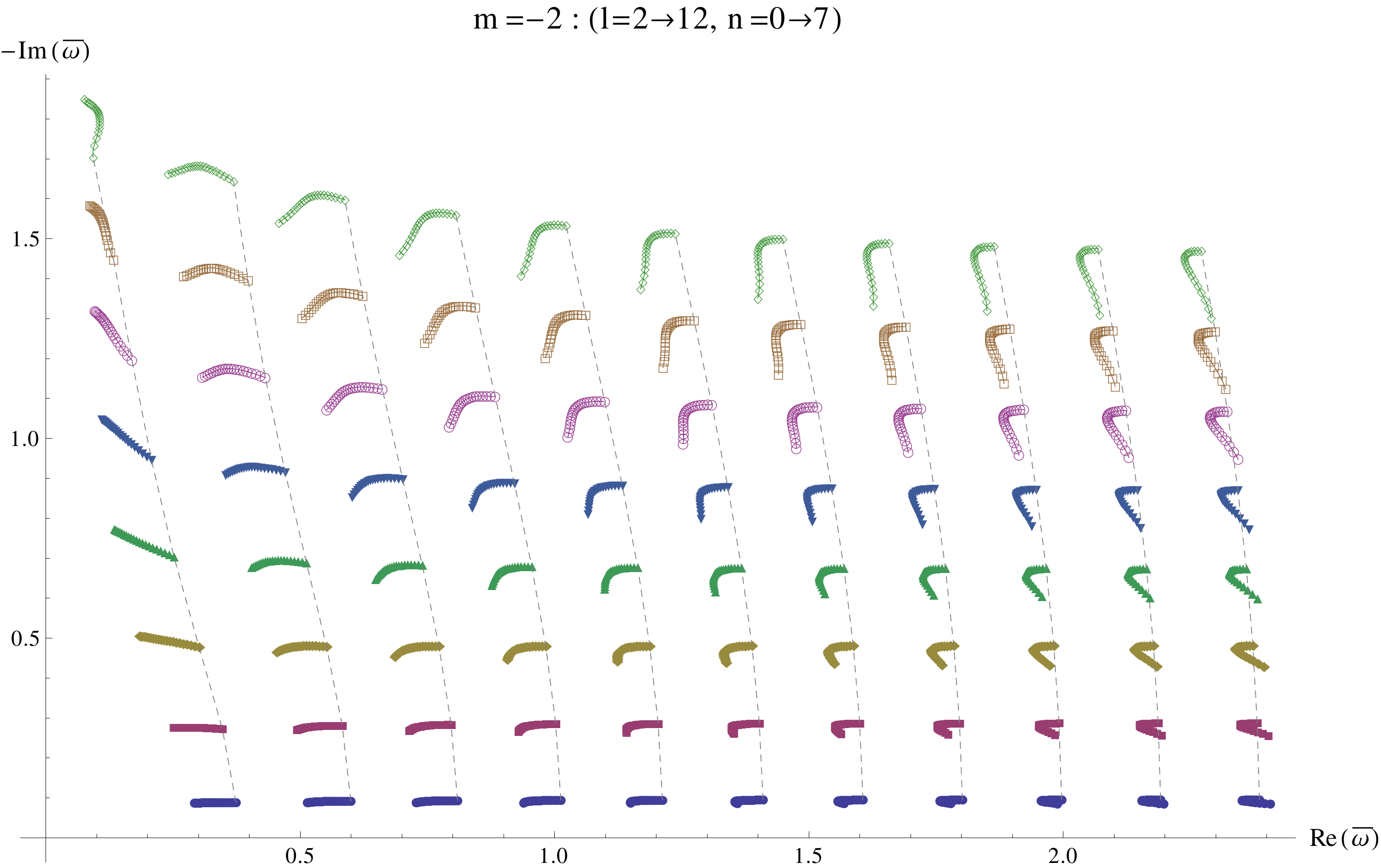}
\caption{\label{fig:AllOmega_m-2} Kerr QNM mode sequences for $m=-2$.
  See Fig.~\ref{fig:AllOmega_m2} for a full description.}
\end{figure*}
The final two mode sets we show are for $m=-1$ and $m=-2$.
Figures~\ref{fig:AllOmega_m-1} and~\ref{fig:AllOmega_m-2} display all
modes for these two sets.  The complex frequencies for modes with
$m<0$ never approach an accumulation point on the positive real axis.
Recall, however, that the behavior we have displayed is for the
positive-frequency modes.  There also exists the second set of
QNMs, see Eq.~(\ref{eq:neg_freq_modes}), for which
$\rm{Re}(\bar\omega)<0$.  For these negative-frequency modes, it
is the modes with $m<0$ that can approach an accumulation point at
$m/2$.


\subsubsection{The extremal limit}
\label{sec:Extreme_Limit}

Examining Figs.~\ref{fig:AllOmega_m2} and \ref{fig:AllOmega_m1}, the
most striking feature in each plot is the accumulation point at
$\bar\omega=m/2$ for the extreme limit of many of the sequences.
While we do not show all of the figures, this behavior is seen for all
of the sets of positive-frequency mode sequences [ie
  $\rm{Re}(\bar\omega)\ge0$] with $m>0$ that we have constructed.  By
symmetry, these accumulation points will also occur for the
negative-frequency mode sequences with $m<0$.  This behavior was first
predicted by Detweiler\cite{detweiler80}.  His analysis was extended
and amended by Cardoso\cite{cardoso04} and Hod\cite{Hod08}, yielding
an approximation for $\bar\omega$ in the limit as $\bar{a}\to1$.
Recently, these modes have been studied extensively by Yang {\em et
  al}\cite{Yang-et-al-2013a,Yang-et-al-2013b}.  Using matched
asymptotic expansions, they derived a formula for $\bar\omega$ in the
extreme limit when the solutions approach an accumulation point.  They
find
\begin{equation}
\label{eq:yang_omega}
\bar\omega \approx \frac{m}2 - \delta\sqrt\frac{\epsilon}{2}
  -i\left(\tilde{n}+\frac12\right)\sqrt\frac{\epsilon}{2},
\end{equation}
where $\epsilon\equiv1-\bar{a}$, $\tilde{n}$ indexes over the set
of overtones that approach the accumulation point, and
\begin{equation}
\label{eq:yang_delta}
  \delta^2\equiv \frac{7m^2}4 - \left(s+\frac12\right)^2
  -\scA{-2}{\ell{m}}{m/2}.
\end{equation}
Equation (\ref{eq:yang_omega}) is supposed to be a good approximation
for $\bar\omega$ except for cases where $|\delta|\sim1$ and for a
small range of $\delta^2<0$ (see Ref.~\cite{Yang-et-al-2013b}).

When the angular Teukolsky equation (\ref{eqn:swSF_DiffEqn}) is solved
for real $c$, the separation constant $\scA{s}{\ell{m}}{c}$ is real,
meaning that $\delta^2$ is real.  This means that $\delta$ will either
be real or pure imaginary, which corresponds to the two types of
behavior seen in the mode sequences that approach the accumulation
points at $\bar\omega=m/2$.  Figure~\ref{fig:AOmega_l2m2}, showing the
case $\{2,2,n\}$, illustrates the behavior when $\delta^2>0$.  In this
case, all of the modes (except for exceptional case of $n=5$) approach
the accumulation point.  The fitting overtone index $\tilde{n}$ in
Eq.~(\ref{eq:yang_omega}) takes on consecutive values starting at
zero, so that in this case the set $n\in\{0,1,2,3,4,6,7\}$ maps to
$\tilde{n}\in\{0,1,2,3,4,5,6\}$.  Figures~\ref{fig:AOmega_l3m2},
\ref{fig:AOmega_l4m2}, and~\ref{fig:AOmega_l5m2}, showing the cases
$\{3\textendash5,2,n\}$, illustrate the behavior when $\delta^2<0$.
In these cases, we see that the first few overtones approach a finite
value for the $\rm{Im}(\bar\omega)$ in the extremal limit, and then the
remaining overtones (except for exceptional cases) approach the
accumulation point.  For example, in Fig.~\ref{fig:AOmega_l4m2} we
find that the overtones $n\in\{3,4,5,7\}$ approach the accumulation
point.  In this case, the fitting overtone index takes on the values
$\tilde{n}\in\{0,1,2,3\}$.

For those mode sequences that approach an accumulation point in the
extremal limit, we have extended the sequence in decadal steps down to
$\epsilon=10^{-8}$ (except for $\{4,3,n\}$ where $\epsilon=10^{-9}$)
and compared our results with Eq.~(\ref{eq:yang_omega}) finding very
good agreement.  In fact, we have fit our results to a version of
Eq.~(\ref{eq:yang_omega}) that has been extended to next to leading
order behavior in $\epsilon$.  We have also constructed fitting
functions for the separation constant
$\scA{-2}{\ell{m}}{\bar{a}\bar\omega}$.  These extended fitting
functions must be defined independently for the cases when
$\delta^2>0$ and $\delta^2<0$.

The fitting functions for the case $\delta^2>0$ are given in
Eqs.~(\ref{eq:fit_pos}).  The real and imaginary parts of $\bar\omega$
are fit using Eqs.~(\ref{eq:fit_pos_ReO}) and~(\ref{eq:fit_pos_ImO}).
\begin{subequations}
\label{eq:fit_pos}
\begin{align}
\label{eq:fit_pos_ReO}
  \rm{Re}[\bar\omega] &= \frac{m}2 -\alpha_1\sqrt{\frac{\epsilon}2} +
                         (\alpha_2 + \alpha_3 \tilde{n})\epsilon \\ 
\label{eq:fit_pos_ImO}
  \rm{Im}[\bar\omega] &= -\left(\tilde{n}+\frac12\right)\left(
                         \sqrt{\frac{\epsilon}2} - \alpha_4\epsilon\right) \\ 
\label{eq:fit_pos_ReA}
  \rm{Re}[{}_sA_{\ell m}] &= \ell(\ell+1)-s(s+1)+\beta_1 \\ 
    &\mbox{}\quad + \beta_2\sqrt{\frac{\epsilon}2} 
      + \left(\beta_3 + \beta_4 \tilde{n} + \beta_5 \tilde{n}^2\right)\epsilon 
      \nonumber\\ 
\label{eq:fit_pos_ImA}
  \rm{Im}[{}_sA_{\ell m}] &= \left(\tilde{n}+\frac12\right)\left(
                           \beta_6\sqrt{\frac{\epsilon}2}+\beta_7\epsilon\right)
\end{align}
\end{subequations}
Comparing to Eq.~(\ref{eq:yang_omega}), we see that our coefficient
$\alpha_1$ should correspond to $\delta$.  The coefficients
$\alpha_2$, $\alpha_3$, and $\alpha_4$ are associated with the next to
leading order behavior in $\epsilon$.  The real and imaginary parts of
$\scA{-2}{\ell{m}}{\bar{a}\bar\omega}$ are fit using
Eqs.~(\ref{eq:fit_pos_ReA}) and~(\ref{eq:fit_pos_ImA}).  For the
separation constant, the coefficients $\beta_2$ and $\beta_6$ are
associated with the leading order behavior in $\epsilon$.  The
coefficient $\beta_1$ defines the limiting behavior
$\scA{-2}{\ell{m}}{m/2}=\ell(\ell+1)-2+\beta_1$, required in
Eq.~(\ref{eq:yang_delta}) to determine the value of $\delta^2$.  The
remaining coefficients $\beta_3$, $\beta_4$, $\beta_5$, and\ $\beta_7$
are associated with its next to leading order behavior for the
separation constant.

\begin{figure*}[htbp!]
\begin{tabular}{cc}
\includegraphics[width=0.5\linewidth,clip]{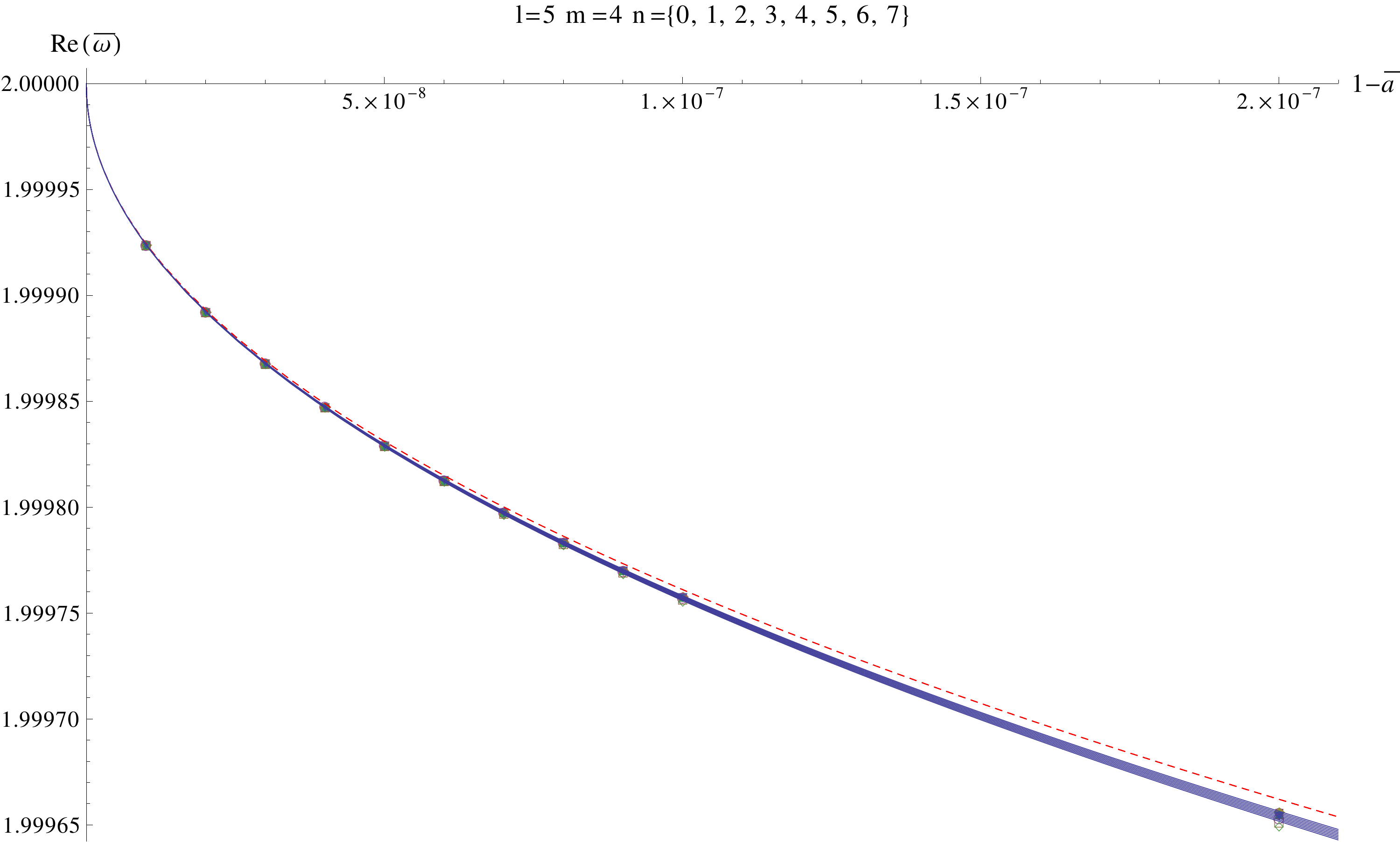} &
\includegraphics[width=0.5\linewidth,clip]{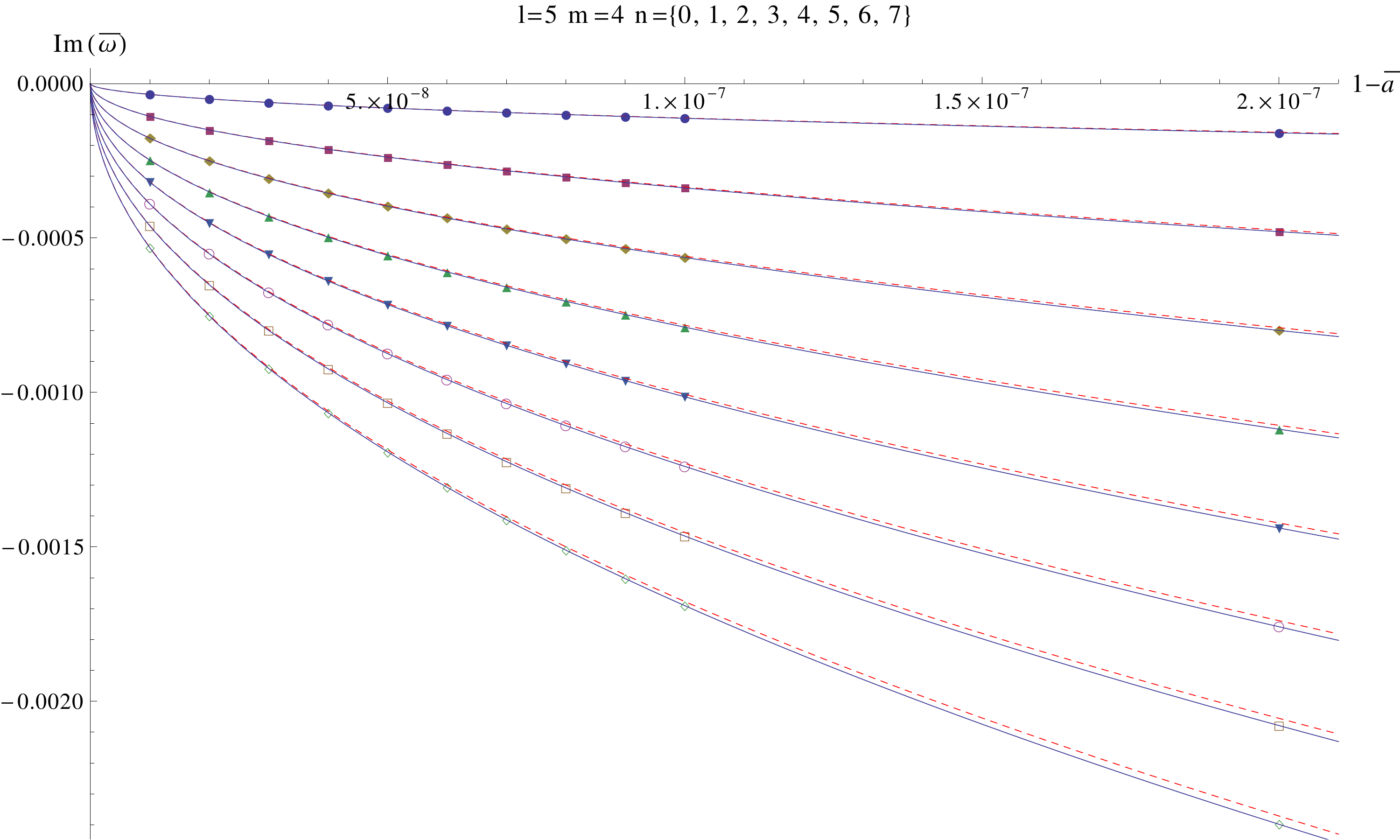}
\end{tabular}
\caption{\label{fig:SupRad_l5m4_O} Near extremal behavior of
  $\bar\omega$ for $\ell=5$, $m=4$.  Symbols in the left plot
  represent data values used to fit the coefficients in
  Eq.~(\ref{eq:fit_pos_ReO}).  The lines represent the resulting fit
  function.  The right plot corresponds to
  Eq.~(\ref{eq:fit_pos_ImO}).  In both plots, the dashed(red) line
  represents the fit function omitting the next to leading order
  behavior in $\epsilon$.}
\end{figure*}

The fit values for all cases with $\delta^2>0$ are shown in
Tables~\ref{tab:fit_pos_Omega} and~\ref{tab:fit_pos_Alm}.  The fitted
values shown in these tables are displayed to six significant digits
unless the fit yielded fewer significant digits.  In this case, the
result is shown to its last significant digit along with the one-sigma
error in this last digit in parentheses.  The excellent agreement of
our results with Eq.~(\ref{eq:yang_omega}) is seen in the agreement
between $\delta$ and $\alpha_1$ in Table~\ref{tab:fit_pos_Omega}.  The
largest discrepancy is seen for the case $\ell=5$, $m=4$ where the
results only agree to three significant figures.  But, in this case,
$\delta\sim1.07$ and we recall that Eq.~(\ref{eq:yang_omega}) was not
claimed to be a good approximation when $|\delta|\sim1$.
Nevertheless, the agreement is still quite good, and we note the clear
tendency for the accuracy of all fit parameters to increase with
increasing $\delta$.  The plots in Fig.~\ref{fig:SupRad_l5m4_O} show
the quality of the fit in the worst case, $\ell=5$ and $m=4$.  The
dependence of the fit to the overtone index occurs at leading order in
$\epsilon$ for $\rm{Im}(\bar\omega)$ and is clearly seen in the right
plot of Fig.~\ref{fig:SupRad_l5m4_O}; however its dependence in the
real part occurs only at next to leading order and is barely visible
in the left plot.  However, the next to leading order behavior in
$\epsilon$ is important for all cases.  To illustrate this, we have
also plotted the fit curves omitting the terms at next to leading
order in $\epsilon$.  The $\tilde{n}$ dependence in
$\rm{Re}(\bar\omega)$ is represented in the $\alpha_3$ coefficient and
we see that its significance is strongly dependent on $\delta$,
increasing with increasing $\delta$.  For the case plotted in
Fig.~\ref{fig:SupRad_l5m4_O}, an even higher-order term with quadratic
dependence on $\tilde{n}$ appears to be important.  In fact, there
is evidence for such a term in all of our fits; however we have not
tried to find a fit with additional terms.  

\begin{table*}[htbp!]
\begin{tabular}{rrcf{7}f{7}f{8}f{8}f{7}}
\hline\hline
$\ell$ & $m$ && \Chead{$\delta_r$} &
\Chead{$\alpha_1$} & \Chead{$\alpha_2$} & \Chead{$\alpha_3$} &
 \Chead{$\alpha_4$} \\
2 & 2 && 2.05093 &
2.05084(3) & 1.64(6) & -0.032(8) & 1.343(5) \\
3 & 3 && 2.79361 &
2.79361 & 2.289(2) & -0.0004(1) & 1.35730(9) \\
4 & 4 && 3.56478 &
3.56478 & 2.79572(9) & -0.00020(1) & 1.34522(2) \\
5 & 4 && 1.07271 &
1.0687(9) & -28.(2) & -3.4(3) & -17.90(6) \\
5 & 5 && 4.35761 &
4.35761 & 3.29989(7) & -0.000142(8) & 1.33085(2) \\
6 & 5 && 2.37521 &
2.37515(1) & -5.50(3) & -0.672(3) & -1.206(1) \\
6 & 6 && 5.16594 &
5.16594 & 3.81003(7) & -0.000111(8) & 1.31809(2) \\
7 & 6 && 3.40439 &
3.40438 & -2.03(4) & -0.0109(4) & 0.2931(3) \\
7 & 7 && 5.98547 &
5.98547 & 4.32747(7) & -0.000091(8) & 1.30757(2) \\
8 & 7 && 4.35924 &
4.35924 & -0.274(1) & -0.0034(1) & 0.74198(9) \\
8 & 8 && 6.81327 &
6.81327 & 4.85207(7) & -0.000078(8) & 1.29905(2) \\
9 & 8 && 5.28081 &
5.28081 & 0.9429(5) & -0.00150(6) & 0.93670(4) \\
10 & 8 && 2.80128 &
2.80099(6) & -22.9(1) & -0.43(2) & -6.456(9) \\
9 & 9 && 7.64735 &
7.64734 & 5.38314(7) & -0.000069(8) & 1.29217(2) \\
10 & 9 && 6.18436 &
6.18436 & 1.92226(3) & -0.00080(3) & 1.03870(2) \\
11 & 9 && 4.05104 &
4.05101 & -11.00(2) & -0.053(2) & -1.568(1) \\
10 & 10 && 8.48628 &
8.48628 & 5.91988(8) & -0.000062(9) & 1.28657(2) \\
11 & 10 && 7.07706 &
7.07706 & 2.7754(2) & -0.00048(2) & 1.09871(1) \\
12 & 10 && 5.14457 &
5.14457(1) & -6.269(5) & -0.0147(6) & -0.2362(5) \\
11 & 11 && 9.32904 &
9.32904 & 6.46155(8) & -0.000056(9) & 1.28198(2) \\
12 & 11 && 7.96274 &
7.96274 & 3.5535(1) & -0.00032(1) & 1.13691 \\
12 & 12 && 10.1749 &
10.1749 & 7.00748(8) & -0.00005(1) & 1.27819(2) \\
\hline\hline
\end{tabular}
\caption{\label{tab:fit_pos_Omega} Values for the coefficients of the
  fitting function for $\bar\omega$ given in
  Eqs.~(\ref{eq:fit_pos_ReO}) and~(\ref{eq:fit_pos_ImO}) appropriate
  for $\delta^2>0$.  $\delta=\delta_r$ obtained from
  Eq.~(\ref{eq:yang_delta}).  All quantities are displayed to six
  significant digits unless the fit yielded fewer significant digits.
  In this case, each result is shown to its last significant digit
  along with its one-sigma error in this last digit printed in
  parentheses.}
\end{table*}

\begin{figure*}[htbp!]
\begin{tabular}{cc}
\includegraphics[width=0.5\linewidth,clip]{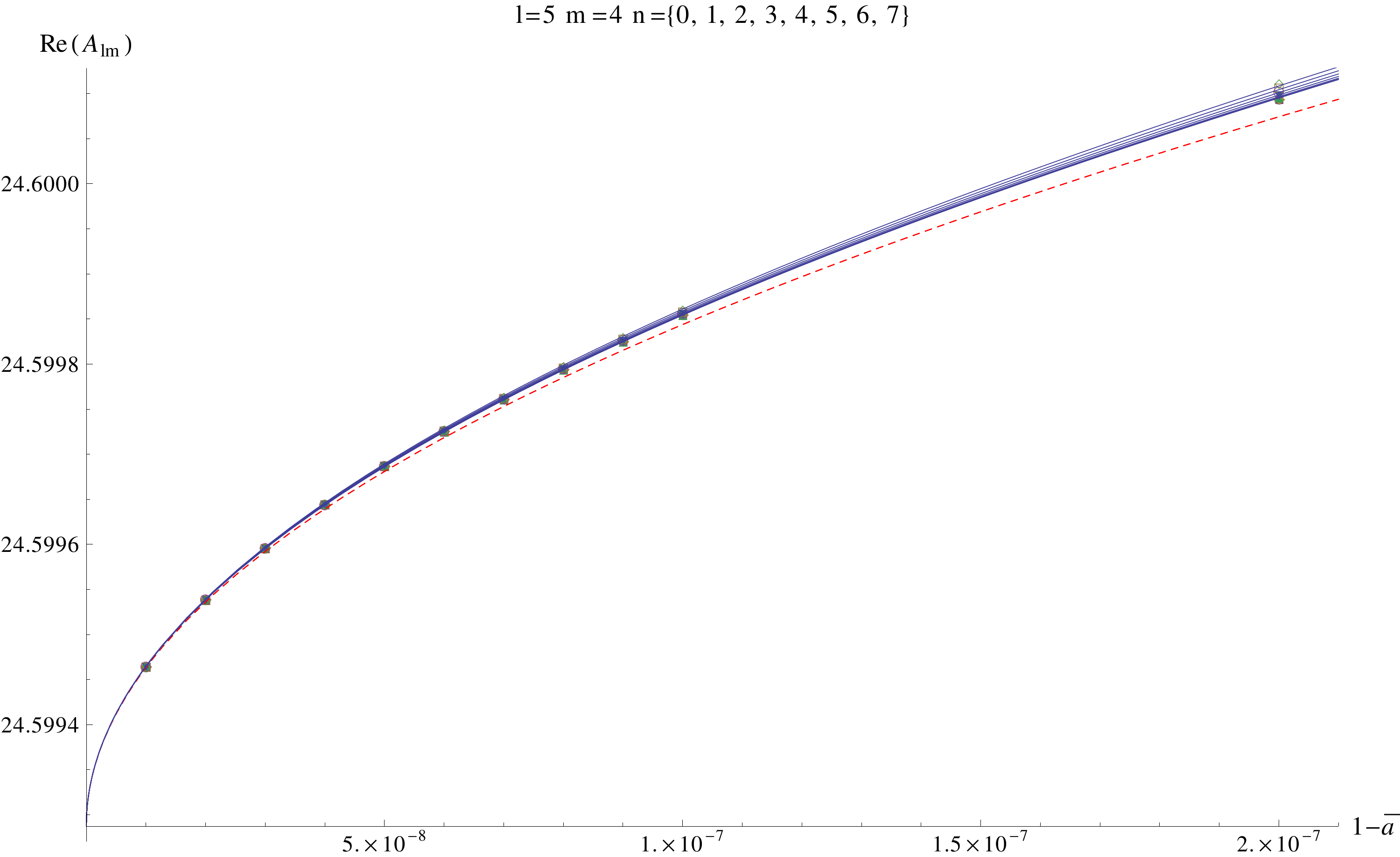} &
\includegraphics[width=0.5\linewidth,clip]{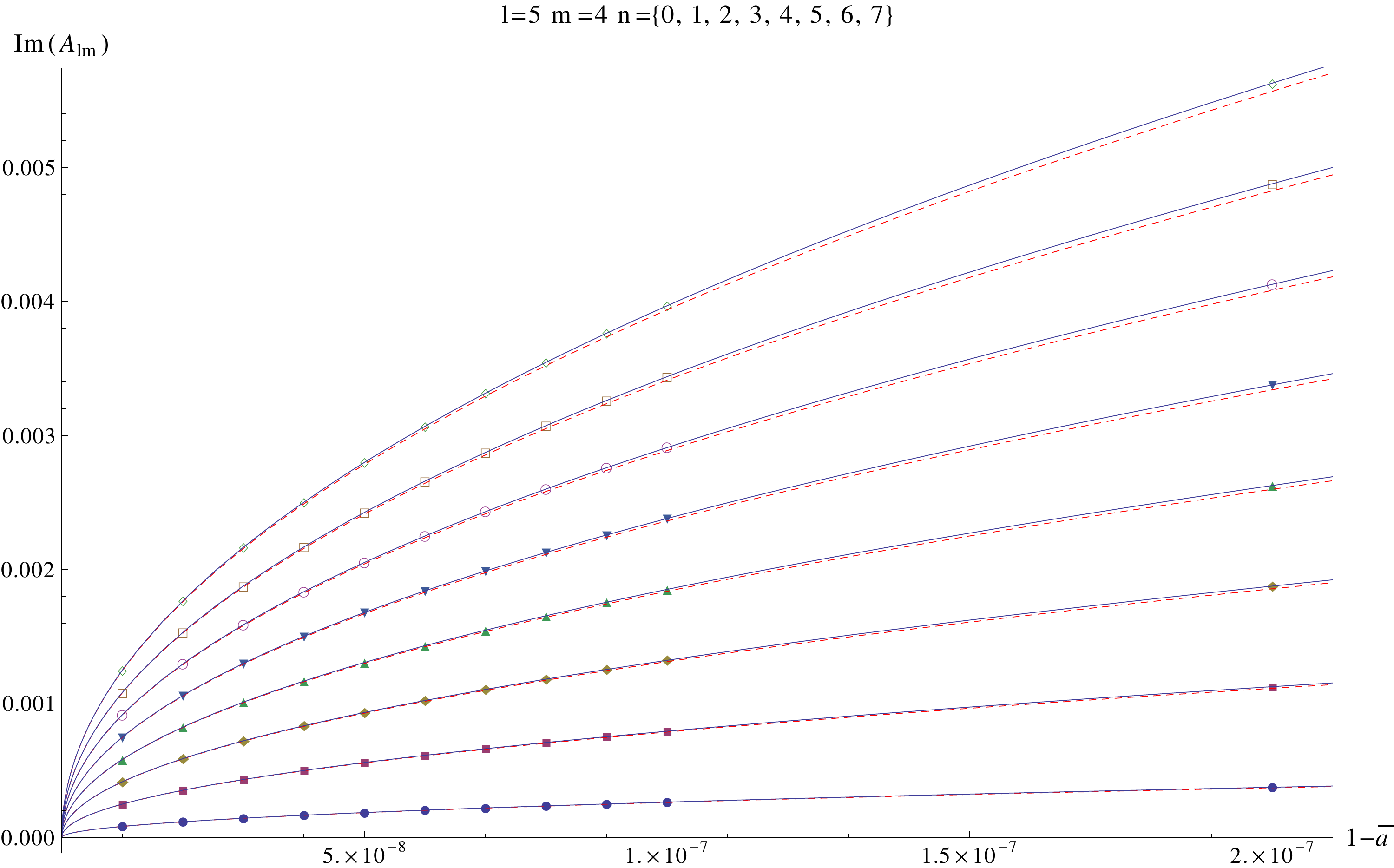}
\end{tabular}
\caption{\label{fig:SupRad_l5m4_A} Near extremal behavior of
  $\scA{-2}{\ell{m}}{\bar{a}\bar\omega}$ for $\ell=5$, $m=4$.  Symbols
  in the left plot represent data values used to fit the coefficients
  in Eq.~(\ref{eq:fit_pos_ReA}).  The solid lines represent the
  resulting fit function.  The right plot corresponds to
  Eq.~(\ref{eq:fit_pos_ImA}).  In both plots, the dashed(red) line
  represents the fit function omitting the next to leading order
  behavior in $\epsilon$.}
\end{figure*}

Table~\ref{tab:fit_pos_Alm} and Fig.~\ref{fig:SupRad_l5m4_A} show the
results for fitting the real and imaginary parts of
$\scA{-2}{\ell{m}}{\bar{a}\bar\omega}$ using
Eqs.~(\ref{eq:fit_pos_ReA}) and~(\ref{eq:fit_pos_ImA}).  We see that
the leading order behavior of $\rm{Im}({}_sA_{\ell{m}})$ is
proportional to the leading order behavior of $\rm{Im}(\bar\omega)$,
including its dependence on $\tilde{n}$.  The same is true for the
next to leading order behavior, although the proportionality constant
is different for the two terms.  As with $\bar\omega$, all of the
$\tilde{n}$ dependence in the real part of
$\scA{-2}{\ell{m}}{\bar{a}\bar\omega}$ occurs in the next to leading
order term, and as can be seen in the left plot of
Fig.~\ref{fig:SupRad_l5m4_A} is barely visible.  The coefficients
$\beta_4$ and $\beta_5$ encode this dependence on the overtone index
and their significance is again strongly correlated with the magnitude
of $\delta$.  Interestingly, for larger values of $\delta$, where
these coefficients are determined with several significant figures, it
seems clear that $\beta_4=\beta_5$, so the dependence on the overtone
index is proportional to $\tilde{n}(\tilde{n}+1)$ [or to
$(\tilde{n}+1)^2$ with a redefinition of $\beta_3$].  Only the case
$\ell=5$, $m=4$ (where $\delta\sim1.07$) seems to violate this scaling.

\begin{table*}[htbp!]
\begin{tabular}{rrcf{7}f{8}f{7}f{8}f{9}f{8}f{7}}
\hline\hline
$\ell$ & $m$ &&
\Chead{$\beta_1$} & \Chead{$\beta_2$} & \Chead{$\beta_3$} & \Chead{$\beta_4$} & 
\Chead{$\beta_5$} & \Chead{$\beta_6$} & \Chead{$\beta_7$} \\
2 & 2 &&
-3.45631 & 8.8067(4) & -3.6(5) & 0.2(1) & 0.50(2) & 
4.29457(3) & -7.81(7) \\
3 & 3 &&
-4.30428 & 10.7854 & -5.82(1) & 0.367(2) & 0.3700(3) &
3.86073 & -7.304(1) \\
4 & 4 &&
-4.95763 & 12.6174 & -6.741(2) & 0.3135(3) & 0.31347(5) &
3.53946 & -6.9957(1) \\
5 & 4 &&
-3.40071 & 2.4883(6) & 106.(8) & -4.(2) & 1.9(2) & 2.3473(2) & 40.6(5) \\
5 & 5 &&
-5.48880 & 14.3063 & -7.692(2) & 0.2692(3) & 0.26919(4) &
3.28306 & -6.7150(1) \\
6 & 5 &&
-4.14163 & 5.65809(8) & 18.6(1) & 0.16(2) & 0.174(3) &
2.38233 & 2.072(5) \\
6 & 6 &&
-5.93698 & 15.8614 & -8.689(1) & 0.2338(3) & 0.23376(4) &
3.07037 & -6.4620(1) \\
7 & 6 &&
-4.83985 & 8.18217(1) & 10.53(1) & 0.143(3) & 0.1425(3) &
2.40344 & -1.6728(7) \\
7 & 7 &&
-6.32583 & 17.2963 & -9.707(1) & 0.2050(3) & 0.20498(4) &
2.88971 & -6.2322(1) \\
8 & 7 &&
-5.50298 & 10.5277 & 6.694(4) & 0.1298(7) & 0.1298(1) &
2.41504 & -2.9234(3) \\
8 & 8 &&
-6.67066 & 18.6259 & -10.724(1) & 0.1812(2) & 0.18128(3) &
2.73376 & -6.0215(1) \\
9 & 8 &&
-6.13691 & 12.7798 & 4.067(1) & 0.1204(3) & 0.12043(4) &
2.42005 & -3.5389(1) \\
10 & 8 &&
-6.09714 & 6.8962(4) & 67.3(5) & 0.3(1) & 0.25(1) &
2.46263(1) & 14.63(3) \\
9 & 9 &&
-6.98190 & 19.8640 & -11.727(1) & 0.1615(2) & 0.16153(3) &
2.59750 & -5.8270(1) \\
10 & 9 &&
-6.74626 & 14.9684 & 1.9491(5) & 0.1126(1) & 0.11262(1) &
2.42037 & -3.90733(6) \\
11 & 9 &&
-6.91089 & 10.2630 & 37.66(7) & 0.13(1) & 0.127(2) &
2.53351 & 2.964(4) \\
10 & 10 &&
-7.26689 & 21.0230 & -12.705(1) & 0.1449(2) & 0.14490(3) &
2.47729 & -5.6465(1) \\
11 & 10 &&
-7.33472 & 17.1075 & 0.0864(2) & 0.10586(5) & 0.105861(6) &
2.41732 & -4.15457(2) \\
12 & 10 &&
-7.71665 & 13.3413 & 26.46(2) & 0.111(4) & 0.1110(5) &
2.59329 & -0.522(2) \\
11 & 11 &&
-7.53100 & 22.1136 & -13.6557(9) & 0.1308(2) & 0.13076(3) &
2.37041 & -5.4786(1) \\
12 & 11 &&
-7.90526 & 19.2047 & -1.6321(3) & 0.09987(5) & 0.099874(7) &
2.41182 & -4.33282(1) \\
12 & 12 &&
-7.77823 & 23.1451 & -14.5753(9) & 0.1186(2) & 0.11864(2) &
2.27472 & -5.32201(9) \\
\hline\hline
\end{tabular}
\caption{\label{tab:fit_pos_Alm} Values for the coefficients of the
  fitting function for $\scA{-2}{\ell{m}}{\bar{a}\bar\omega}$ given in
  Eqs.~(\ref{eq:fit_pos_ReA}) and~(\ref{eq:fit_pos_ImA}) appropriate
  for $\delta^2>0$.  See Table~\ref{tab:fit_pos_Omega} for additional
  details.}
\end{table*}

The fitting functions for the case $\delta^2<0$ are given in
Eqs.~(\ref{eq:fit_neg}).  The real and imaginary parts of $\bar\omega$
are fit using Eqs.~(\ref{eq:fit_neg_ReO}) and~(\ref{eq:fit_neg_ImO}):
\begin{subequations}
\label{eq:fit_neg}
\begin{align}
\label{eq:fit_neg_ReO}
  \rm{Re}[\bar\omega] &= \frac{m}2 + (\alpha_1 + \alpha_2 \tilde{n})\epsilon \\
\label{eq:fit_neg_ImO}
  \rm{Im}[\bar\omega] &= -\left(\alpha_3+\tilde{n}+\frac12\right)
                                                \sqrt{\frac{\epsilon}2}
                         + \left(\alpha_4+\alpha_5\tilde{n}\right)\epsilon \\
\label{eq:fit_neg_ReA}
  \rm{Re}[{}_sA_{\ell m}] &= \ell(\ell+1)-s(s+1)+\beta_1 \\
           &\mbox{}\quad
  + \left(\beta_2 + \beta_3 \tilde{n} + \beta_4 \tilde{n}^2\right)\epsilon
   \nonumber\\
\label{eq:fit_neg_ImA}
  \rm{Im}[{}_sA_{\ell m}] &= \left(\beta_5+\tilde{n}+\frac12\right)\beta_6
          \sqrt{\frac{\epsilon}2} + \left( \beta_7 +\beta_8 \tilde{n} \right)
          \epsilon
\end{align}
\end{subequations}
In this case, we see that the coefficient $\alpha_3$ in
Eq.~(\ref{eq:fit_neg_ImO}) corresponds to the imaginary part of
$\delta$ in Eq.~(\ref{eq:yang_omega}).  The difference between 
Eqs.~(\ref{eq:fit_pos}) and (\ref{eq:fit_neg}) is not simply the
switch of $\delta$ from real to imaginary.  If that were the case,
then we should find $\alpha_4=\alpha_5/2$ in
Eq.~(\ref{eq:fit_neg_ImO}).  The fit values for all cases with
$\delta^2<0$ are shown in Tables~\ref{tab:fit_neg_Omega}
and~\ref{tab:fit_neg_Alm}.  We clearly see in
Table~\ref{tab:fit_neg_Omega} that the fit value for $\alpha_3$ is in
excellent agreement with $\rm{Im}(\delta)$, and that
$\alpha_4\ne\alpha_5/2$.  Note that there are several cases where the
fit data for a given $(\ell,m)$ pair includes only a single overtone.
In these cases, the $\alpha_2$ and $\alpha_5$ coefficients cannot be
determined.

\begin{figure*}[htbp!]
\begin{tabular}{cc}
\includegraphics[width=0.5\linewidth,clip]{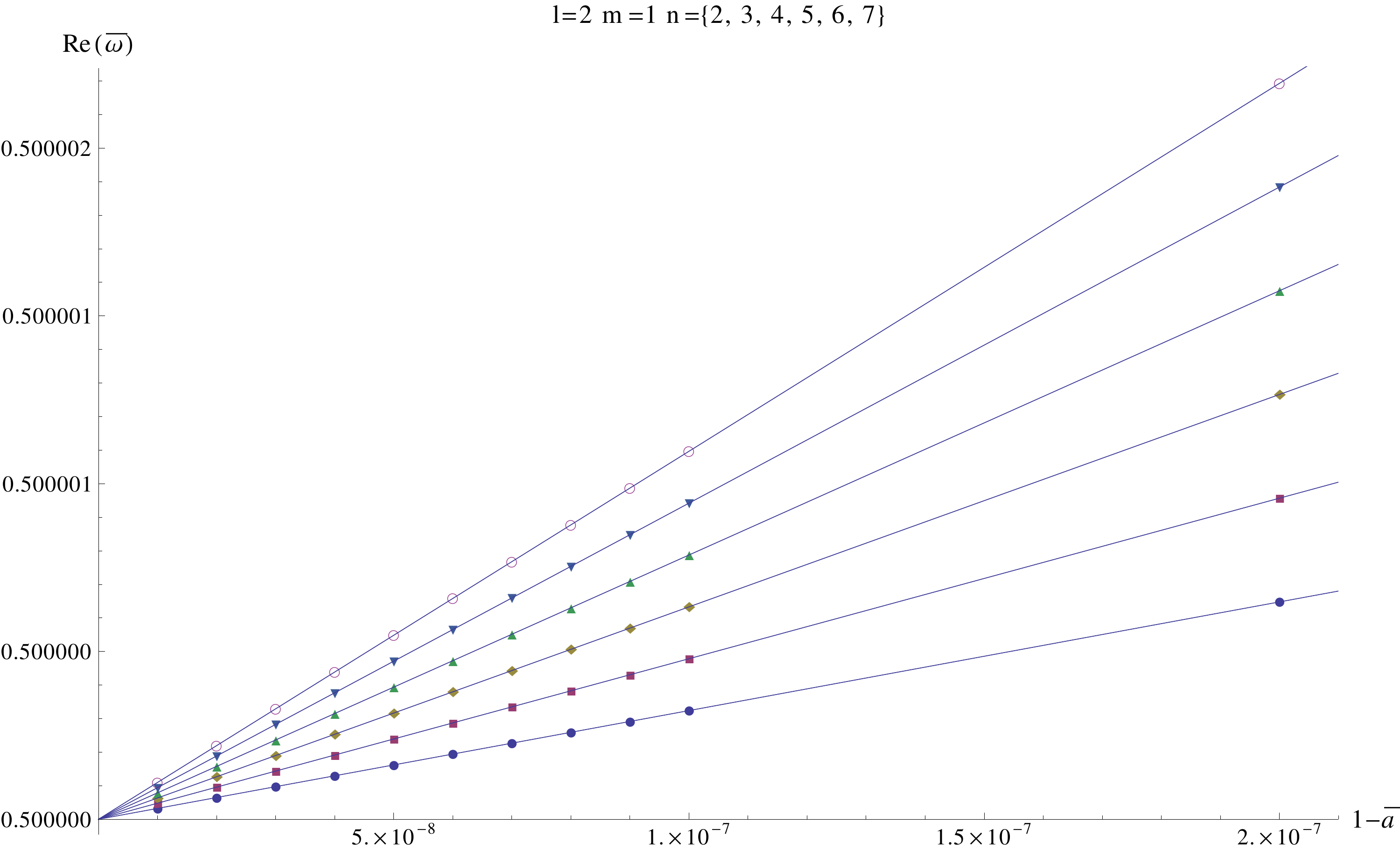} &
\includegraphics[width=0.5\linewidth,clip]{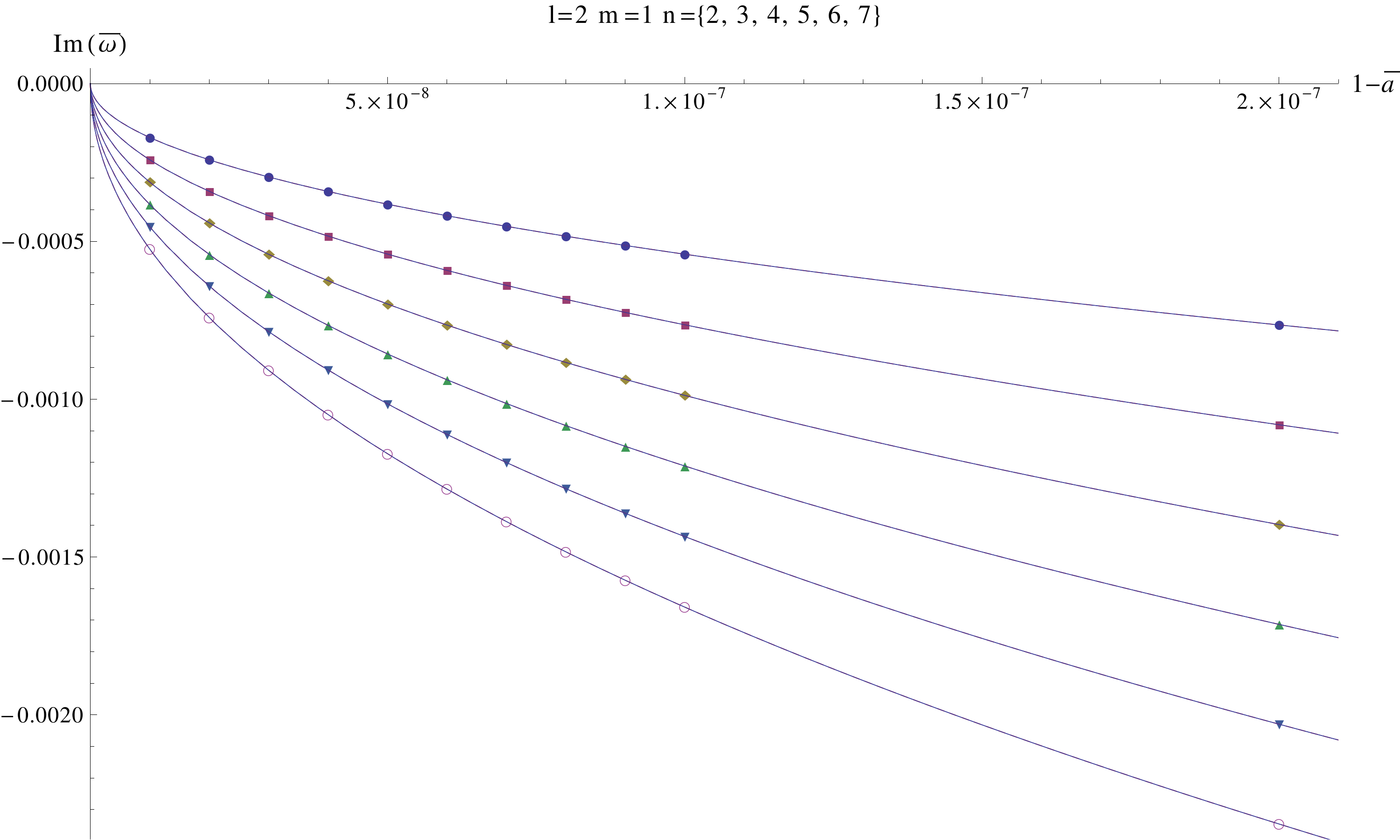}
\end{tabular}
\caption{\label{fig:SupRad_l2m1_O} Near extremal behavior of
  $\bar\omega$ for $\ell=2$, $m=1$.  Symbols in the left plot
  represent data values used to fit the coefficients in
  Eq.~(\ref{eq:fit_neg_ReO}).  The lines represent the resulting fit
  function.  The right plot corresponds to Eq.~(\ref{eq:fit_neg_ImO}).
  In the right plot, the nearly imperceptible dashed(red) line
  represents the fit function omitting the next to leading order
  behavior in $\epsilon$.}
\end{figure*}

\begin{figure*}[htbp!]
\begin{tabular}{cc}
\includegraphics[width=0.5\linewidth,clip]{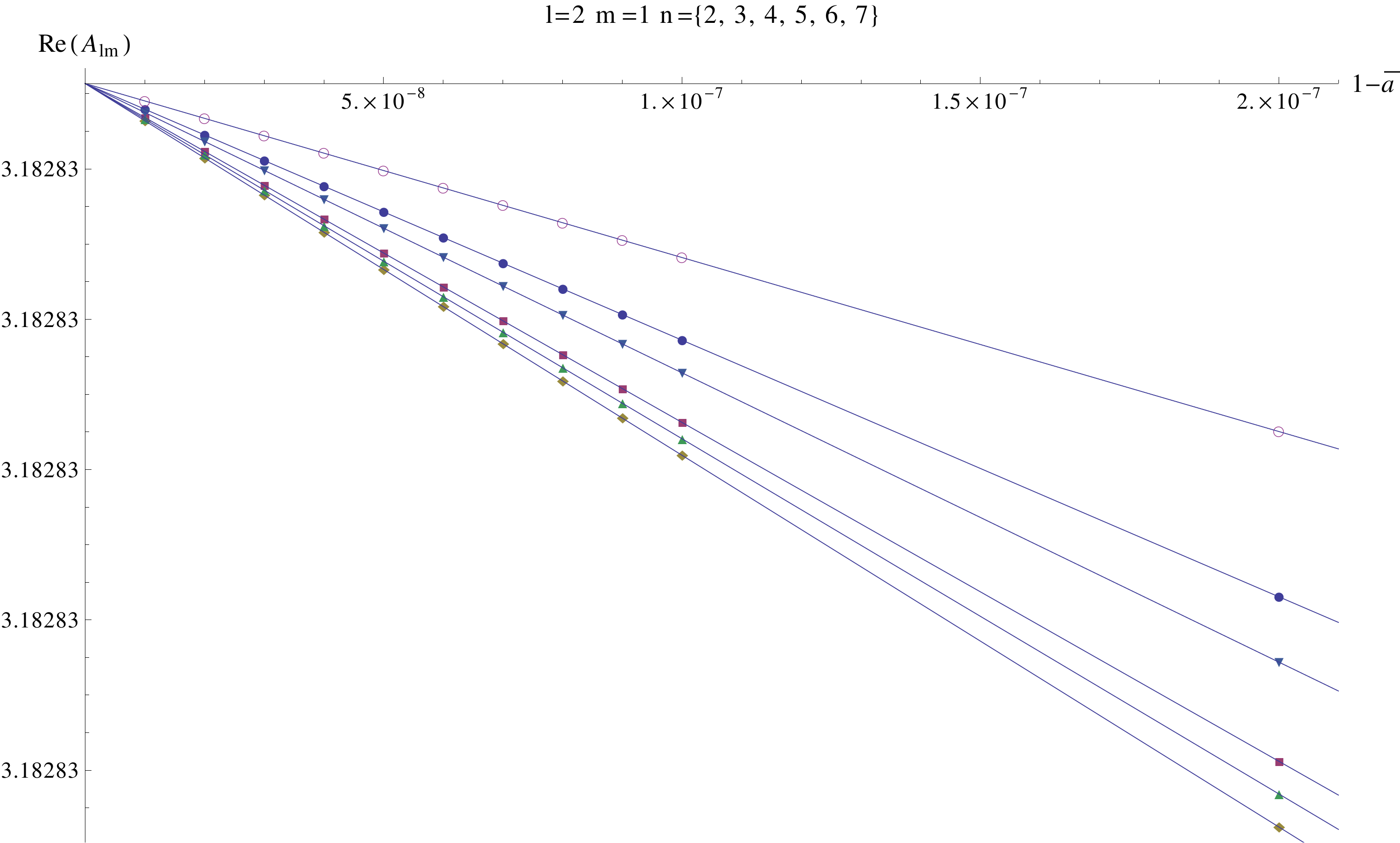} &
\includegraphics[width=0.5\linewidth,clip]{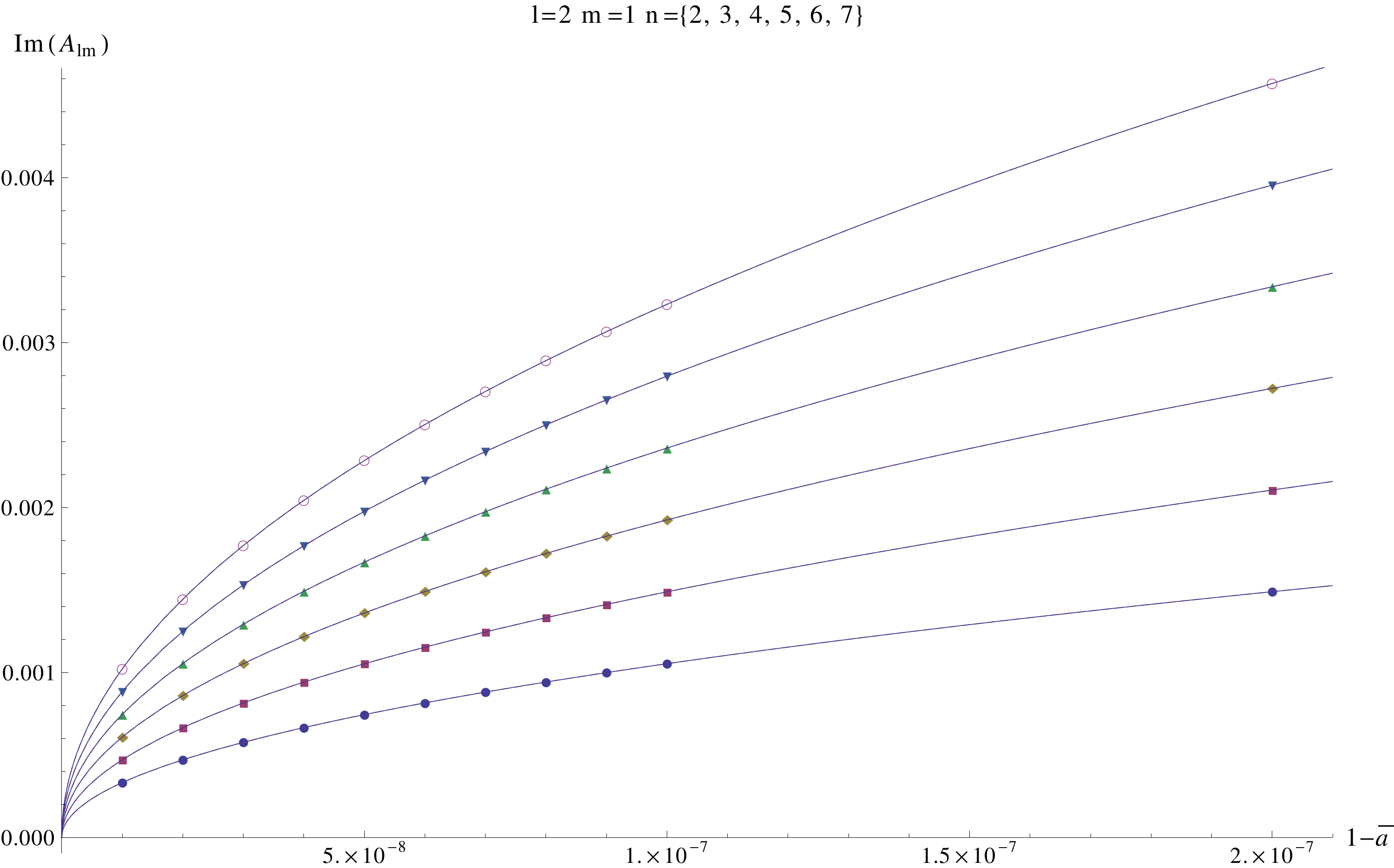}
\end{tabular}
\caption{\label{fig:SupRad_l2m1_A} Near extremal behavior of
  $\scA{-2}{\ell{m}}{\bar{a}\bar\omega}$ for $\ell=2$, $m=1$.  Symbols
  in the left plot represent data values used to fit the coefficients
  in Eq.~(\ref{eq:fit_neg_ReA}).  The lines represent the resulting
  fit function.  The right plot corresponds to
  Eq.~(\ref{eq:fit_neg_ImA}).  In the right plot, the nearly
  imperceptible dashed(red) line represents the fit function omitting
  the next to leading order behavior in $\epsilon$.}
\end{figure*}

Figures~\ref{fig:SupRad_l2m1_O} and~\ref{fig:SupRad_l2m1_A} present
the fits for $\ell=2$ and $m=1$, for which $\delta^2<0$.  This example
is chosen because it clearly illustrates the necessity of including a
quadratic dependence on $\tilde{n}$ in Eq.~(\ref{eq:fit_neg_ReA}).
The most striking difference we see between
Figs.~\ref{fig:SupRad_l2m1_O} and~\ref{fig:SupRad_l5m4_O} is the
strong dependence on the overtone index $\tilde{n}$ in the
$\rm{Re}(\bar\omega)$.  What had been the next to leading order
behavior for $\delta^2>0$ is now the leading order behavior when
$\delta^2<0$.  The same is true for the separation constant.  As seen
in Eq.~(\ref{eq:fit_neg_ReA}), the leading order behavior has a
quadratic dependence on $\tilde{n}$.  This is similar to the next to
leading order behavior seen in Eq.~(\ref{eq:fit_pos_ReA}); however, in
this case $\beta_3\ne\beta_5$.  Also, the necessity of the quadratic
dependence can be clearly seen in Fig.~\ref{fig:SupRad_l2m1_A}.
Finally, as we found for $\delta^2>0$, the leading order behavior of
$\rm{Im}({}_sA_{\ell{m}})$ is proportional to the leading order
behavior of $\rm{Im}(\bar\omega)$.  We see this from the fact that, in
Table~\ref{tab:fit_neg_Alm}, the fitting coefficient $\beta_5$ is in
excellent agreement with $\rm{Im}(\delta)$.  Recalling from above that
a small set of the fits were made with only a single overtone, we note
that in these cases we cannot determine $\beta_3$ and $\beta_4$.
Furthermore, we cannot separately fit for both $\beta_5$ and
$\beta_6$.  In these cases we have fixed $\beta_5=\rm{Im}{\delta}$ and
fit only for $\beta_6$.  The fit values for $\beta_5$ and $\beta_6$
are marked with an asterisk when this modified fitting has been used.
There were also two cases where only two overtones were available for
fits.  For these cases, $\beta_3$ and $\beta_4$ cannot be
distinguished and these fit values are also marked with an asterisk.

\begin{table*}[htbp!]
\begin{tabular}{rrcf{6}f{8}f{9}f{7}f{7}f{6}}
\hline\hline
$\ell$ & $m$ && \Chead{$\delta_i$} &
\Chead{$\alpha_1$} & \Chead{$\alpha_2$} & \Chead{$\alpha_3$} &
\Chead{$\alpha_4$} & \Chead{$\alpha_5$} \\
2 & 1 && 1.91907 &
3.23813(3) & 1.54514(1) & 1.91906 & -0.021(3) & -0.0109(4) \\
3 & 1 && 3.17492 &
2.11224 & 0.710824(1) & 3.17492 & -0.0061(4) & -0.0022(1) \\
4 & 1 && 4.26749 &
1.82009 & \Cempty & 4.26749 & -0.0048(3) & \Cempty \\
3 & 2 && 1.87115 &
7.2471(9) & 3.4744(2) & 1.87108(1) & -0.12(2) & -0.061(2) \\
4 & 2 && 3.47950 &
4.19465(1) & 1.30534 & 3.47950 & -0.0106(8) & -0.0040(2) \\
5 & 2 && 4.72816 &
3.61332 & 0.882398(1) & 4.72816 & -0.0067(3) & -0.0018(2) \\
4 & 3 && 1.37578 &
22.66(2) & 12.807(4) & 1.37549(4) & -1.6(3) & -0.87(4) \\
5 & 3 && 3.54313 &
6.80319(2) & 2.05358(1) & 3.54312 & -0.024(2) & -0.0094(4) \\
6 & 3 && 4.98492 &
5.59000 & 1.29263 & 4.98492 & -0.0103(6) & -0.0029(2) \\
7 & 3 && 6.24553 &
5.13084 & \Cempty & 6.24552 & -0.0077(4) & \Cempty \\
6 & 4 && 3.38736 &
10.6913(1) & 3.26423(5) & 3.38733(1) & -0.075(7) & -0.029(1) \\
7 & 4 && 5.07533 &
7.93057(1) & 1.78114(1) & 5.07532 & -0.018(1) & -0.0053(3) \\
8 & 4 && 6.47378 &
7.07896 & \Cempty & 6.47378 & -0.0100(5) & \Cempty \\
7 & 5 && 2.98127 &
18.146(2) & 5.9243(6) & 2.98114(1) & -0.32(3) & -0.131(5) \\
8 & 5 && 5.01168 &
10.9114 & 2.43320(1) & 5.01167 & -0.033(2) & -0.0101(5) \\
9 & 5 && 6.57480 &
9.30775 & 1.66898(1) & 6.57480 & -0.0152(7) & -0.0036(4) \\
8 & 6 && 2.19168 &
43.25(9) & 16.68(2) & 2.1906(1) & -2.9(3) & -1.45(3) \\
9 & 6 && 4.78975 &
15.0733(1) & 3.41635(4) & 4.78972 & -0.077(6) & -0.024(1) \\
10 & 6 && 6.55627 &
11.9630 & 2.12050(1) & 6.55626 & -0.0252(1) & -0.0063(4) \\
11 & 6 && 8.06162 &
10.7711 & \Cempty & 8.06162 & -0.0153(8) & \Cempty \\
10 & 7 && 4.38687 &
21.7120(5) & 5.1575(2) & 4.38680(1) & -0.21(2) & -0.068(3) \\
11 & 7 && 6.41836 &
15.2830 & 2.71488(1) & 6.41835 & -0.042(2) & -0.0108(7) \\
12 & 7 && 8.07005 &
13.2593 & \Cempty & 8.07005 & -0.021(1) & \Cempty \\
11 & 8 && 3.74604 &
34.980(6) & 9.159(2) & 3.74569(3) & -0.91(8) & -0.32(1) \\
12 & 8 && 6.15394 &
19.6999(1) & 3.56155(4) & 6.15392 & -0.084(5) & -0.022(1) \\
12 & 9 && 2.70389 &
79.21(6) & 25.64(2) & 2.7015(3) & -7.2(8) & -3.20(7) \\
\hline\hline
\end{tabular}
\caption{\label{tab:fit_neg_Omega} Values for the coefficients of the
  fitting function for $\bar\omega$ given in
  Eqs.~(\ref{eq:fit_neg_ReO}) and~(\ref{eq:fit_neg_ImO}) appropriate
  for $\delta^2<0$.  $\delta=i\delta_i$ obtained from
  Eq.~(\ref{eq:yang_delta}). See Table~\ref{tab:fit_pos_Omega} for
  additional details.}
\end{table*}

\begin{table*}[htbp!]
\begin{tabular}{rrcf{6}f{6}f{8}f{9}f{10}f{8}f{7}f{7}f{7}}
\hline\hline
$\ell$ & $m$ && \Chead{$\delta_i$} &
\Chead{$\beta_1$} & \Chead{$\beta_2$} & \Chead{$\beta_3$} & \Chead{$\beta_4$} & 
\Chead{$\beta_5$} & \Chead{$\beta_6$} & \Chead{$\beta_7$} & \Chead{$\beta_8$} \\
2 & 1 && 1.91907 &
-0.817168 & -3.41373(8) & -1.42243(6) & 0.32846(1) &
1.91908(1) & 1.94872 & 0.000(7) & 0.044(2) \\
3 & 1 && 3.17492 &
-0.419864 & 0.636795(1) & 0.514828(1) & 0.167730(1) &
3.17493 & 1.01001 & 0.0046(5) & 0.0053(4) \\
4 & 1 && 4.26749 &
-0.288496 & 2.86068 & \Cempty & \Cempty &
4.26749\mbox{${}^*$} & 0.750315\mbox{${}^*$} & 0.0043(2) & \Cempty \\
3 & 2 && 1.87115 &
-1.74881 & -12.322(2) & -6.493(1) & 0.2325(2) &
1.87121(2) & 2.18260 & -0.05(5) & 0.25(1) \\
4 & 2 && 3.47950 &
-1.14307 & -2.24180 & -0.685669(7) & 0.154790(2) &
3.47951 & 1.46904 & 0.010(2) & 0.01341(9) \\
5 & 2 && 4.72816 &
-0.894475 & 1.27953 & 0.400252\mbox{${}^*$} & 0.400252\mbox{${}^*$} &
4.72817 & 1.24091 & 0.0091(6) & 0.0062(9) \\
4 & 3 && 1.37578 &
-2.60722 & -47.50(4) & -28.71(3) & 0.215(4) & 
1.37604(7) & 2.28731(3) & -1.3(8) & 3.7(2) \\
5 & 3 && 3.54313 &
-1.94623 & -6.97433(4) & -2.44406(4) & 0.146252(9) &
3.54315(1) & 1.76588 & 0.019(5) & 0.034(2) \\
6 & 3 && 4.98492 &
-1.65057 & -1.78614(1) & -0.332540(8) & 0.158689(4) &
4.98493 & 1.60394 & 0.0158(1) & 0.012(1) \\
7 & 3 && 6.24553 &
-1.49341 & 2.43042(1) & \Cempty & \Cempty &
6.24553\mbox{${}^*$} & 1.54481\mbox{${}^*$} & 0.0157(9) & \Cempty \\
6 & 4 && 3.38736 &
-2.77576 & -15.1115(3) & -5.3855(3) & 0.13901(5) &
3.38741(1) & 1.98024 & 0.04(2) & 0.111(6) \\
7 & 4 && 5.07533 &
-2.49104 & -6.48399(1) & -1.67232(2) & 0.151562(5) &
5.07534 & 1.88766 & 0.027(3) & 0.023(2) \\
8 & 4 && 6.47378 &
-2.34013 & -1.30084(1) & \Cempty & \Cempty &
6.47378\mbox{${}^*$} & 1.87150\mbox{${}^*$} & 0.024(1) & \Cempty \\
7 & 5 && 2.98127 &
-3.61202 & -31.920(4) & -11.791(3) & 0.1344(6) &
2.98143(3) & 2.14367(1) & 0.15(8) & 0.51(3) \\
8 & 5 && 5.01168 &
-3.38311 & -13.4092(1) & -3.55546(5) & 0.14478(2) &
5.01169 & 2.11696 & 0.051(6) & 0.045(3) \\
9 & 5 && 6.57480 &
-3.27195 & -6.54422(1) & -0.570367(2)\mbox{${}^*$} & -0.570367(2)\mbox{${}^*$} &
6.57481(1) & 2.14543(1) & 0.037(2) & 0.019(3) \\
8 & 6 && 2.19168 &
-4.44655 & -88.9(2) & -38.1(1) & 0.28(2) &
2.1924(3) & 2.2721(1) & 1.(1) & 5.1(3) \\
9 & 6 && 4.78975 &
-4.30830 & -23.9745(2) & -6.4163(2) & 0.13857(5) &
4.78979(1) & 2.30668 & 0.10(2) & 0.109(6) \\
10 & 6 && 6.55627 &
-4.26538 & -13.6545 & -2.87125(4) & 0.15400(2) &
6.55628(1) & 2.37885 & 0.059(4) & 0.033(3) \\
11 & 6 && 8.06162 &
-4.26020 & -6.76295(1) & \Cempty & \Cempty &
8.06162\mbox{${}^*$} & 2.45322\mbox{${}^*$} & 0.049(3) & \Cempty \\
10 & 7 && 4.38687 &
-5.25539 & -41.747(1) & -11.427(1) & 0.1334(3) & 
4.38697(2) & 2.46644(1) & 0.26(4) & 0.31(2) \\
11 & 7 && 6.41836 &
-5.30470 & -23.3442(1) & -4.9645(1) & 0.14755(5) &
6.41838(1) & 2.58040 & 0.100(8) & 0.058(6) \\
12 & 7 && 8.07005 &
-5.37425 & -14.4186 & \Cempty & \Cempty &
8.07005\mbox{${}^*$} & 2.68892\mbox{${}^*$} & 0.071(4) & \Cempty \\
11 & 8 && 3.74604 &
-6.21720 & -78.26(2) & -22.80(1) & 0.136(3) &
3.74647(9) & 2.60270(3) & 0.8(2) & 1.48(7) \\
12 & 8 && 6.15394 &
-6.37897 & -37.0045(2) & -7.9329(3) & 0.14171(8) &
6.15399(1) & 2.75634 & 0.18(2) & 0.120(9) \\
12 & 9 && 2.70389 &
-7.18896 & -200.7(2) & -69.3(1) & 0.18(2) &
2.7059(7) & 2.7190(3) & 4.(3) & 13.9(8) \\
\hline\hline
\end{tabular}
\caption{\label{tab:fit_neg_Alm} Values for the coefficients of the
  fitting function for $\scA{-2}{\ell{m}}{\bar{a}\bar\omega}$ given in
  Eqs.~(\ref{eq:fit_neg_ReA}) and~(\ref{eq:fit_neg_ImA}) appropriate
  for $\delta^2<0$.  $\delta=i\delta_i$ obtained from
  Eq.~(\ref{eq:yang_delta}).  Special situations where results are
  marked with an asterisk(*) are discussed in the text.  See
  Table~\ref{tab:fit_pos_Omega} for additional details.}
\end{table*}

\subsubsection{Zero-frequency modes}
Another interesting regime for QNMs occurs when modes approach the
negative imaginary axis.  These modes seem to share a connection with
the ``algebraically special''
modes\cite{couch-newman-1973,wald-1973,chandra-1984}, sometimes
referred to as TTMs.  These will be
discussed in some detail in Sec.~\ref{sec:polynomial_solutions}.  The
most easily accessible examples of such behavior is found in the
various $n=8$ overtones of $\ell=2$.  As $\bar{a}\to0$, we find that
these modes approach the negative imaginary axis and, in fact, appear
to approach the frequency of the algebraically special mode of
Schwarzschild\cite{leaver-1985,onozawa-1997} at $\bar\omega=-2i$.
There has been some controversy as to whether such QNMs can actually
exist with zero frequency (see the introduction of
Ref.~\cite{van_den_brink-2000} and Appendix~A of
Ref.~\cite{berti-QNM-2009} for additional details, and
Sec.~\ref{sec:a0_Kerr_modes} below for additional discussion).  In a
very detailed treatment, Maassen~van~den Brink\cite{van_den_brink-2000} claims
that, for the Zerilli equation, the algebraically special modes are
simultaneously a QNM and a TTM${}_{\rm{L}}$.  Furthermore, he shows
that a set of standard Kerr QNMs branch from each of these
algebraically special modes.  For the case of $\ell=2$ he finds
(translated to our notation)
\begin{align}
\label{eq:vdB_limit}
  \bar\omega_{2m8}(\bar{a}) = -2i - &\frac{8269544}{700009}m\bar{a}
       + \frac{436576}{41177}i\bar{a}^2  \nonumber\\
       & + O(m\bar{a}^2) +O(\bar{a}^4).
\end{align}
For positive-frequency modes, Eq.~(\ref{eq:vdB_limit}) only
applies to $m\le0$.  For the cases with $m>0$, Maassen~van~den
Brink\cite{van_den_brink-2000} suggested several possible behaviors,
and in a later paper\cite{van_den_brink-2003} he and coauthors
suggested that the $m>0$ modes might be related to a pair of
``unconventional damped modes'' that seem to exist on the unphysical
sheet lying behind a branch cut.  Of this pair, one set of modes is
associated with the positive-frequency modes and the other with
the negative-frequency modes. The unconventional QNMs were
predicted to be at
\begin{equation}
\label{eq:vdB_ucQNM}
  \bar\omega_\pm=\mp0.013 + (0.0016-2)i,
\end{equation}
with $\bar\omega_+$ serving as the $\bar{a}\to0$ limit for the
positive-frequency $m=1$ and $2$ modes.  The authors suggest that
the modes will move through the NIA as $\bar{a}$ is tuned.

To our knowledge, the best numerical results to
date\cite{onozawa-1997,berticardoso-2003} have not followed these
sequences to very small values of $\bar{a}$ except perhaps for the
$m=0$ case.  It is, in fact, quite challenging to obtain solutions
close to the imaginary axis.  Using the methods described in
Secs.~\ref{sec:Radial_TE} and~\ref{sec:Angular_TE} we have obtained
solutions down to $\bar{a}=10^{-6}$ for $m\in\{0,-1,-2\}$.
\begin{figure*}[htbp!]
\begin{tabular}{cc}
\includegraphics[width=0.5\linewidth,clip]{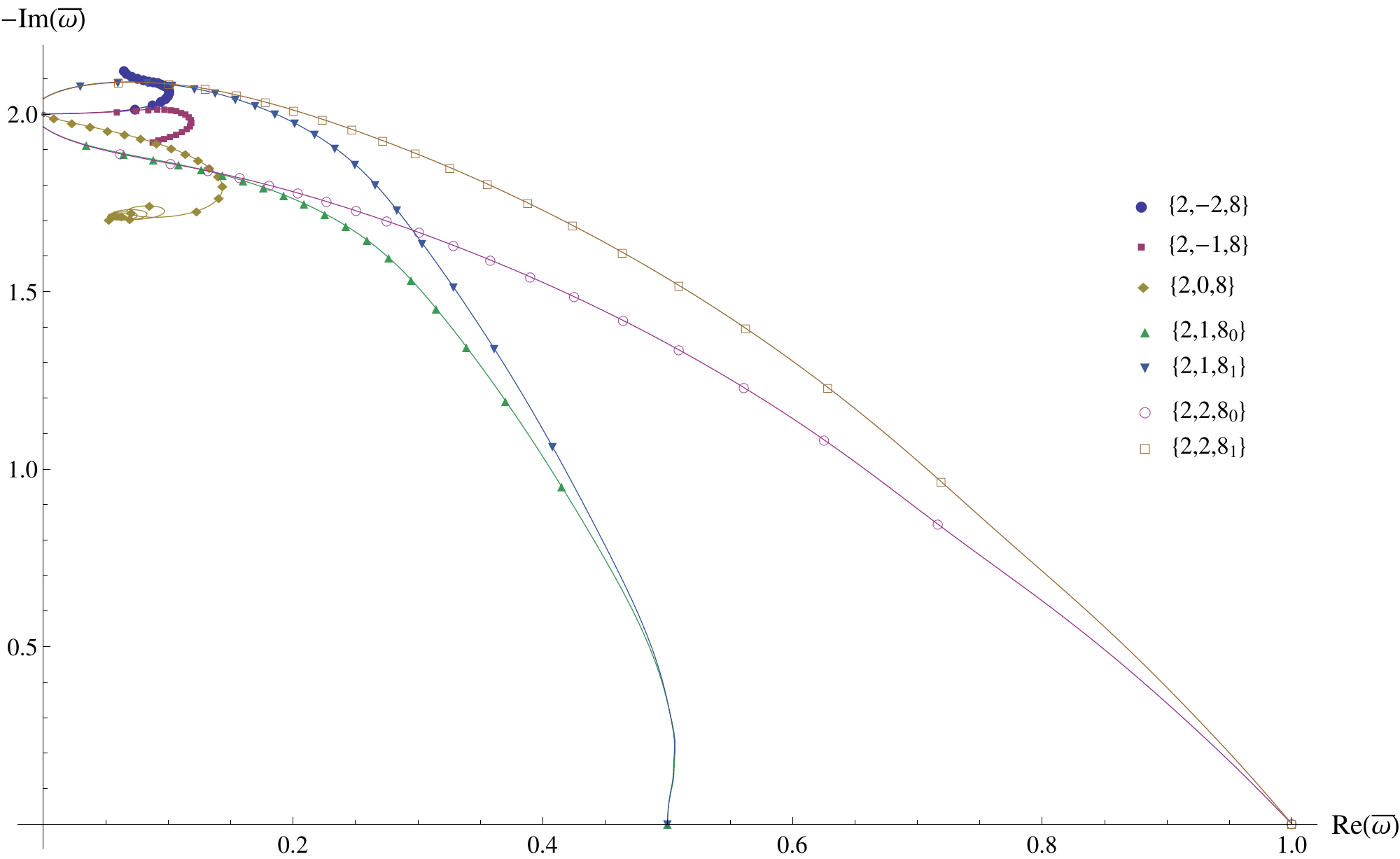} &
\includegraphics[width=0.5\linewidth,clip]{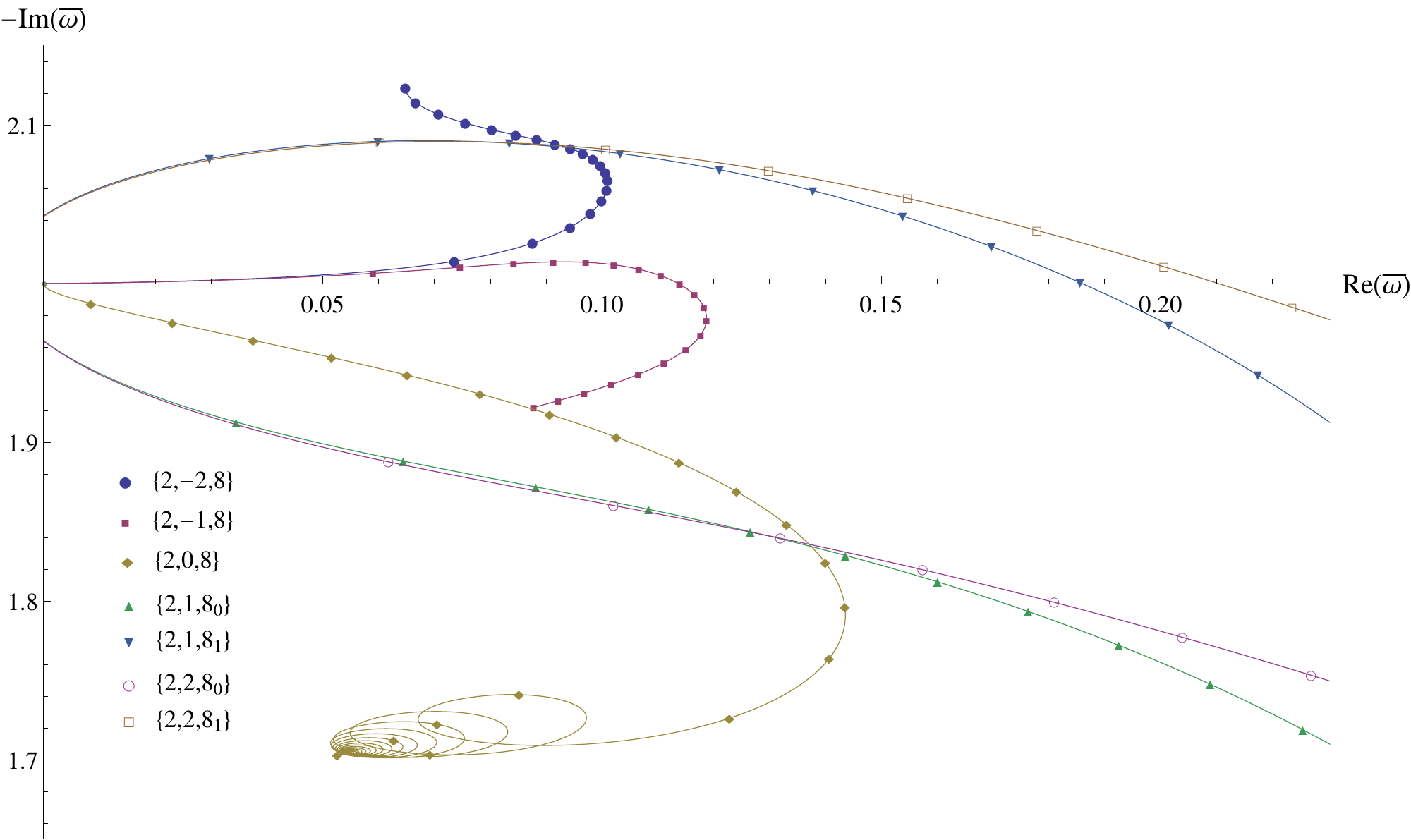}
\end{tabular}
\caption{\label{fig:NIA_l2n8} Behavior of $\bar\omega$ for all $n=8$
  overtones of $\ell=2$.  The left plot shows the full extent of all
  sequences with $-2\le m\le2$ and clearly shows that the $m=1$ and
  $m=2$ modes are both multiplets.  The multiplets are distinguished
  by a subscript on the overtone index.  The right plot shows a
  close-up of the behavior near the NIA.  Note that the $m=1,2$ modes
  for the $n=8_0$ multiplet approach the NIA near $\bar\omega=-1.964$
  while the modes for the $n=8_1$ multiplet approach the NIA near
  $\bar\omega=-2.042$.}
\end{figure*}
The right plot in Fig.~\ref{fig:NIA_l2n8} shows our results for
$\bar\omega$.  While we have not been able to find solutions for these
modes for $\bar{a}=0$, our results are consistent with these modes
beginning at $\bar{a}=0$ at $\bar\omega=-2i$.  For both $m=1$ and $2$,
we find a {\em pair of solutions} associated with $n=8$.  This was
first seen in Ref.~\cite{berticardoso-2003} (see the left panel of
their Fig.~10).  The left plot in Fig.~\ref{fig:NIA_l2n8} shows both
pairs of solutions including their approach to the accumulation points
at $\bar\omega=m/2$.  We refer to these multiple solutions as
``overtone multiplets'' and distinguish them via a subscript on the
overtone.  The use of multiplets is appropriate as confirmed by the
behavior of these modes near the accumulation points.  For both the
$m=1$ and $2$ cases, the $n=8_0$ and $8_1$ mode sequences approach the
accumulation points at $\bar\omega=m/2$ {\em between} the sequences
that extend from the $n=7$ and $n=9$(not shown in this paper)
Schwarzschild QNMs.  Interestingly, these mode sequences do not seem
to start at $\bar{a}=0$.  Instead, they begin at the NIA at a finite
value for $\bar{a}$.  Table~\ref{tab:NIA_l2n8} lists the starting
values for $\bar\omega$ and $\bar{a}$ for each of these overtone
multiplets.
\begin{table}
\begin{tabular}{ccf{6}cf{6}}
\hline\hline
Mode && \Chead{$\bar\omega$} && \Chead{$\bar{a}$} \\
$\{2,2,8_0\}$ && -1.96384i && 0.0034826 \\
$\{2,1,8_0\}$ && -1.96407i && 0.0068819 \\
$\{2,2,8_1\}$ && -2.04223i && 0.0053279 \\
$\{2,1,8_1\}$ && -2.04259i && 0.0108327 \\
\hline\hline
\end{tabular}
\caption{\label{tab:NIA_l2n8} Initial values for $\bar\omega$ and
  $\bar{a}$ for the $\ell=2$ multiplets of $m=1$ and $2$.  The
  ``first'' multiplet is labeled by $n=8_0$ and the ``second'' by
  $n=8_1$.}
\end{table}

We have obtained $\bar{a}\to0$ fits of our numerical results for the
$\{2,m,8\}$ mode sequences.  The results are
\begin{subequations}
\label{eq:AS_fit}
\begin{align}
\label{eq:AS_fit_m2}
  \bar\omega_{2\minus28}(\bar{a})=&-2i + 23.29(4)\bar{a} - 4220(15)i\bar{a}^2, \\
\label{eq:AS_fit_m1}
  \bar\omega_{2\minus18}(\bar{a})=&-2i + 11.70(1)\bar{a} - 1080(2)i\bar{a}^2, \\
\label{eq:AS_fit_0}
  \bar\omega_{208}(\bar{a})=&-2i + 10.60i\bar{a}^2 + 7781\bar{a}^4, \\
\label{eq:AS_fit_10}
  \bar\omega_{218_0}(\bar{a})=&-0.007323 + 1.235\bar{a} - 2.065\bar{a}^{3/2}
  \nonumber\\&\mbox{}
      -1.989i + 5.554i\bar{a} - 24.00i\bar{a}^{3/2}, \\
\label{eq:AS_fit_11}
  \bar\omega_{218_1}(\bar{a})=&-0.008364 + 0.7306\bar{a} +0.3983(1)\bar{a}^{3/2}
  \nonumber\\&\mbox{}
      -2.013i -4.367i\bar{a} +15.91i\bar{a}^{3/2}, \\
\label{eq:AS_fit_20}
  \bar\omega_{228_0}(\bar{a})=&-0.007315 + 2.460\bar{a} - 6.096\bar{a}^{3/2}
  \nonumber\\&\mbox{}
      -1.989i + 11.02i\bar{a} - 66.83i\bar{a}^{3/2}, \\
\label{eq:AS_fit_21}
  \bar\omega_{228_1}(\bar{a})=&-0.008375 + 1.476\bar{a} +1.312(2)\bar{a}^{3/2}
  \nonumber\\&\mbox{}
      -2.013i -8.797(2)i\bar{a} + 45.70(2)i\bar{a}^{3/2},
\end{align}
\end{subequations}
where we display results to four significant digits unless the fit
yielded fewer.  In that case, we also display in parentheses the
one-sigma uncertainty in the fit.  

In Ref.~\cite{berticardoso-2003}, the authors compare their numerical
results to the predictions of Maassen~van~den
Brink\cite{van_den_brink-2000,van_den_brink-2003}.  They found that
``none of the QNMs we numerically found seem to agree with the
analytic prediction when the rotation rate $a$ is small,'' referring
to the prediction of Eq.~(\ref{eq:vdB_limit}).  Their nonagreement
is likely due to not having data for sufficiently small $\bar{a}$.  We
find that Eqs.~(\ref{eq:AS_fit_m2}), (\ref{eq:AS_fit_m1}),
and~(\ref{eq:AS_fit_0}) show excellent agreement with
Eq.~(\ref{eq:vdB_limit}).  The coefficients for the terms that are
linear in $\bar{a}$ are in good agreement with
$m\times8269544/700009$.  The imaginary term that is quadratic in
$\bar{a}$ in Eq.~(\ref{eq:AS_fit_0}) is in good agreement with
$436576/41177$.  Coefficients for the higher-order terms in these
three equations are not determined in Eq.~(\ref{eq:vdB_limit}), but
are of the prescribed form.  Interestingly, while
Eq.~(\ref{eq:vdB_limit}) predicts that
$\bar\omega_{2\minus28}(\bar{a})=\bar\omega_{2\minus18}(2\bar{a})$ up
to linear order, we find that it is a good approximation including the
quadratic term in $\bar{a}$.

Our results for the $m=1$ and $2$ multiplets seem to follow the general
predictions of Ref.~\cite{van_den_brink-2003}, except that the
multiplets suggest that there are {\em two} pairs of unconventional
modes in the Schwarzschild limit, not the single pair they found.  The
first, associated with Eqs.~(\ref{eq:AS_fit_10}) and
(\ref{eq:AS_fit_20}) extrapolates to
\begin{equation}
  \bar\omega \approx -0.0073 + (0.011 - 2)i,
\end{equation}
while the second, associated with Eqs.~(\ref{eq:AS_fit_11}) and
(\ref{eq:AS_fit_21}) extrapolates to
\begin{equation}
  \bar\omega \approx -0.0084 + (-0.013 - 2)i.
\end{equation}
Again, our results for $m=1$ and $2$ are significantly different from
the related extrapolated results in Ref.~\cite{berticardoso-2003}.
Interestingly, we also find that, to a good approximation,
$\bar\omega_{228_{0,1}}(\bar{a})=\bar\omega_{218_{0,1}}(2\bar{a})$ for
$\bar{a}\to0$.

\subsection{Polynomial solutions of the Teukolsky equation}
\label{sec:polynomial_solutions}

In Sec.~\ref{sec:Heun_polynomials} we described briefly the
construction of confluent Heun polynomials.  As a concrete example,
consider the case of the TTM modes.  It is customary to refer to modes
which travel away from the black-hole horizon and out at infinity as
TTM${}_{\rm{L}}$ modes, and those traveling into the black-hole
horizon and in from infinity as TTM${}_{\rm{R}}$ modes.  These were
studied in detail first by Chandrasekhar\cite{chandra-1984}, and later
by Andersson\cite{andersson-1994} and Maassen~van~den
Brink\cite{van_den_brink-2000}.  We will show that these modes can be
derived by means of confluent Heun polynomials.

Recall that the radial Teukolsky equation can be put in the form of
the confluent Heun equation in eight different ways depending on the
choice of the parameters $\{\zeta,\xi,\eta\}$ [see
Eqs.~(\ref{eq:Teukolsky_Heun_parameters})].  The parameter $\xi$
affects the behavior of the solution at the black-hole horizon.  If we
choose $\xi=\xi_+$ then the unphysical solution of modes propagating
out of the black hole is represented by the local solution of
Eq.~(\ref{eq:local_sol_z1b}), or if we choose $\xi=\xi_\minus$, then
the same unphysical solution is represented by
Eq.~(\ref{eq:local_sol_z1a}).  Both of these behave at the event
horizon like the second form of Eq.~(\ref{eq:R_local_z1}).  Similarly,
$\zeta_+$ and Eq.~(\ref{eq:local_sol_zinfa}), or $\zeta_\minus$ and
Eq.~(\ref{eq:local_sol_zinfb}) both represent modes traveling out at
infinity as seen in the first form of Eq.~(\ref{eq:R_local_zinf}).
These two parameters yield four different ways of writing the same physical
solutions representing a TTM${}_{\rm{L}}$ mode.  However, confluent
Heun polynomials are simultaneously Frobenius solutions at all three
singular points.  The choice of $\eta$ affects the behavior of the
solution at the Cauchy horizon and we see that a TTM${}_{\rm{L}}$
solution can have one of the two different behaviors near the Cauchy
horizon given by Eq.~(\ref{eq:R_local_z0}).

The two possible solutions with different behaviors at the Cauchy
horizon can each be constructed by any of the 4 different
decompositions of the radial Teukolsky equation, but they each reduce
to the same underlying condition that must be satisfied if a
polynomial solution is to exist.  That condition, that the second
parameter of the local solution [either Eq.~(\ref{eq:local_a_sol})
or (\ref{eq:local_r_sol})] must be a nonpositive integer $-q$,
has the following forms for possible TTM${}_{\rm{L}}$ modes:
\begin{equation}\label{eq:NP_boundary_set_a}
  -q= \left\{\begin{array}{cc}
            \alpha & (\bar\zeta_+,\xi_+,\eta_\pm), \\
            \alpha+1-\delta & (\bar\zeta_+,\xi_\minus,\eta_\pm), \\
            -\alpha+\gamma+\delta & (\bar\zeta_\minus,\xi_+,\eta_\pm), \\
            \alpha+1-\gamma & (\bar\zeta_+,\xi_+,\eta_\mp).
  \end{array}\right. 
\end{equation}
The combinations of $\alpha$, $\gamma$, and $\delta$ are simply the
second parameter of the appropriate version of
Eqs.~(\ref{eq:all_local_sol}). In parentheses are the corresponding
choices for the parameters $\{\zeta,\xi,\eta\}$.  Each line on the
right-hand side of Eq.~(\ref{eq:NP_boundary_set_a}) reduces to the same
two conditions:
\begin{subequations}
\begin{align}
\label{eq:NP_boundary_set_b1}
  q &= -1-2s \quad\mbox{or} \\
\label{eq:NP_boundary_set_b2}
  q &= -1-s
      -i\frac{2\bar\omega(1-\sqrt{1-\bar{a}^2})-\bar{a}m}{\sqrt{1-\bar{a}^2}},
\end{align}
\end{subequations}
depending on the choice of $\eta_\pm$.  The upper sign choice for
$\eta$ in Eq.~(\ref{eq:NP_boundary_set_a}) corresponds to
Eq.~(\ref{eq:NP_boundary_set_b1}) and the lower sign choice to
Eq.~(\ref{eq:NP_boundary_set_b2}).  The possible solutions with $q$
given by Eq.~(\ref{eq:NP_boundary_set_b1}) correspond to the solution
that behaves like $z^{-i\sigma_\minus}$ at the Cauchy horizon
($z\to0$).  For $q$ given by Eq.~(\ref{eq:NP_boundary_set_b2}),
possible solutions behaves like $z^{-s+i\sigma_\minus}$ at the Cauchy
horizon.

Consider the case where $q$ is given by
Eq.~(\ref{eq:NP_boundary_set_b1}).  The requirement that $q$ must be a
non-negative integer does not fix $\bar\omega$, but does require that
$s\le-\frac12$ and that $s$ is either an integer or a half integer in
magnitude.  The requirement on $s$ is sensible given that we require
$s\le0$ for solutions of the radial Teukolsky equation to represent
outgoing waves at infinity.  The requirement on $s$ is necessary, but
not sufficient to guarantee the existence of polynomial solutions.  In
conjunction with this condition, the $\Delta_{q+1}=0$ condition (see
Sec.~\ref{sec:Heun_polynomials}) is necessary and sufficient to
guarantee the existence of polynomial solutions, and will in fact
determine $\bar\omega$ when solutions exist.

For gravitational perturbations, $s=-2$ giving $q=3$.  The matrix for
the $\Delta_{q+1}=0$ condition can be constructed from any of the six local
solutions in Eqs.~(\ref{eq:all_local_sol}), but it is easiest to use
Eq.~(\ref{eq:local_sol_z0a}).  Written as an eigenvalue problem for
the quantity $\sigma_{n3}$ (not to be confused with $\sigma_\pm$), the
determinant of the matrix takes the form 
\begin{widetext}
\begin{equation}\label{eq:gen_poly_matrix}
  \left|\begin{array}{cccc}
    -\sigma_{n3} & 
    1-2i\bar\omega+i\frac{2\bar\omega-\bar{a}m}{\sqrt{1-\bar{a}^2}} &
    0 & 0 \\
    -12i\bar\omega\sqrt{1-\bar{a}^2} &
    -2 + 4i\bar\omega\left(1-\sqrt{1-\bar{a}^2}\right) - \sigma_{n3} &
    -4i\bar\omega + 2i\frac{2\bar\omega-\bar{a}m}{\sqrt{1-\bar{a}^2}} &
    0 \\
    0 & -8i\bar\omega\sqrt{1-\bar{a}^2} &
    -2 + 8i\bar\omega\left(1-\sqrt{1-\bar{a}^2}\right) - \sigma_{n3} &
    -3 - 6i\bar\omega + 3i\frac{2\bar\omega-\bar{a}m}{\sqrt{1-\bar{a}^2}} \\
    0 & 0 & -4i\bar\omega\sqrt{1-\bar{a}^2} &
    12i\bar\omega\left(1-\sqrt{1-\bar{a}^2}\right) - \sigma_{n3}
  \end{array}\right|=0.
\end{equation}
\end{widetext}
For the nonspinning case $\bar{a}=0$, we find the four eigenvalues to be
\begin{subequations}\label{eq:TTML_a0_eigenvalues}
\begin{align}
\label{eq:TTML_a0_eigenvalues_03}
  \sigma_{03} &= -1 - \sqrt{1+12i\bar\omega}, \\
\label{eq:TTML_a0_eigenvalues_13}
  \sigma_{13} &= -1 - \sqrt{1-12i\bar\omega}, \\
\label{eq:TTML_a0_eigenvalues_23}
  \sigma_{23} &= -1 + \sqrt{1-12i\bar\omega}, \\
\label{eq:TTML_a0_eigenvalues_33}
  \sigma_{33} &= -1 + \sqrt{1+12i\bar\omega}.
\end{align}
\end{subequations}
These eigenvalues are associated with Eq.~(\ref{eq:local_sol_z0a}), so
we must also compute the corresponding value of $\sigma$ associated
with Eq.~(\ref{eq:local_sol_z0a}).  For $\bar{a}=0$, we find
$\sigma=\scA{-2}{\ell{m}}{0}=\ell(\ell+1)-2$.  Polynomial solutions
exist when $\sigma_{nq}=\sigma$.  Since $\sigma>0$, both $\sigma_{03}$
and $\sigma_{13}$ are not allowed.
Equations.~(\ref{eq:TTML_a0_eigenvalues_23})
and~(\ref{eq:TTML_a0_eigenvalues_33}) yield
\begin{equation}\label{eq:TTML_a0_omega}
  \bar\omega=\pm\frac{i}{12}(\ell-1)\ell(\ell+1)(\ell+2).
\end{equation}
$\sigma_{23}$ has the positive sign and represents an unstable
solution.  $\sigma_{33}$, with the negative sign, represents the
physical solution.  These frequencies are the well-known algebraically
special frequencies of Schwarzschild.  Since these appear frequently,
we will define them as
\begin{equation}\label{eqn:AS_Omega}
\bar\Omega\equiv-\frac{i}{12}(\ell-1)\ell(\ell+1)(\ell+2).
\end{equation}
Associated with these frequencies are the confluent Heun polynomial
solutions
\begin{equation}\label{eq:TTML_a0_H_sol1}
  Hc_{33}(i\bar\Omega,-1,-1+4i\bar\Omega;z) = z^2-\frac{l(l+1)}3z^3
\end{equation}
which corresponds to the physical solution
\begin{equation}\label{eq:TTML_a0_R_sol1}
  R(z) = z^2(z-1)^{2i\bar\Omega}\left(1-\frac{l(l+1)}3z\right)e^{2i\bar\Omega z}.
\end{equation}
We note that Eq.~(\ref{eq:TTML_a0_H_sol1}) does not have the expected
behavior for $z\to0$.  The reason for this is that for $\bar{a}=0$,
$\gamma=-1$ and the characteristic exponents for the singular point at
$z=0$ (Cauchy horizon) differ by an integer.  Only the series solution
corresponding to the larger exponent must be in the form of a simple
polynomial.  The situation is even more interesting, because
$\delta=-1+4i\bar\Omega$ so the characteristic exponents for the
singular point at $z=1$ (event horizon) also differ by an integer.
Rewriting Eq.~(\ref{eq:TTML_a0_H_sol1}) as a local solution at the
$z=1$ regular-singular point, we find that it corresponds to the
larger characteristic exponent and the physical solution of
Eq.~(\ref{eq:TTML_a0_R_sol1}) has the desired behavior for a
TTM${}_{\rm L}$ mode.  However, we will return to further consider the
behavior at $z=1$ below.

Returning to the $\Delta_{q+1}=0$ condition of
Eq.~(\ref{eq:gen_poly_matrix}), for the general case with
$\bar{a}\ne0$, the corresponding value of $\sigma$ is
\begin{equation}\label{eq:gen_sigma_val}
  \sigma = \scA{-2}{\ell{m}}{\bar{a}\bar\omega} 
    + \bar{a}^2\bar\omega^2 + \bar\omega\left(
       6i\left(1-\sqrt{1-\bar{a}^2}\right) - 2\bar{a}m\right).
\end{equation}
While we could, in principle, follow a similar procedure and equate
the eigenvalues for the general case with $\sigma$, it is easier to
insert this value for $\sigma$ into the matrix and then simply require
that the determinant vanish.  The result can be written as
\begin{align}\label{eq:Starobinsky_const}
  0=\lambdabar^2(\lambdabar+2)^2 
  &+ 8\lambdabar\bar{a}\bar\omega\left(
      6(\bar{a}\bar\omega+m) -5\lambdabar(\bar{a}\bar\omega-m)\right)
      \nonumber\\ \mbox{}&
      + 144\bar\omega^2\left(1+\bar{a}^2(\bar{a}\bar\omega-m)^2\right),
\end{align}
where
\begin{equation}
  \lambdabar=\lambdabar_\minus\equiv \scA{-2}{\ell{m}}{\bar{a}\bar\omega}
     + \bar{a}^2\bar\omega^2 - 2m\bar{a}\bar\omega,
\end{equation}
and the right-hand side of Eq.~(\ref{eq:Starobinsky_const}) is just the
magnitude squared of the Starobinsky constant\cite{chandra-1984}.

We will examine the general solutions associated with
Eq.~(\ref{eq:NP_boundary_set_b1}) below, but first, we consider the
other possible solution with $q$ given by
Eq.~(\ref{eq:NP_boundary_set_b2}).  This case is somewhat more
complicated.  With $s=-2$ we find that, in general, only $q=0$ can
have a stable solution with $\rm{Im}(\bar\omega)<0$.  The associated
matrix is one-dimensional and trivial, with vanishing eigenvalue.  In
this case, since $\bar\omega$ is fixed, we must find that an
eigenvalue equals $\sigma$ for that frequency in order for a
polynomial solution to exist.  But in this case, $\sigma$ is a
function of $\bar{a}$, $m$, and
$\scA{-2}{\ell{m}}{\bar{a}\bar\omega}$.  The $\Delta_{q+1}=0$
condition is not satisfied and no solution exists for arbitrary
$\bar{a}$.  However, if $\bar{a}=0$, and $s=-2$, we find $q=1$ for any
$\bar\omega$, and the $\Delta_{q+1}=0$ condition leads to the two
eigenvalues
\begin{equation}
\sigma_{01}=1-\sqrt{1+12i\bar\omega},\quad
\sigma_{11}=1+\sqrt{1+12i\bar\omega}.
\end{equation}
With $\sigma_{01}$ not allowed because it is negative for stable
solutions, we find from $\sigma_{11}$ that $\bar\omega=\bar\Omega$
again.  The associated confluent Heun polynomial solution is
\begin{equation}\label{eq:TTML_a0_H_sol2}
  Hc_{11}(i\bar\Omega,3,-1+4i\bar\Omega;z) = 1-\frac{l(l+1)}3z.
\end{equation}
Interestingly, we find that the corresponding physical solution is
identical to Eq.~(\ref{eq:TTML_a0_R_sol1}) which supposedly has a
different boundary condition on the Cauchy horizon.  The reason for
this is, of course, that the characteristic exponents for the singular
point at $z=0$ (Cauchy horizon) differ by an integer.  For
$\bar{a}=0$, the polynomial solutions corresponding to both
Eqs.~(\ref{eq:NP_boundary_set_b1}) and (\ref{eq:NP_boundary_set_b2})
both yield $\bar\omega=\bar\Omega$ and each is finding the series
solution corresponding to the larger characteristic exponent.

Returning to the solutions corresponding to
Eq.~(\ref{eq:NP_boundary_set_b1}), when $\bar{a}\ne0$,
$\gamma=-1-i\sigma_\minus$ which will not be an integer in general.
Therefore a polynomial solution that behaves like
$z^{-i\sigma_\minus}$ at the Cauchy horizon will exist for the
TTM${}_{\rm{L}}$ case when $a\ne0$.  These solutions are the
well-known algebraically special solutions which have been studied
in the past.  The associated polynomial solution is a simple cubic
function of the form $P_0+P_1z+P_2z^2+z^3$, where the $P_n$ are
complicated functions of $\ell$, $m$, $\bar{a}$, $\bar\omega$, and
$\scA{-2}{\ell{m}}{\bar{a}\bar\omega}$ whose exact form is not
important to us here.  But we do find that $P_{0,1}\to0$ as
$\bar{a}\to0$ allowing for a smooth transition to
Eq.~(\ref{eq:TTML_a0_H_sol1}), and the general form of the physical
solution is
\begin{equation}
\label{eqn:Gen_AS_TTML}
  R(z) = z^{-i\sigma_\minus}(z-1)^{i\sigma_+}
      (P_0+P_1z+P_2z^2+z^3)e^{i(\bar{r}_+-\bar{r}_\minus)\bar\omega z}.
\end{equation}

The TTM${}_{\rm{R}}$ modes can be found in an analogous manner.  We
require the physical boundary condition at the event horizon to permit
only modes that travel into the black hole.  At infinity, we allow
only incoming waves.  In this case, the condition that the second
parameter of the local solution [either Eq.~(\ref{eq:local_a_sol}) or
(\ref{eq:local_r_sol})] must be a nonpositive integer $-q$ has the
following forms:
\begin{equation}\label{eq:PN_boundary_set_a}
  -q= \left\{\begin{array}{cc}
            \alpha & (\bar\zeta_\minus,\xi_\minus,\eta_\pm), \\
            \alpha+1-\delta & (\bar\zeta_\minus,\xi_+,\eta_\pm), \\
            -\alpha+\gamma+\delta & (\bar\zeta_+,\xi_\minus,\eta_\pm), \\
            \alpha+1-\gamma & (\bar\zeta_\minus,\xi_\minus,\eta_\mp).
  \end{array}\right. 
\end{equation}
As before, these all reduce to a pair of conditions:
\begin{subequations}
\begin{align}\label{eq:PN_boundary_set_b1}
  q &=-1+s+i\frac{2\bar\omega(1-\sqrt{1-\bar{a}^2})-\bar{a}m}{\sqrt{1-\bar{a}^2}} \quad\mbox{or} \\ \label{eq:PN_boundary_set_b2}
  q &= -1+2s.
\end{align} 
\end{subequations}
Because of the boundary condition at infinity we must consider only
positive values for $s$.  For the case of gravitational waves with
$s=2$, we find that the possible solutions given by
Eq.~(\ref{eq:PN_boundary_set_b1}) yield no physical solutions, even
for $\bar{a}=0$.  The possible solutions given by
Eq.~(\ref{eq:PN_boundary_set_b2}) correspond to $q=3$ and, when
$\bar{a}=0$ yields only one physical solution with
$\bar\omega=\bar\Omega$.  The associated solution is
\begin{align}
  Hc_{33}(-&i\bar\Omega,-1,-1-4i\bar\Omega;z) \nonumber \\ &=
  1 + (\ell-1)(\ell+2)z \\ \mbox{} & + \frac13(\ell-1)^2(\ell+2)^2z^2\left(
  1 + \frac{\ell(\ell+1)}3 z\right) \nonumber
\end{align}
which corresponds to the physical solution
\begin{align}\label{eq:TTMR_a0_R_sol1}
R(z)=z^{-2}&(z-1)^{-2-2i\bar\Omega}\times \nonumber \\ \biggl[
  1& + (\ell-1)(\ell+2)z \\\mbox{}& + \frac13(\ell-1)^2(\ell+2)^2z^2\left(
  1 + \frac{\ell(\ell+1)}3 z\right)\biggr]e^{-2i\bar\Omega z}.\nonumber
\end{align}
When $\bar{a}\ne0$, we again find that polynomial solutions are allowed
when Eq.~(\ref{eq:Starobinsky_const}) is satisfied, except that in this
case, 
\begin{equation}
  \lambdabar=\lambdabar_+\equiv \scA{2}{\ell{m}}{\bar{a}\bar\omega}
     + \bar{a}^2\bar\omega^2 - 2m\bar{a}\bar\omega+4.
\end{equation}
However, since
$\scA{-s}{\ell{m}}{\bar{a}\bar\omega}=\scA{s}{\ell{m}}{\bar{a}\bar\omega}+2s$,
we find $\lambdabar_+=\lambdabar_\minus$ yielding the known result
that the TTM${}_{\rm L}$ and TTM${}_{\rm R}$ algebraically special
modes share the same frequency spectrum.

The general form of the physical solution is
\begin{align}
\label{eqn:Gen_AS_TTMR}
  R(z) = z^{-2+i\sigma_\minus}(z&-1)^{-2-i\sigma_+} \times \\
     & (1+P_1z+P_2z^2+P_3z^3)e^{-i(\bar{r}_+-\bar{r}_\minus)\bar\omega z}.\nonumber
\end{align}
In this case, there is no unusual behavior for the $P_n$ coefficients
as $\bar{a}\to0$.  The exact forms of the $P_n$ can be found, but again 
are not important for this paper.

\begin{figure}
\includegraphics[width=\linewidth,clip]{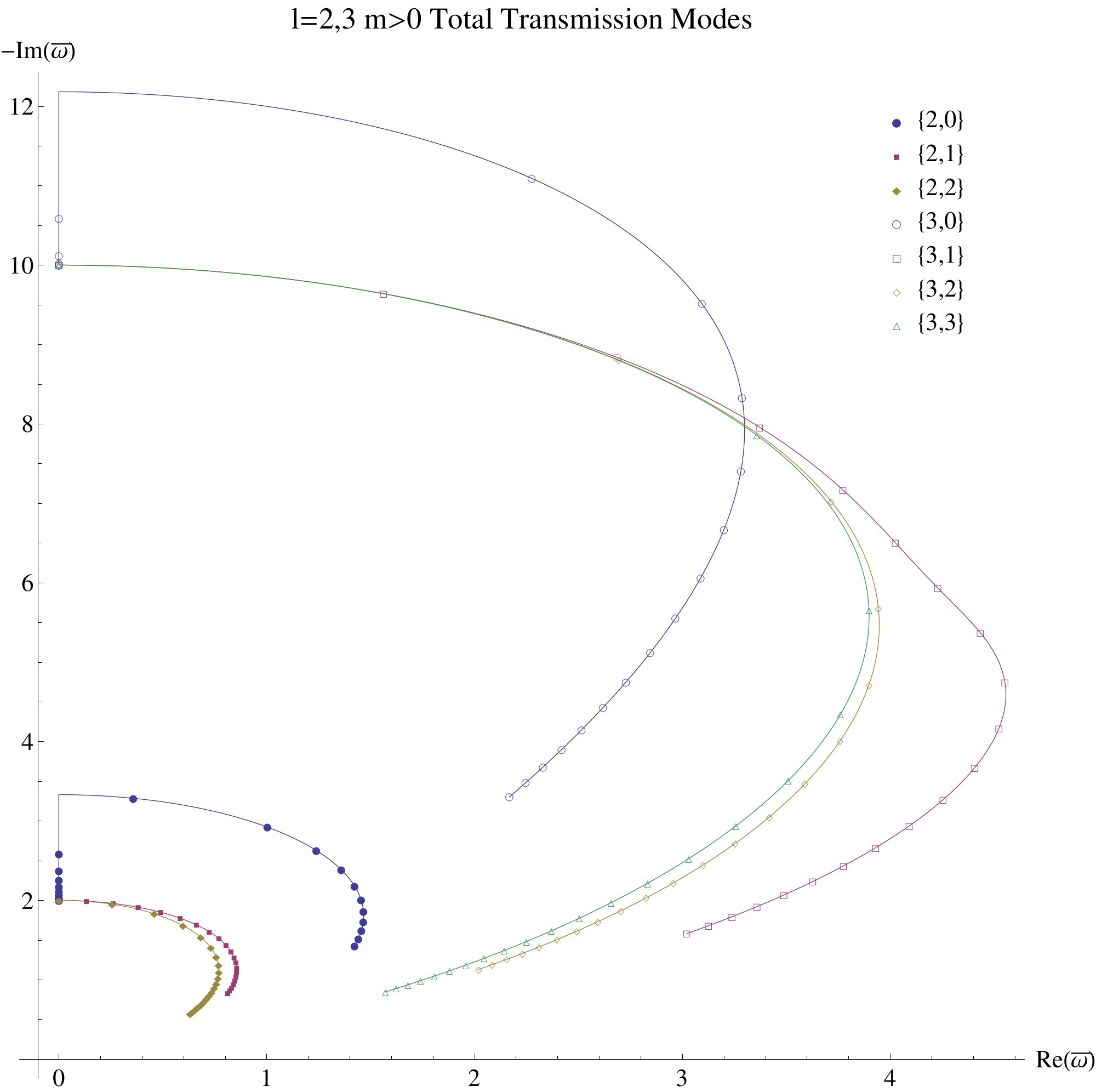}
\caption{\label{fig:TTMs} Kerr algebraically special TTM mode
  sequences for $\ell=2$, $m=0$--2 and for $\ell=3$, $m=0$--3. Note
  that the imaginary axis is inverted and shifted to the left so the
  $m=0$ behavior can be more easily seen.  Each sequence covers a
  range $0\le\bar{a}\le1$, with markers on each sequence denoting a
  change in $\bar{a}$ of $0.05$.}
\end{figure}
We can determine the TTMs of Kerr numerically by simultaneously
solving the $\Delta_{q+1}=0$ condition as given by
Eq.~(\ref{eq:Starobinsky_const}) along with the angular Teukolsky
equation, Eq.~(\ref{eqn:swSF_DiffEqn}).
Equation~(\ref{eq:Starobinsky_const}) is an eight-order polynomial in
$\bar\omega$ whose roots can be found by many methods, and we solve
Eq.~(\ref{eqn:swSF_DiffEqn}) by the spectral eigenvalue method
described in Sec.~\ref{sec:sectral_EV_method}.  In
Fig.~\ref{fig:TTMs}, we plot the $m\ge0$ modes for $\ell=2$ and $3$.
The modes with $m<0$ all have $\rm{Re}(\bar\omega)<0$ and are the
negative frequency counterparts of the $m>0$ modes as given by
Eq.~(\ref{eq:neg_freq_modes}).  The $m=0$ cases are especially
interesting.  The solutions for both $\ell=2$ and $3$ begin, as
$\bar{a}$ increases from zero, along the negative imaginary axis.  At
a critical value of $\bar{a}$, the solution splits into positive
frequency and negative frequency modes again obeying
Eq.~(\ref{eq:neg_freq_modes}).  For $\ell=2$ the critical rotation
rate is $\bar{a}=0.4944459549$ where $\bar\omega=-3.3308102325i$.  For
$\ell=3$ it is $\bar{a}=0.2731627006$ where
$\bar\omega=-12.182211380i$.

\subsubsection{Kerr modes in the $\bar{a}=0$ limit}
\label{sec:a0_Kerr_modes}

The existence, or not, of modes (QNM, TTM${}_{\rm L}$, or TTM${}_{\rm
  R}$) at the special frequencies $\bar\omega=\bar\Omega$ has been the
source of considerable confusion (see
Refs.~\cite{van_den_brink-2000,berti-QNM-2009} for summaries).  In
summary, it seems that most
authors\cite{chandra-1984,leaver-1985,andersson-1994,onozawa-1997}
have taken the algebraically special solutions of Kerr to represent
the TTMs, with the $s=-2$ solutions representing the TTM${}_{\rm L}$
case and $s=2$ the TTM${}_{\rm R}$ case, even in the static limit
$\bar{a}=0$ where the frequencies are given by $\bar\Omega$.  At the
same time, numerical solutions\cite{leaver-1985,onozawa-1997} seem to
suggest that certain QNMs of Kerr approach $\bar\Omega$ as
$\bar{a}\to0$, although numerical difficulties have made it impossible
to explore these cases exactly at $\bar{a}=0$.
Onozawa\cite{onozawa-1997} has suggested that the positive- and
negative-frequency modes cancel at $\bar{a}=0$ leading to no QNMs at
the special frequencies $\bar\Omega$. However, Maassen~van~den
Brink\cite{van_den_brink-2000} comes to a different conclusion, that
there are no TTM${}_{\rm R}$ solutions at $\bar\Omega$ and that a
certain set of solutions are simultaneously TTM${}_{\rm L}$s and QNMs
at $\bar\Omega$.  His arguments are sophisticated, dense, and
difficult to follow but ultimately seem to hinge on how one defines an
ingoing solution at the event horizon (see Secs.~III, VII.A and VII.B
of Ref.~\cite{van_den_brink-2000}).

Using the results derived above for the TTM modes, we hope to bring
some additional clarity to the confusion.  In the
Appendix, we present a standard
derivation\cite{teukolsky-1973} of the asymptotic behavior of solutions
at the event horizon and at infinity.  Assuming these definitions
remain valid at $\bar{a}=0$ and $\bar\omega=\bar\Omega$, then
Eq.~(\ref{eq:TTML_a0_R_sol1}) represents a valid TTM${}_{\rm L}$
solution, and Eq.~(\ref{eq:TTMR_a0_R_sol1}) a valid TTM${}_{\rm R}$
solution, at $\bar{a}=0$ and $\bar\omega=\bar\Omega$.  The latter is
in disagreement with Maassen~van~den Brink\cite{van_den_brink-2000}.

To examine the behavior of the QNM at the special frequencies
$\bar\Omega$, we could again look for Heun polynomial solutions but
with the $\{\zeta,\xi,\eta\}$ parameters set to prefer QNM solutions.
We have done this, but because the characteristic exponents at both
the $z=0$ and $z=1$ regular singular points differ by an integer we
obtain the same TTM${}_{\rm L}$ solution obtained above for
$\bar{a}=0$.  However, we can use this TTM${}_{\rm L}$ solution to
obtain a second, linearly independent solution of the radial Teukolsky
equation at the same frequencies $\bar\omega=\bar\Omega$.  For
simplicity, we will use the Heun polynomial of
Eq.~(\ref{eq:TTML_a0_H_sol2}) obtained from the parameter choice
$\{\zeta_+,\xi_+,\eta_\minus\}$.  Referring to this solution as
$y_1(z)$, we seek a solution $y_2(z)=v(z)y_1(z)$.  Using standard
methods, we find
\begin{equation}
v^\prime(z) = z^{-3}(z-1)^{1-4i\bar\Omega}\left(1-\frac{\ell(\ell+1)}3z\right)^{-2}
             e^{-4i\bar\Omega z}.
\end{equation}
The corresponding physical solution will have the form
\begin{equation}\label{eqn:a0_R_sol2}
R_2(z)=z^2(z-1)^{2i\bar\Omega}\left(1-\frac{\ell(\ell+1)}3z\right)e^{2i\bar\Omega z}
   \int{v^\prime(z){\rm d}z}.
\end{equation}

Expanding about $z=1$ and integrating, we find
\begin{align}
  v(z) =& \frac{f^{(4i\bar\Omega-2)}_{(1)}}{(4i\bar\Omega-2)!}\ln(z-1)\\
  & + (z-1)^{2-4i\bar\Omega}\sum_{n=0}^\prime{
      \frac{f^{(n)}_{(1)}}{n!(n+2-4i\bar\Omega)}(z-1)^n},\nonumber
\end{align}
where the prime denotes that the value of $n=4i\bar\Omega-2$ is
omitted from the sum, and $f^{(n)}_{(1)}=d^nf(z)/dz^n|_{z=1}$ where
$f(z)\equiv (z-1)^{4i\bar\Omega-1}v^\prime(z)$.  We immediately see
that
\begin{equation}
\lim_{z\to1}R_2(z) \sim (z-1)^{2-2i\bar\Omega},
\end{equation}
and this solution obeys the boundary condition at the event horizon
for a QNM [see Eq.~(\ref{eqn:HorizonBehavior}) or
(\ref{eq:R_local_z1}).

For the boundary at infinity, we find that 
\begin{equation}
\lim_{z\to\infty}v(z) = -e^{-4i\bar\Omega z}z^{-4i\bar\Omega}\left(
    \frac{9}{4i\bar\Omega\ell^2(\ell+1)^2z^4} +  \mathcal{O}(z^{-5})\right),
\end{equation}
giving us
\begin{equation}
\lim_{z\to\infty}R_2(z) \sim z^{-1-2i\bar\Omega}e^{-2i\bar\Omega z}.
\end{equation}
However, this is the behavior of an incoming wave [see
Eq.~(\ref{eqn:InfinityBehavior}) or (\ref{eq:R_local_zinf})], the
incorrect behavior for a QNM.  In fact, the two boundary conditions
suggest that $R_2(z)$ is a TTM${}_{\rm R}$ (note, however that
$s=-2$).  What is clear is that, based on the definitions for incoming
and outgoing boundary conditions given in the Appexdix,
neither Eq.~(\ref{eq:TTML_a0_R_sol1}) nor (\ref{eqn:a0_R_sol2})
are a QNM.  Since, for each $\ell$, there can be only two linearly
independent solutions of the radial Teukolsky equation for
$\bar{a}=0$, $\bar\omega=\bar\Omega(\ell)$ and $s=-2$, we must
conclude that no QNM solutions coincide with the algebraically special
solutions of Schwarzschild.

\section{Summary and Discussion}

We have developed an independent code, built as a {\em Mathematica} package,
that can solve the Teukolsky master equation for the QNMs of Kerr.
While we use the well-known continued fraction method for solving the
radial equation, we have implemented and tested a spectral method
for solving the angular equation for the spin-weighted spheroidal
function.  Using this code, we have performed an extensive,
high-accuracy survey of the gravitational QNMs of Kerr.
Figures~\ref{fig:AllOmega_m2}--\ref{fig:AllOmega_m-2} display our
results for $m\in\{-2,-1,0,1,2\}$, $\ell=2\to12$, and $n=0\to7$.  These
results are quantitatively similar to prior works, improving on prior
results only in accuracy.

As a quantitative demonstration of the accuracy and precision of our
code, we have carefully explored those QNMs that, in the extreme Kerr
limit, approach accumulation points at $\bar\omega = m/2$.  We have
fit these data to extended versions [Eqs.~(\ref{eq:fit_pos})
  and~(\ref{eq:fit_neg})] of the predicted leading order behavior of
$\bar\omega$ [Eq.~(\ref{eq:yang_omega})] as $\bar{a}\to1$.
Equations~(\ref{eq:fit_pos}) and~(\ref{eq:fit_neg}) also include
fitting functions for $\scA{-2}{\ell{m}}{\bar{a}\bar\omega}$.  The
high quality of our fitting functions is illustrated in
Figs.~\ref{fig:SupRad_l5m4_O}--\ref{fig:SupRad_l2m1_A} and in
Tables~\ref{tab:fit_pos_Omega}--\ref{tab:fit_neg_Alm}.  While our
immediate purpose in reporting these details is to help validate our
code, Eqs.~(\ref{eq:fit_pos}) and~(\ref{eq:fit_neg}) and the data
in the corresponding tables may be beneficial to anyone trying to 
find higher-order approximations of the behavior of $\bar\omega$ and
$\scA{-2}{\ell{m}}{\bar{a}\bar\omega}$ near the accumulation points.

One of the most interesting topics associated with QNMs is their
behavior at the NIA and, in particular, near the algebraically special
frequencies $\bar\Omega$.  We first explored this numerically for the
simplest case when $\ell=2$.  For this case, the $n=8$ overtones
approach $\bar\omega=-2i$ in the Schwarzschild limit, and the two
plots in Fig.~\ref{fig:NIA_l2n8} show our results.  Maassen~van~den
Brink\cite{van_den_brink-2000} predicted the behavior of the $m\le0$
(for positive-frequency) modes and suggested how the corresponding
$m>0$ modes could behave\cite{van_den_brink-2000,van_den_brink-2003}.
In contrast to the lower precision solutions considered in
Ref.~\cite{berticardoso-2003}, our numerical results for the $m\le0$
modes agree very well with Maassen~van~den Brink's formula,
Eq.~(\ref{eq:vdB_limit}).  For small $\bar{a}$, our results are fit
well by Eqs.~(\ref{eq:AS_fit_m2})--(\ref{eq:AS_fit_0}).  As previously
found in Ref.~\cite{berticardoso-2003}, we also find that the $m>0$
modes exist as a pair of overtone multiplets.  That is, there are
$n=8_0$ overtones for the $m=1$, and $2$ modes that appear to emerge
from a single limit frequency given by Eq.~(\ref{eq:AS_fit_20}), and
$n=8_1$ overtones for the same values of $m$ that appear to emerge
from a different limit frequency given by Eq.~(\ref{eq:AS_fit_21}).
In the language of Ref.~\cite{van_den_brink-2003}, these limit
frequencies could be unconventional poles on the unphysical sheet
beyond the branch cut at the NIA.  As $\bar{a}$ increases from zero,
the QNMs first appear on the physical sheet at the NIA at some finite
value of $\bar{a}$.  Table~\ref{tab:NIA_l2n8} displays our numerical
values for these modes where they emerge from the NIA.  While our
results for $m>0$ are similar to the general behavior described in
Ref.~\cite{van_den_brink-2003} neither of our limit frequencies agrees
well with their estimates.  Furthermore they only find a single limit
frequency while the numerical solutions show that there are two.

Throughout this work, we have made use of the theory of confluent Heun
functions.  It is particularly useful in clarifying how we can ensure
that a particular numerical solution will obey some desired pair of
boundary conditions.  This was, of course, done originally by
Leaver\cite{leaver-1985,leaver-1986} for QNMs without explicitly
casting the Teukolsky equations as confluent Heun equations, but
confluent Heun theory does provide a very useful framework.  And
perhaps most important, it provides a clear means for finding
polynomial solutions.  We have given only a few examples of finding
these polynomial solutions, but we believe that they may be very
useful in understanding the various perturbations of the Kerr
space-time.

We rederived the algebraically special TTMs of Kerr as examples of
the method for finding confluent Heun polynomial solutions.  Without
controversy\cite{van_den_brink-2000}, we can say that the polynomial
solution of Eq.~(\ref{eqn:Gen_AS_TTML}) represents TTM${}_{\rm L}$
modes when $\bar{a}>0$, and the polynomial solution of
Eq.~(\ref{eqn:Gen_AS_TTMR}) represents TTM${}_{\rm R}$ modes when
$\bar{a}>0$.  However, for $\bar{a}=0$ the situation is not so simple.
In the limit that $\bar{a}\to0$, Eqs.~(\ref{eqn:Gen_AS_TTML})
and~(\ref{eqn:Gen_AS_TTMR}) smoothly become, respectively,
Eqs.~(\ref{eq:TTML_a0_R_sol1}) and~(\ref{eq:TTMR_a0_R_sol1}).  These
$\bar{a}=0$ solutions still satisfy the simple definitions of ingoing
and outgoing waves at their boundaries derived in the
Appendix, but according to Maassen~van~den
Brink\cite{van_den_brink-2000} the definitions of ingoing and outgoing
waves at the event horizon becomes more complicated in the
Schwarzschild limit.  He concludes that no TTM${}_{\rm R}$ mode exists
for $\bar{a}=0$ and that Eq.~(\ref{eq:TTML_a0_R_sol1}) is {\em both} a
QNM and a TTM${}_{\rm L}$.  These conclusions are rather
counterintuitive, and his reasoning is very complicated.  As we write
this, we do not fully understand all of the subtleties in his
arguments, and we are not yet convinced that his conclusions are
correct.  Our primary concern at this time is that the definitions of
ingoing and outgoing solutions at the horizons [see
  Eqs.~(\ref{eqn:HorizonBehavior_Tortose})
  and~(\ref{eqn:HorizonBehavior})] are given in terms of
Boyer-Lindquist coordinates.  This coordinate system is, itself,
singular at the event horizon.  As Teukolsky
showed\cite{teukolsky-1973} by transforming to ingoing Kerr
coordinates and performing a spin rotation to obtain a nonsingular
null tetrad, the effect of this Boyer-Lindquist singular behavior is a
factor of $\Delta^{-s}$ in the behavior at the boundary.  Terms of
this form play an important role in Maassen~van~den Brink's arguments about
the behavior at the boundary and it is not clear to us if he has
correctly handled this subtlety.


\appendix
\section{ASYMPTOTIC BEHAVIOR}
\label{sec:Asymptotics}
In order to explore the behavior of solutions near the horizon $r_+$
and near infinity, it is useful to rewrite the radial equation
in terms of the tortoise coordinate $r^*$.  Following
Teukolsky's\cite{teukolsky-1973} notation and approach, we define
\begin{align}\label{eqn:tortoise_coord:DiffEqn}
  \frac{dr^*}{dr} &\equiv \frac{r^2+a^2}{\Delta}, \\
  Y(r^*) &\equiv \Delta^{s/2}(r^2+a^2)^{1/2}R(r),
\end{align}
where we assume $r=r(r^*)$.
In the asymptotic limit $r\to\infty$ ($r^*\to\infty$),
Eq.~(\ref{eqn:radialR:Diff_Eqn}) becomes
\begin{equation}
\frac{d^2Y(r^*)}{d{r^*}^2} + \left(\omega^2 
     + 2\frac{is\omega}{r}\right)Y(r^*) \approx 0,
\end{equation}
which has the two solutions
\begin{equation}
   Y(r^*) \sim r^{\pm{s}}e^{\mp{i}\omega{r^*}}.
\end{equation}
In terms of the original radial function, and given the sign choice in
Eq.~(\ref{eq:Teukolsky_separation_form}), we find
\begin{equation}
  \lim_{r\to\infty}R(r) \sim \left\{\begin{array}{lcl}
      \frac{e^{-i\omega{r^*}}}{r} &:& \mbox{ingoing wave}, \\
      \frac{e^{i\omega{r^*}}}{r^{2s+1}} &:& \mbox{outgoing wave}. \\
   \end{array}\right.
\end{equation}
In the asymptotic limit $r\to r_+$ ($r^*\to-\infty$),
Eq.~(\ref{eqn:radialR:Diff_Eqn}) becomes
\begin{equation}
\frac{d^2Y(r^*)}{d{r^*}^2} + \left(k 
    - is\frac{(r_+-M)}{2Mr_+}\right)^2Y(r^*) \approx 0,
\end{equation}
where $k\equiv\omega-ma/(2Mr_+)$,
which has the two solutions
\begin{equation}\label{eqn:rplusYlimit}
   Y(r^*) \sim e^{\pm{i}k{r^*}}\left[e^{\frac{(r_+-M)}{Mr_+}r^*}\right]^{\pm{s}/2}.
\end{equation}
In order to simplify the asymptotic limit, we
need to integrate Eq.~(\ref{eqn:tortoise_coord:DiffEqn}) for the
Tortoise coordinates.  The result is
\begin{equation}\label{eqn:Tortose_coord}
  r^* = r + \frac{2M}{r_+-r_\minus}\left[
    r_+\ln\left(\frac{r-r_+}{M}\right) 
    - r_\minus\ln\left(\frac{r-r_\minus}{M}\right)\right].
\end{equation}
Now, we let $r=r_++\epsilon$ giving us
\begin{equation}
   r^*(\epsilon) = r_+ + \epsilon + \frac{2M}{r_+-r_\minus}\left[
    r_+\ln\left(\frac{\epsilon}{M}\right) 
    - r_\minus\ln\left(\frac{\epsilon +r_+-r_\minus}{M}\right)\right].
\end{equation}
It is then straightforward to show
\begin{equation}
  e^{\frac{(r_+-M)}{Mr_+}r^*(\epsilon)} 
       \propto \Delta + \mathcal{O}(\epsilon^2),
\end{equation}
giving us
\begin{equation}
   Y(r^*) \sim \Delta^{\pm{s}/2}e^{\pm{i}k{r^*}}.
\end{equation}
Then, in terms of the original radial function
\begin{equation}
\label{eqn:HorizonBehavior_Tortose}
  \lim_{r^*\to-\infty}R(r^*) \sim \left\{\begin{array}{lcl}
      e^{ik{r^*}} &:& \mbox{ingoing (out of BH)}, \\
     \Delta^{-s}e^{-ik{r^*}} &:& \mbox{outgoing (into BH)}. \\
   \end{array}\right.
\end{equation}
The latter solution, representing waves flowing into the black hole is
the physically relevant condition.  Note that the unusual term of
$\Delta^{-s}$ in the physical condition is present because the
Boyer-Lindquist coordinates used in Eq.~(\ref{eq:Kerr_Metric}) are
singular at the horizon.  A change of coordinates (cf.\ 
Ref.~\cite{teukolsky-1973}) to a nonsingular coordinate system removes
this behavior (but creates a factor of $\Delta^{s}$ in the unphysical
condition).

Returning to the standard radial coordinate $r$ via
Eq.~(\ref{eqn:Tortose_coord}) gives us the behavior near spatial
infinity as
\begin{equation}
\label{eqn:InfinityBehavior}
  \lim_{r\to\infty}R(r) \sim \left\{\begin{array}{lcl}
      r^{-1-2i\omega M}e^{-i\omega r} &:& \mbox{ingoing wave}, \\
      r^{-1-2s+2i\omega M}e^{i\omega r} &:& \mbox{outgoing wave}, \\
   \end{array}\right.
\end{equation}
where we have used $\lim_{r\to\infty}e^{i\omega{r^*}}\propto
r^{2i\omega M}e^{i\omega r}$.  Near the horizon, $r=r_+$ we find
\begin{equation}
\label{eqn:HorizonBehavior}
  \lim_{r\to r_+}R(r) \sim \left\{\begin{array}{lcl}
      (r - r_+)^{i\sigma_+} &:& \mbox{ingoing (out of BH)}, \\
      (r - r_+)^{-s-i\sigma_+} &:& \mbox{outgoing (into BH)}. \\
   \end{array}\right.
\end{equation}
where we have used $\lim_{r\to r_+}e^{ik{r^*}} \propto (r-r_+)^{i\sigma_+}$.


\begin{thebibliography}{10}%
\makeatletter
\providecommand \@ifxundefined [1]{%
 \ifx #1\undefined \expandafter \@firstoftwo
 \else \expandafter \@secondoftwo
\fi
}%
\providecommand \@ifnum [1]{%
 \ifnum #1\expandafter \@firstoftwo
 \else \expandafter \@secondoftwo
\fi
}%
\providecommand \enquote [1]{``#1''}%
\providecommand \bibnamefont  [1]{#1}%
\providecommand \bibfnamefont [1]{#1}%
\providecommand \citenamefont [1]{#1}%
\providecommand\href[0]{\@sanitize\@href}%
\providecommand\@href[1]{\endgroup\@@startlink{#1}\endgroup\@@href}%
\providecommand\@@href[1]{#1\@@endlink}%
\providecommand \@sanitize [0]{\begingroup\catcode`\&12\catcode`\#12\relax}%
\@ifxundefined \pdfoutput {\@firstoftwo}{%
 \@ifnum{\z@=\pdfoutput}{\@firstoftwo}{\@secondoftwo}%
}{%
 \providecommand\@@startlink[1]{\leavevmode}%
 \providecommand\@@endlink[0]{}%
}{%
 \providecommand\@@startlink[1]{%
  \leavevmode
  \pdfstartlink
   attr{/Border[0 0 1 ]/H/I/C[0 1 1]}%
   user{/Subtype/Link/A<</Type/Action/S/URI/URI(#1)>>}%
  \relax
 }%
 \providecommand\@@endlink[0]{\pdfendlink}%
}%
\providecommand \url  [0]{\begingroup\@sanitize \@url }%
\providecommand \@url [1]{\endgroup\@href {#1}{\urlprefix}}%
\providecommand \urlprefix [0]{URL }%
\providecommand \Eprint[0]{\href }%
\@ifxundefined \urlstyle {%
  \providecommand \doi [1]{doi:\discretionary{}{}{}#1}%
}{%
  \providecommand \doi [0]{doi:\discretionary{}{}{}\begingroup
  \urlstyle{rm}\Url }%
}%
\providecommand \doibase [0]{http://dx.doi.org/}%
\providecommand \Doi[1]{\href{\doibase#1}}%
\providecommand \bibAnnote [3]{%
  \BibitemShut{#1}%
  \begin{quotation}\noindent
    \textsc{Key:}\ #2\\\textsc{Annotation:}\ #3%
  \end{quotation}%
}%
\providecommand \bibAnnoteFile [2]{%
  \IfFileExists{#2}{\bibAnnote {#1} {#2} {\input{#2}}}{}%
}%
\providecommand \typeout [0]{\immediate \write \m@ne }%
\providecommand \selectlanguage [0]{\@gobble}%
\providecommand \bibinfo [0]{\@secondoftwo}%
\providecommand \bibfield [0]{\@secondoftwo}%
\providecommand \translation [1]{[#1]}%
\providecommand \BibitemOpen[0]{}%
\providecommand \bibitemStop [0]{}%
\providecommand \bibitemNoStop [0]{.\EOS\space}%
\providecommand \EOS [0]{\spacefactor3000\relax}%
\providecommand \BibitemShut [1]{\csname bibitem#1\endcsname}%
\bibitem{berti-QNM-2009}%
  \BibitemOpen
  \bibfield{author}{%
  \bibinfo {author} {\bibfnamefont{E.}~\bibnamefont{Berti}}, \bibinfo {author}
  {\bibfnamefont{V.}~\bibnamefont{Cardoso}},\ and\ \bibinfo {author}
  {\bibfnamefont{A.~O.}\ \bibnamefont{Starinets}},\ }%
  \bibfield{journal}{%
  \bibinfo {journal} {Classical Quantum Gravity}\ }%
  \textbf{\bibinfo {volume} {26}},\ \bibinfo {pages} {163001} (\bibinfo
  {year} {2009})%
  \bibAnnoteFile{NoStop}{berti-QNM-2009}%
\bibitem{nollert-qnm-1999}%
  \BibitemOpen
  \bibfield{author}{%
  \bibinfo {author} {\bibfnamefont{H.-P.}\ \bibnamefont{Nollert}},\ }%
  \bibfield{journal}{%
  \bibinfo {journal} {Classical Quantum Gravity}\ }%
  \textbf{\bibinfo {volume} {16}},\ \bibinfo {pages} {R159} (\bibinfo {year}
  {1999})%
  \bibAnnoteFile{NoStop}{nollert-qnm-1999}%
\bibitem{chandra75}%
  \BibitemOpen
  \bibfield{author}{%
  \bibinfo {author} {\bibfnamefont{S.}~\bibnamefont{Chandrasekhar}}\ and\
  \bibinfo {author} {\bibfnamefont{S.}~\bibnamefont{Detweiler}},\ }%
  \bibfield{journal}{%
  \bibinfo {journal} {Proc. R. Soc. A}\ }%
  \textbf{\bibinfo {volume} {344}},\ \bibinfo {pages} {441} (\bibinfo {year}
  {1975})%
  \bibAnnoteFile{NoStop}{chandra75}%
\bibitem{piranstark86}%
  \BibitemOpen
  \bibfield{author}{%
  \bibinfo {author} {\bibfnamefont{T.}~\bibnamefont{Piran}}\ and\ \bibinfo
  {author} {\bibfnamefont{R.~F.}\ \bibnamefont{Stark}},\ }%
  in\ \emph{\bibinfo {booktitle} {Dynamical Spacetimes and Numerical
  Relativity}},\ \bibinfo {editor} {edited by\ \bibinfo {editor}
  {\bibfnamefont{J.~M.}\ \bibnamefont{Centrella}}}\ (\bibinfo {publisher}
  {Cambridge University Press},\ \bibinfo {address} {Cambridge, England},\
  \bibinfo {year} {1986})\ pp.\ \bibinfo {pages} {40--73}%
  \bibAnnoteFile{NoStop}{piranstark86}%
\bibitem{leaver-1985}%
  \BibitemOpen
  \bibfield{author}{%
  \bibinfo {author} {\bibfnamefont{E.~W.}\ \bibnamefont{Leaver}},\ }%
  \bibfield{journal}{%
  \bibinfo {journal} {Proc. R. Soc. A}\ }%
  \textbf{\bibinfo {volume} {402}},\ \bibinfo {pages} {285} (\bibinfo 
  {year} {1985})%
  \bibAnnoteFile{NoStop}{leaver-1985}%
\bibitem{leaver-1986}%
  \BibitemOpen
  \bibfield{author}{%
  \bibinfo {author} {\bibfnamefont{E.~W.}\ \bibnamefont{Leaver}},\ }%
  \bibfield{journal}{%
  \bibinfo {journal} {J.~Math. Phys. (N.Y.)}\ }%
  \textbf{\bibinfo {volume} {27}},\ \bibinfo {pages} {1238} (\bibinfo 
  {year} {1986})%
  \bibAnnoteFile{NoStop}{leaver-1986}%
\bibitem{nollert-1993}%
  \BibitemOpen
  \bibfield{author}{%
  \bibinfo {author} {\bibfnamefont{H.-P.}\ \bibnamefont{Nollert}},\ }%
  \bibfield{journal}{%
  \bibinfo {journal} {Phys. Rev. D}\ }%
  \textbf{\bibinfo {volume} {47}},\ \bibinfo {pages} {5253} (\bibinfo 
  {year} {1993})%
  \bibAnnoteFile{NoStop}{nollert-1993}%
\bibitem{onozawa-1997}%
  \BibitemOpen
  \bibfield{author}{%
  \bibinfo {author} {\bibfnamefont{H.}~\bibnamefont{Onozawa}},\ }%
  \bibfield{journal}{%
  \bibinfo {journal} {Phys. Rev. D}\ }%
  \textbf{\bibinfo {volume} {55}},\ \bibinfo {pages} {3593} (\bibinfo 
  {year} {1997})%
  \bibAnnoteFile{NoStop}{onozawa-1997}%
\bibitem{Mashhoon-1983}%
  \BibitemOpen
  \bibfield{author}{%
  \bibinfo {author} {\bibfnamefont{B.}~\bibnamefont{Mashhoon}},\ }%
  in\ \emph{\bibinfo {booktitle} {Proceedings of the Third Marcel Grossmann
  Meeting on Recent Developments of General Relativity, Shanghai, 1982}},\
  \bibinfo {editor} {edited by\ \bibinfo {editor}
  {\bibfnamefont{H.}~\bibnamefont{Ning}}}\ (\bibinfo {publisher}
  {North-Holland, Amsterdam},\ \bibinfo {year} {1983})%
  \bibAnnoteFile{NoStop}{Mashhoon-1983}%
\bibitem{ValeriaMashhoon-1084}%
  \BibitemOpen
  \bibfield{author}{%
  \bibinfo {author} {\bibfnamefont{V.}~\bibnamefont{Ferrari}}\ and\ \bibinfo
  {author} {\bibfnamefont{B.}~\bibnamefont{Mashhoon}},\ }%
  \bibfield{journal}{%
  \Doi{10.1103/PhysRevD.30.295}{\bibinfo {journal} {Phys. Rev. D}}\ }%
  \textbf{\bibinfo {volume} {30}},\ \bibinfo {pages} {295} (\bibinfo 
  {year} {1984})%
  \bibAnnoteFile{NoStop}{ValeriaMashhoon-1084}%
\bibitem{SchutzWill-1985}%
  \BibitemOpen
  \bibfield{author}{%
  \bibinfo {author} {\bibfnamefont{B.~F.}\ \bibnamefont{Schutz}}\ and\ \bibinfo
  {author} {\bibfnamefont{C.~M.}\ \bibnamefont{Will}},\ }%
  \bibfield{journal}{%
  \Doi{10.1086/184453}{\bibinfo {journal} {Astrophys. J. Lett.}}\ }%
  \textbf{\bibinfo {volume} {291}},\ \bibinfo {pages} {L33} (\bibinfo 
  {year} {1985})%
  \bibAnnoteFile{NoStop}{SchutzWill-1985}%
\bibitem{Yang-et-al-2012}%
  \BibitemOpen
  \bibfield{author}{%
  \bibinfo {author} {\bibfnamefont{H.}~\bibnamefont{Yang}}, \bibinfo {author}
  {\bibfnamefont{D.~A.}\ \bibnamefont{Nichols}}, \bibinfo {author}
  {\bibfnamefont{F.}~\bibnamefont{Zhang}}, \bibinfo {author}
  {\bibfnamefont{A.}~\bibnamefont{Zimmerman}}, \bibinfo {author}
  {\bibfnamefont{Z.}~\bibnamefont{Zhang}},\ and\ \bibinfo {author}
  {\bibfnamefont{Y.}~\bibnamefont{Chen}},\ }%
  \bibfield{journal}{%
  \Doi{10.1103/PhysRevD.86.104006}{\bibinfo {journal} {Phys. Rev. D}}\ }%
  \textbf{\bibinfo {volume} {86}},\ \bibinfo {pages} {104006} (\bibinfo 
  {year} {2012})%
  \bibAnnoteFile{NoStop}{Yang-et-al-2012}%
\bibitem{Dolan-2010}%
  \BibitemOpen
  \bibfield{author}{%
  \bibinfo {author} {\bibfnamefont{S.~R.}\ \bibnamefont{Dolan}},\ }%
  \bibfield{journal}{%
  \Doi{10.1103/PhysRevD.82.104003}{\bibinfo {journal} {Phys. Rev. D}}\ }%
  \textbf{\bibinfo {volume} {82}},\ \bibinfo {pages} {104003} (\bibinfo 
  {year} {2010})%
  \bibAnnoteFile{NoStop}{Dolan-2010}%
\bibitem{Hod-2012}%
  \BibitemOpen
  \bibfield{author}{%
  \bibinfo {author} {\bibfnamefont{S.}~\bibnamefont{Hod}},\ }%
  \bibfield{journal}{%
  \bibinfo {journal} {Phys. Lett. B}\ }%
  \textbf{\bibinfo {volume} {715}},\ \bibinfo {pages} {348} (\bibinfo {year}
  {2012})%
  \bibAnnoteFile{NoStop}{Hod-2012}%
\bibitem{teukolsky-1973}%
  \BibitemOpen
  \bibfield{author}{%
  \bibinfo {author} {\bibfnamefont{S.~A.}\ \bibnamefont{Teukolsky}},\ }%
  \bibfield{journal}{%
  \bibinfo {journal} {Astrophys. J.}\ }%
  \textbf{\bibinfo {volume} {185}},\ \bibinfo {pages} {635} (\bibinfo 
  {year} {1973})%
  \bibAnnoteFile{NoStop}{teukolsky-1973}%
\bibitem{detweiler80}%
  \BibitemOpen
  \bibfield{author}{%
  \bibinfo {author} {\bibfnamefont{S.}~\bibnamefont{Detweiler}},\ }%
  \bibfield{journal}{%
  \bibinfo {journal} {Astrophys. J.}\ }%
  \textbf{\bibinfo {volume} {239}},\ \bibinfo {pages} {292} (\bibinfo 
  {year} {1980})%
  \bibAnnoteFile{NoStop}{detweiler80}%
\bibitem{cardoso04}%
  \BibitemOpen
  \bibfield{author}{%
  \bibinfo {author} {\bibfnamefont{V.}~\bibnamefont{Cardoso}},\ }%
  \bibfield{journal}{%
  \bibinfo {journal} {Phys. Rev. D}\ }%
  \textbf{\bibinfo {volume} {70}},\ \bibinfo {pages} {127502} (\bibinfo
  {year} {2004})%
  \bibAnnoteFile{NoStop}{cardoso04}%
\bibitem{Yang-et-al-2013a}%
  \BibitemOpen
  \bibfield{author}{%
  \bibinfo {author} {\bibfnamefont{H.}~\bibnamefont{Yang}}, \bibinfo {author}
  {\bibfnamefont{F.}~\bibnamefont{Zhang}}, \bibinfo {author}
  {\bibfnamefont{A.}~\bibnamefont{Zimmerman}}, \bibinfo {author}
  {\bibfnamefont{D.~A.}\ \bibnamefont{Nichols}}, \bibinfo {author}
  {\bibfnamefont{E.}~\bibnamefont{Berti}},\ and\ \bibinfo {author}
  {\bibfnamefont{Y.}~\bibnamefont{Chen}},\ }%
  \bibfield{journal}{%
  \bibinfo {journal} {Phys. Rev. D}\ }%
  \textbf{\bibinfo {volume} {87}},\ \bibinfo {pages} {041502(R)} (\bibinfo
  {year} {2013})%
  \bibAnnoteFile{NoStop}{Yang-et-al-2013a}%
\bibitem{Yang-et-al-2013b}%
  \BibitemOpen
  \bibfield{author}{%
  \bibinfo {author} {\bibfnamefont{H.}~\bibnamefont{Yang}}, \bibinfo {author}
  {\bibfnamefont{A.}~\bibnamefont{Zimmerman}}, \bibinfo {author}
  {\bibfnamefont{A.}~\bibnamefont{Zengino\u{g}lu}}, \bibinfo {author}
  {\bibfnamefont{F.}~\bibnamefont{Zhang}}, \bibinfo {author}
  {\bibfnamefont{E.}~\bibnamefont{Berti}},\ and\ \bibinfo {author}
  {\bibfnamefont{Y.}~\bibnamefont{Chen}},\ }%
  \bibfield{journal}{%
  \bibinfo {journal} {Phys. Rev. D}\ }%
  \textbf{\bibinfo {volume} {88}},\ \bibinfo {pages} {044047} (\bibinfo
  {year} {2013})%
  \bibAnnoteFile{NoStop}{Yang-et-al-2013b}%
\bibitem{chandra-1984}%
  \BibitemOpen
  \bibfield{author}{%
  \bibinfo {author} {\bibfnamefont{S.}~\bibnamefont{Chandrasekhar}},\ }%
  \bibfield{journal}{%
  \bibinfo {journal} {Proc. R. Soc. A}\ }%
  \textbf{\bibinfo {volume} {392}},\ \bibinfo {pages} {1} (\bibinfo 
  {year} {1984})%
  \bibAnnoteFile{NoStop}{chandra-1984}%
\bibitem{andersson-1994}%
  \BibitemOpen
  \bibfield{author}{%
  \bibinfo {author} {\bibfnamefont{N.}~\bibnamefont{Andersson}},\ }%
  \bibfield{journal}{%
  \bibinfo {journal} {Classical Quantum Gravity}\ }%
  \textbf{\bibinfo {volume} {11}},\ \bibinfo {pages} {L39} (\bibinfo {year}
  {1994})%
  \bibAnnoteFile{NoStop}{andersson-1994}%
\bibitem{van_den_brink-2000}%
  \BibitemOpen
  \bibfield{author}{%
  \bibinfo {author} {\bibfnamefont{A.}~\bibnamefont{Maassen van~den Brink}},\
  }%
  \bibfield{journal}{%
  \bibinfo {journal} {Phys. Rev. D}\ }%
  \textbf{\bibinfo {volume} {62}},\ \bibinfo {pages} {064009} (\bibinfo
  {year} {2000})%
  \bibAnnoteFile{NoStop}{van_den_brink-2000}%
\bibitem{Fiziev-2010}%
  \BibitemOpen
  \bibfield{author}{%
  \bibinfo {author} {\bibfnamefont{P.~P.}\ \bibnamefont{Fiziev}},\ }%
  \bibfield{journal}{%
  \bibinfo {journal} {Classical Quantum Gravity}\ }%
  \textbf{\bibinfo {volume} {27}},\ \bibinfo {pages} {135001} (\bibinfo {year}
  {2010})%
  \bibAnnoteFile{NoStop}{Fiziev-2010}%
\bibitem{Heun-eqn}%
  \BibitemOpen
  \emph{\bibinfo {title} {Heun's Differential Equations}},\ edited by\ \bibinfo
  {editor} {\bibfnamefont{A.}~\bibnamefont{Ronveaux}}\ (\bibinfo {publisher}
  {Oxford University, New York},\ \bibinfo {year} {1995})%
  \bibAnnoteFile{NoStop}{Heun-eqn}%
\bibitem{Fiziev-2009b}%
  \BibitemOpen
  \bibfield{author}{%
  \bibinfo {author} {\bibfnamefont{R.~S.}\ \bibnamefont{Borissov}}\ and\
  \bibinfo {author} {\bibfnamefont{P.~P.}\ \bibnamefont{Fiziev}},\ }%
  \bibfield{journal}{%
  \bibinfo {journal} {Bulg. J. Phys.}\ }%
  \textbf{\bibinfo {volume} {37}},\ \bibinfo {pages} {65} (\bibinfo {year}
  {2010})%
  \bibAnnoteFile{NoStop}{Fiziev-2009b}%
\bibitem{hughes-2000}%
  \BibitemOpen
  \bibfield{author}{%
  \bibinfo {author} {\bibfnamefont{S.~A.}\ \bibnamefont{Hughes}},\ }%
  \bibfield{journal}{%
  \Doi{10.1103/PhysRevD.61.084004}{\bibinfo {journal} {Phys. Rev. D}}\ }%
  \textbf{\bibinfo {volume} {61}},\ \bibinfo {pages} {084004} (\bibinfo 
  {year} {2000})%
  \bibAnnoteFile{NoStop}{hughes-2000}%
\bibitem{Hod08}%
  \BibitemOpen
  \bibfield{author}{%
  \bibinfo {author} {\bibfnamefont{S.}~\bibnamefont{Hod}},\ }%
  \bibfield{journal}{%
  \bibinfo {journal} {Phys. Rev. D}\ }%
  \textbf{\bibinfo {volume} {78}},\ \bibinfo {pages} {084035} (\bibinfo
  {year} {2008})%
  \bibAnnoteFile{NoStop}{Hod08}%
\bibitem{berticardoso-2003}%
  \BibitemOpen
  \bibfield{author}{%
  \bibinfo {author} {\bibfnamefont{E.}~\bibnamefont{Berti}}, \bibinfo {author}
  {\bibfnamefont{V.}~\bibnamefont{Cardoso}}, \bibinfo {author}
  {\bibfnamefont{K.~D.}\ \bibnamefont{Kokkotas}},\ and\ \bibinfo {author}
  {\bibfnamefont{H.}~\bibnamefont{Onozawa}},\ }%
  \bibfield{journal}{%
  \bibinfo {journal} {Phys. Rev. D}\ }%
  \textbf{\bibinfo {volume} {68}},\ \bibinfo {pages} {124018} (\bibinfo
  {year} {2003})%
  \bibAnnoteFile{NoStop}{berticardoso-2003}%
\bibitem{Fiziev-2009c}%
  \BibitemOpen
  \bibfield{author}{%
  \bibinfo {author} {\bibfnamefont{P.~P.}\ \bibnamefont{Fiziev}},\ }%
  \bibfield{journal}{%
  \Doi{10.1103/PhysRevD.80.124001}{\bibinfo {journal} {Phys. Rev. D}}\ }%
  \textbf{\bibinfo {volume} {80}},\ \bibinfo {pages} {124001} (\bibinfo 
  {year} {2009})%
  \bibAnnoteFile{NoStop}{Fiziev-2009c}%
\bibitem{Gautschi-1967}%
  \BibitemOpen
  \bibfield{author}{%
  \bibinfo {author} {\bibfnamefont{W.}~\bibnamefont{Gautschi}},\ }%
  \bibfield{journal}{%
  \bibinfo {journal} {SIAM Rev.}\ }%
  \textbf{\bibinfo {volume} {9}},\ \bibinfo {pages} {24} (\bibinfo 
  {year} {1967})%
  \bibAnnoteFile{NoStop}{Gautschi-1967}%
\bibitem{numrec_c}%
  \BibitemOpen
  \bibfield{author}{%
  \bibinfo {author} {\bibfnamefont{W.~H.}\ \bibnamefont{Press}}, \bibinfo
  {author} {\bibfnamefont{S.~A.}\ \bibnamefont{Teukolsky}}, \bibinfo {author}
  {\bibfnamefont{W.~T.}\ \bibnamefont{Wetterling}},\ and\ \bibinfo {author}
  {\bibfnamefont{B.~P.}\ \bibnamefont{Flannery}},\ }%
  \emph{\bibinfo {title} {Numerical Recipes in C}},\ \bibinfo {edition} {2nd}\
  ed.\ (\bibinfo {publisher} {Cambridge University Press},\ \bibinfo {address}
  {Cambridge, England},\ \bibinfo {year} {1992})%
  \bibAnnoteFile{NoStop}{numrec_c}%
\bibitem{sasakinakamura-1982}%
  \BibitemOpen
  \bibfield{author}{%
  \bibinfo {author} {\bibfnamefont{M.}~\bibnamefont{Sasaki}}\ and\ \bibinfo
  {author} {\bibfnamefont{T.}~\bibnamefont{Nakamura}},\ }%
  \bibfield{journal}{%
  \bibinfo {journal} {Prog. Theor. Phys.}\ }%
  \textbf{\bibinfo {volume} {67}},\ \bibinfo {pages} {1788} (\bibinfo 
  {year} {1982})%
  \bibAnnoteFile{NoStop}{sasakinakamura-1982}%
\bibitem{fackerell-1976}%
  \BibitemOpen
  \bibfield{author}{%
  \bibinfo {author} {\bibfnamefont{E.~D.}\ \bibnamefont{Fackerell}}\ and\
  \bibinfo {author} {\bibfnamefont{R.~G.}\ \bibnamefont{Crossman}},\ }%
  \bibfield{journal}{%
  \bibinfo {journal} {J.~Math. Phys. (N.Y.)}\ }%
  \textbf{\bibinfo {volume} {18}},\ \bibinfo {pages} {1849} (\bibinfo 
  {year} {1977})%
  \bibAnnoteFile{NoStop}{fackerell-1976}%
\bibitem{blanco-1997}%
  \BibitemOpen
  \bibfield{author}{%
  \bibinfo {author} {\bibfnamefont{M.~A.}\ \bibnamefont{Blanco}}, \bibinfo
  {author} {\bibfnamefont{M.}~\bibnamefont{Fl{\'o}rez}},\ and\ \bibinfo
  {author} {\bibfnamefont{M.}~\bibnamefont{Bermejo}},\ }%
  \bibfield{journal}{%
  \bibinfo {journal} {J. of Mol. Struct.: THEOCHEM}\ }%
  \textbf{\bibinfo {volume} {419}},\ \bibinfo {pages} {19} (\bibinfo {year}
  {1997})%
  \bibAnnoteFile{NoStop}{blanco-1997}%
\bibitem{Glampedakis01}%
  \BibitemOpen
  \bibfield{author}{%
  \bibinfo {author} {\bibfnamefont{K.}~\bibnamefont{Glampedakis}}\ and\
  \bibinfo {author} {\bibfnamefont{N.}~\bibnamefont{Andersson}},\ }%
  \bibfield{journal}{%
  \bibinfo {journal} {Phys. Rev. D}\ }%
  \textbf{\bibinfo {volume} {64}},\ \bibinfo {pages} {104021} (\bibinfo
  {year} {2001})%
  \bibAnnoteFile{NoStop}{Glampedakis01}%
\bibitem{couch-newman-1973}%
  \BibitemOpen
  \bibfield{author}{%
  \bibinfo {author} {\bibfnamefont{W.~E.}\ \bibnamefont{Couch}}\ and\ \bibinfo
  {author} {\bibfnamefont{E.~T.}\ \bibnamefont{Newman}},\ }%
  \bibfield{journal}{%
  \bibinfo {journal} {J.~Math. Phys. (N.Y.)}\ }%
  \textbf{\bibinfo {volume} {14}},\ \bibinfo {pages} {285} (\bibinfo 
  {year} {1973})%
  \bibAnnoteFile{NoStop}{couch-newman-1973}%
\bibitem{wald-1973}%
  \BibitemOpen
  \bibfield{author}{%
  \bibinfo {author} {\bibfnamefont{R.~M.}\ \bibnamefont{Wald}},\ }%
  \bibfield{journal}{%
  \bibinfo {journal} {J.~Math. Phys. (N.Y.)}\ }%
  \textbf{\bibinfo {volume} {14}},\ \bibinfo {pages} {1453} (\bibinfo 
  {year} {1973})%
  \bibAnnoteFile{NoStop}{wald-1973}%
\bibitem{van_den_brink-2003}%
  \BibitemOpen
  \bibfield{author}{%
  \bibinfo {author} {\bibfnamefont{P.~T.}\ \bibnamefont{Leung}}, \bibinfo
  {author} {\bibfnamefont{A.}~\bibnamefont{Maassen van~den Brink}}, \bibinfo
  {author} {\bibfnamefont{K.~W.}\ \bibnamefont{Mak}},\ and\ \bibinfo {author}
  {\bibfnamefont{K.}~\bibnamefont{Young}},\ }%
  \bibfield{journal}{%
  \bibinfo {journal} {Classical Quantum Gravity}\ }%
  \textbf{\bibinfo {volume} {20}},\ \bibinfo {pages} {L217} (\bibinfo 
  {year} {2003})%
  \bibAnnoteFile{NoStop}{van_den_brink-2003}%
\end{thebibliography}
\end{document}